\documentclass[14pt]{article}

\usepackage{times}
\usepackage{arxiv}

\usepackage[utf8]{inputenc} 
\usepackage[T1]{fontenc}    
\usepackage{hyperref}       
\usepackage{url}            
\usepackage{booktabs}       
\usepackage{amsfonts}       
\usepackage{nicefrac}       
\usepackage{lipsum}		
\usepackage{graphicx}
\usepackage{doi}
\usepackage{listings}

\usepackage[sort]{natbib}

\usepackage{xcolor}
\usepackage{amsmath}

\makeatletter
\newcommand*{\rom}[1]{\expandafter\@slowromancap\romannumeral #1@}

\usepackage{xargs}
\usepackage{geometry}
\usepackage{fancyhdr} 
\usepackage{lastpage} 
\usepackage{extramarks} 
\usepackage[most]{tcolorbox} 
\usepackage{graphicx} 
\usepackage{xcolor} 

\usepackage[export]{adjustbox}
\usepackage[parfill]{parskip}
\usepackage{amssymb}
\newlength{\minuslength}
\usepackage[ruled,vlined]{algorithm2e}
\usepackage{tikz}
\newcommand\norm[1]{\left\lVert#1\right\rVert}
\SetKwRepeat{Do}{do}{while}
\usepackage{caption} 
\usepackage{subcaption}
\usepackage{multirow}
\usepackage{setspace}
\setcitestyle{numbers}
\setcitestyle{square}
\setcitestyle{citesep={,}}

\title{INLA$^+$ - Approximate Bayesian inference for non-sparse models using HPC}

\author{ \href{https://orcid.org/0000-0003-1587-3288}{\includegraphics[scale=0.06]{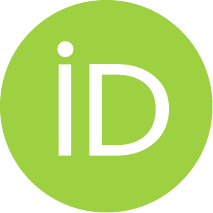}\hspace{1mm}Esmail Abdul-Fattah}\thanks{Corresponding Author} \\
	CEMSE Division\\
	King Abdullah University of Science and Technology\\
	Thuwal, 23955, Makkah\\
	\texttt{esmail.abdulfattah@kaust.edu.sa} \\
        \And
	\href{https://orcid.org/0000-0002-4334-2057}Janet van Niekerk \\
	CEMSE Division\\
	King Abdullah University of Science and Technology\\
	Thuwal, 23955, Makkah\\
	\texttt{janet.vanNiekerk@kaust.edu.sa}
        \And
	\href{https://orcid.org/0000-0002-0222-1881}H{\aa}vard Rue \\
	CEMSE Division\\
	King Abdullah University of Science and Technology\\
	Thuwal, 23955, Makkah\\
	\texttt{haavard.rue@kaust.edu.sa} \\
}

\renewcommand{\shorttitle}{Approximate Bayesian inference based on non-sparse matrices}

\hypersetup{
pdftitle={A template for the arxiv style},
pdfsubject={q-bio.NC, q-bio.QM},
pdfauthor={David S.~Hippocampus, Elias D.~Striatum},
pdfkeywords={First keyword, Second keyword, More},
}

\begin{document}

\maketitle

\begin{abstract}

The integrated nested Laplace approximations (INLA) method  has become a widely utilized tool for researchers and practitioners seeking to perform approximate Bayesian inference across various fields of application. To address the growing demand for incorporating more complex models and enhancing the method's capabilities, this paper introduces a novel framework that leverages dense matrices for performing approximate Bayesian inference based on INLA across multiple computing nodes using HPC. When dealing with non-sparse precision or covariance matrices, this new approach scales better compared to the current INLA method, capitalizing on the computational power offered by multiprocessors in shared and distributed memory architectures available in contemporary computing resources and specialized dense matrix algebra. To validate the efficacy of this approach, we conduct a simulation study then apply it to analyze cancer mortality data in Spain, employing a three-way spatio-temporal interaction model.

\end{abstract}


\section{Introduction}

Larger-scale Bayesian inference problems are becoming more popular due to the complexity of statistical models or the abundance of data. High-dimensional integrals are the main computational obstacle in Bayesian analysis. Various inference architectures including exact
(analytical or sampling-based) or approximate inferential methods, like the integrated nested Laplace approximation (INLA) \citep{Rue2009ApproximateBI}, have been proposed. INLA has become a state-of-the-art method to perform approximate Bayesian inference for Latent Gaussian Models (LGMs). Since 2009, INLA has been widely used in many applications due to its efficiency, speed, and user-friendly interface.
Latent Gaussian models are generalized additive mixed models (GAMMs) where the latent field is assigned a Gaussian prior and the data are assumed conditionally independent given the parameter space (latent field $\pmb x$ and hyperparameters $\pmb \theta$). For data $\pmb y$ with data generating model $f(.)$, covariates $\pmb z$ and linear predictor $\pmb \eta$, an LGM is defined as follows:

\begin{eqnarray*}
    y_i|\pmb x,\pmb\theta &\sim& f(\eta_i, \pmb\theta),\quad
    \pmb\eta = \pmb A \pmb x\\
    \pmb x |\pmb\theta &\sim& \mathcal{N}(\pmb 0, \pmb Q_{\text{prior}}^{-1}(\pmb\theta))\\
    \pmb\theta &\sim& \pi(\pmb\theta),
\end{eqnarray*}
where $\pi(.)$ can assume any form and $\pmb A$ is the design matrix based on covariates $\pmb z$. In most cases $\pmb Q_{\text{prior}}(\pmb\theta)$ is naturally a sparse matrix. The computational efficiency of the INLA method stems from performing Bayesian inference based on the sparse precision matrices of LGMs. Due to the additive nature of the linear predictor of an LGM, the resulting precision matrix of the model is sparse even if the model considers spline effects, temporal effects, spatial effects, survival models with frailties and many more. The abundance of statistical models that can be classified as LGMs has been a driving force in the success of INLA as a feasible alternative for Bayesian inference.  

\begin{figure}[ht]
\centering
\begin{subfigure}{0.5\textwidth}
  \centering
  \includegraphics[scale=0.2]{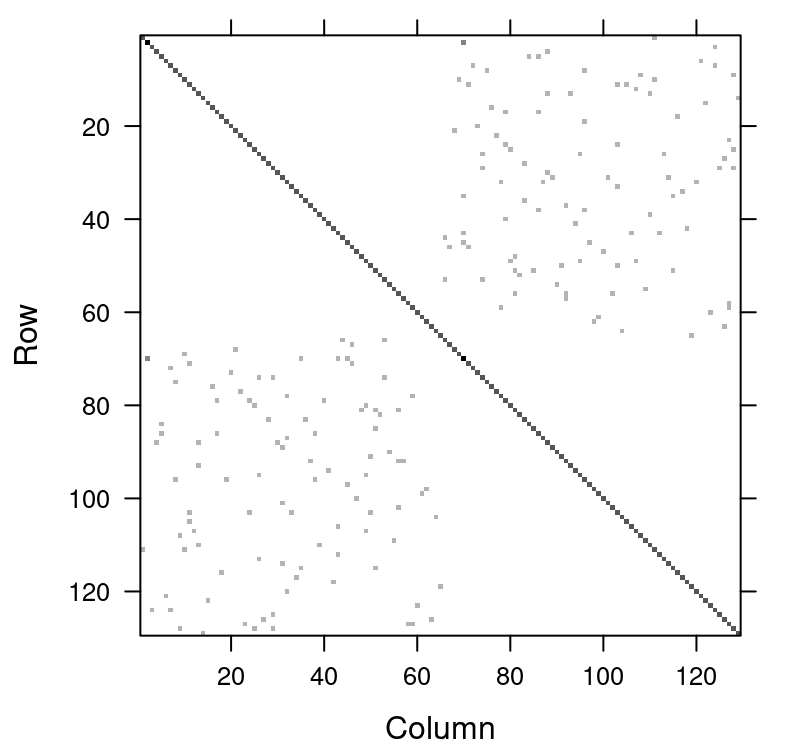}
\end{subfigure}%
\begin{subfigure}{.5\textwidth}
  \centering
  \includegraphics[scale=0.2]{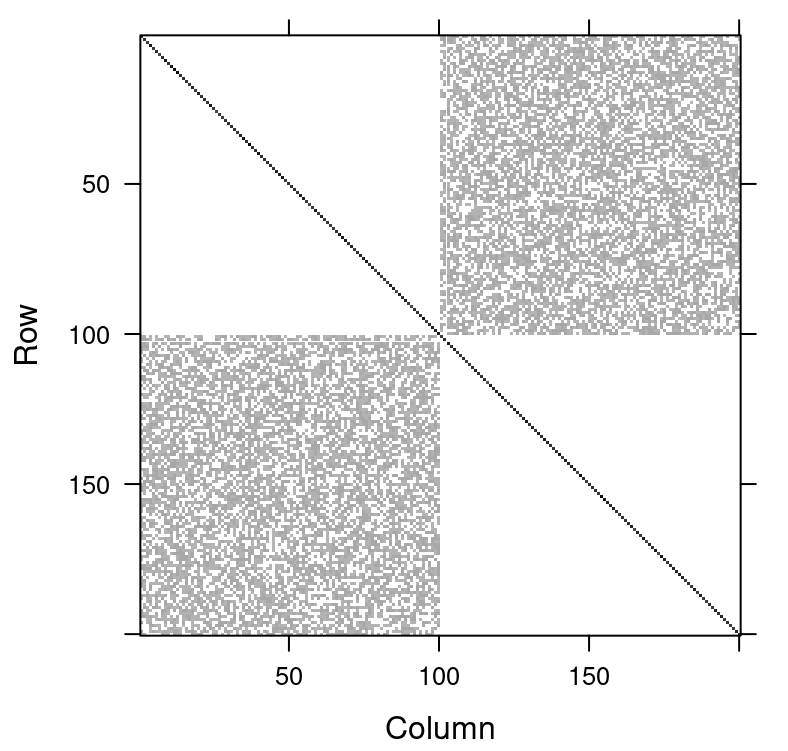}
\end{subfigure}
\caption{Visual depiction of the posterior precision matrix of the latent field for two different models. The patterns describes the non-zero terms computed from the second-order partial derivatives of the likelihood function with respect to the latent field.}
\label{fig:prec.patterns}
\end{figure}

Nevertheless, the current implementation of INLA is not designed to efficiently accommodate models with dense precision matrices, even when they arise from relatively simple models. The success of INLA is rooted in the assumption of sparsity, which arises from the conditional independence property of the Gaussian Markov random field (GMRF) used in the methodology. This property allows for efficient computations by exploiting the sparsity pattern of the precision matrix and sparse matrix algebra. However, this assumption breaks down when the Hessian of the likelihood function, denoted as $\pi(\pmb y|\pmb x)$, with respect to the latent field $\pmb x$ is not sparse, indicating that there are strong dependencies or interactions between the latent field components.\\
Consider a simple model where the Gaussian observations $\pmb y$ of size $n = 100$ depends on two latent variables, $\pmb \alpha$ and $\pmb \gamma$, along with an error term $\pmb \epsilon$ that follows a Gaussian distribution of precision $\tau$ (say 1),
\begin{equation}
    y_{k} = \alpha_{i} + \gamma_{j} + \epsilon_{k}, \quad \alpha_i \sim \mathcal{N}(0, 1), \quad \gamma_j \sim \mathcal{N}(0, 1), \quad \epsilon_{k} \sim \mathcal{N}(0, \tau^{-1}), \quad k = 1, \ldots, n
\end{equation}
where $i$ and $j$, each with $n$ elements, are drawn with replacement from the indices 1 to $m \leq n$. The size $s$ of the latent field $\pmb x = (\pmb \alpha, \pmb \gamma)$ is determined by the sum of the unique elements of $i$ and $j$. The log posterior distribution of the latent field is
\begin{equation}
\begin{split}
        \log \pi(\pmb x| \pmb y) &\approx \text{constant} + \log \pi(\pmb y|\pmb x) + \log \pi(\pmb x) \\
        &\approx \text{constant} - \dfrac{\tau}{2} (\pmb y - \pmb A \pmb x)^T(\pmb y - \pmb A \pmb x) - \dfrac{1}{2} \pmb x^T \pmb x,
\end{split}
\end{equation}
where $\pmb A$ is a design matrix. We compute the posterior mean $\pmb x^*$ by solving the system of equations:
\begin{equation}
    \pmb x^* = {\pmb Q^{*}}^{-1} \pmb A^T \pmb y = (\pmb A^T \pmb A + \tau \pmb I_{s})^{-1} \pmb A^T \pmb y,
    \label{system}
\end{equation}
where $\pmb I_{s}$ is identity matrix of size $s$. We present a visual depiction of the posterior precision matrix $\pmb Q^*$ in Figure \ref{fig:prec.patterns} for $m = 100$ (left sub-figure) and $m = 10$ (right sub-figure). This graphical representation provides insights into the sparsity pattern of the posterior precision matrix $\pmb Q^*$, and highlights that dense solvers are necessary for solving the system of equations described in Equation \ref{system} for $m = 10$. The dense pattern comes from the term $\pmb A^T \pmb A$ that represents minus the second-order partial derivatives of the likelihood function with respect to the latent field. It is not appropriate to perform Bayesian inference assuming that the precision matrices are sparse and they are not, like the case we have in this example.

Currently, some examples of LGMs with dense matrices are crossed random effects models \citep{papaspiliopoulos2020scalable} and disease mapping models with space-time interactions \citep{goicoa2018spatio}, amongst others. Crossed effects models simultaneously relate the response variable to multiple categorical predictors. These predictors, representing various factors or grouping variables that intersect, necessitate the use of dense matrices. In a related context, covariance matrices often display dense patterns and offer another avenue for performing Bayesian inference with dense matrices. Disease mapping models with space-time interactions have precision matrices that are highly rank-deficient, resulting in identifiability issues, for which some constraints on the latent field is necessary. Conversely, we can use the Moore-Penrose inverse of the rank-deficient matrix to circumvent the constraints, but this results in a dense covariance matrix.

In this paper we introduce a framework that is based, in part, on the INLA methodology, but optimized for LGMs with dense matrices. We call the new approach INLA$^+$ and present the details in Section \ref{sec:prop}. This proposal allows the use of more complex likelihoods and generalized precision matrices \citep{Stringer2020ApproximateBI}. Moreover, we implement INLA$^+$, in a high-performance computing (HPC) environment using the Open Multi-Processing (OpenMP) \citep{Chapman2007UsingO} and Message Passing Interface (MPI) \citep{Pacheco1996ParallelPW} frameworks to utilize multiprocessors on shared and distributed memory \citep{Shalf2020TheFO}.  There are several advantages of designing this new framework, of which we highlight a few:

\begin{enumerate}
    \item Handling complex relationships: dense matrices can capture intricate and complex relationships among variables. By accommodating dense matrices, we enable the modeling of sophisticated dependency structures among the latent field components. 
    \item Scalability: as the size of the data or the complexity of the problem increases, a single node may not be sufficient to handle the workload. Implementing the code on multiple nodes allows for the workload to be distributed among the nodes, enabling the processing of larger datasets or more complex problems.
    \item Performance: implementing the code on multiple nodes can also improve performance, as the workload is distributed among the nodes, enabling parallel processing. This can result in faster processing times, especially for compute-intensive tasks.
    \item Flexibility: running the new framework on multiple nodes provides greater flexibility for users. Researchers can choose the number of nodes to use based on their specific needs and the available resources. This enables them to customize their computing environment to optimize for speed or cost-effectiveness.
\end{enumerate}

In Section \ref{sec:mpi}, we discuss the concepts of OpenMP and MPI parallelization. In Section \ref{sec:examples}, we demonstrate with examples the necessity of Bayesian computing based on dense matrices using INLA. Then in Section \ref{sec:sim}, we emphasize the benefits of the new approach with a simulation study, showing reduced computational cost and enhanced adaptability for executing the code across multiple nodes as well as within a single node. In Section \ref{sec:app}, we fit count data as an application of INLA$^+$ using a three-way spatio-temporal disease mapping interaction model. Finally, we conclude with a summary of the key takeaways from our exploration of INLA based on dense matrices. 

\section{Bayesian computing based on dense matrices using INLA\texorpdfstring{$^*$}{*}}\label{sec:prop}

The INLA and INLA$^+$ methods are specifically designed to handle latent Gaussian models (LGMs), which belong to a distinct class of models. LGMs are structured hierarchically, comprising of three levels: the hyperparameter distribution $\pi(\pmb \theta)$, the Gaussian latent field $\pi(\pmb x | \pmb \theta)$, and the likelihood model $\pi(\pmb y|\pmb x, \pmb \theta)$. The data $\pmb y$ is assumed to follow a particular distribution family, where the linear predictor $\pmb \eta$ in this family is linked to the expected value of the response $\pmb y$ through a chosen link function $g(.)$. Consequently, the expected value of the response is expressed as E$(\pmb y) = g^{-1}(\pmb \eta)$. The form of the linear predictor is determined as
\begin{equation}
    \eta_i = \pmb A_{i,.} \pmb x = \alpha + \displaystyle \sum_{j=1}^{n_\beta}  \beta_j z_{ji} + \displaystyle \sum_{k=1}^{n_f} f^{(k)}(u_{ki}) + \epsilon_i, 
\end{equation}
where $\pmb A$ is a mapping matrix, $\pmb x$ is the latent field containing the fixed and random effects, $\alpha$ is an overall intercept, $\{\beta_j\}_{j=1}^{n_\beta}$ are coefficients of some covariates $ \{z_j\}_{j=1}^{n_\beta} $, functions $ f^{(k)} $ define $ n_f $ random effects on some vector of covariates $ \{u_k\}_{k=1}^{n_f} $, and $ \epsilon_i $ is the error term that might come from the likelihood. 

\subsection{Methodology}
The main aims of INLA and INLA$^+$ are identical, but they differ in their underlying assumptions about matrix sparsity, and how \emph{computations} are done within the stages. While INLA assumes sparse matrices and uses sparse matrix algebra, INLA$^+$ is based on dense matrices. The primary goal of inference in both methods is to approximate the marginal posteriors of hyperparameters and latent field elements. In this section, we summarize the key steps of Bayesian inference with INLA based on \citep{Rue2009ApproximateBI, Martins2013BayesianCW}, while highlighting the parallel strategies employed in INLA$^+$, as described in the next section. \\

\noindent \textbf{\underline{Stage 1:}} \\

\noindent In this stage, the goal is to approximate the marginal posterior distribution of $\pmb \theta$,
\begin{equation}
\pi(\theta_{i}|\pmb y) = \displaystyle\int \pi(\pmb \theta|\pmb y)~ d\pmb \theta_{-i} \propto \displaystyle\int_{\pmb x} \displaystyle\int_{\pmb \theta_{-i}} \pi(\pmb y|\pmb x,\pmb \theta) \pi(\pmb x|\pmb \theta) \pi(\pmb \theta)  d\pmb \theta_{-i} d\pmb x.
\end{equation}
This involves constructing the approximation $\tilde{\pi}(\pmb \theta|\pmb y)$ (\ref{mode-approx}) to find the mode $\pmb \theta^*$, capturing some of the asymmetries of $\tilde{\pi}(\pmb \theta|\pmb y)$ using scaling parameters on each direction of each axis of $\pmb \theta$ \citep{Martins2013BayesianCW}, and then finding the marginal posterior $\pi(\pmb \theta_i|\pmb y)$ for each $i$ by integrating out some sequence of points of the hyperparameter using the scaling parameters, (see Appendix \ref{Appendix A.1} for details).

The optimal value of the hyperparameter $\pmb \theta$, with dimension $t$, is determined using the BFGS algorithm \citep{nocedal2006numerical}. The gradient, required for each optimization iteration, is computed using either the first-order forward or the central difference method. To compute the Hessian of the hyperparameter's marginal posterior at the optimal $\pmb \theta^*$, up to $2(t^2 + t)$ evaluations of $\tilde{\pi}(\pmb \theta|\pmb y)$ are necessary. \\

\noindent \textbf{\underline{Stage 2:}} \\ 

\noindent In this stage, the objective is to approximate the conditional distribution of the latent field $\pi(x_i|\pmb \theta, \pmb y)$ to get the approximation $\tilde{\pi}(x_{i}|\pmb y)$ of the marginal posterior
\begin{equation}
    \label{marginallatentxapprox}
    \pi(x_{i}|\pmb y) = \displaystyle\int \pi(x_i|\pmb \theta, \pmb y) \pi(\pmb \theta|\pmb y) ~ d\pmb \theta \propto \displaystyle\int_{\pmb x_{-i}} \displaystyle\int_{\pmb \theta} \pi(\pmb y|\pmb x,\pmb \theta) \pi(\pmb x|\pmb \theta) \pi(\pmb \theta)  d\pmb \theta d\pmb x_{-i}.
\end{equation}
This involves exploring a set of hyperparameters and their respective weights using the grid or Central Composite Design (CCD) strategy \citep{Box1951OnTE}, approximating the conditional distribution, and integrating out the evaluation points of $\pmb \theta$ to obtain the approximated marginal posterior, (see Appendix \ref{Appendix A.2} for details).

During the exploration of the hyperparameter space, the marginal posteriors of the latent field are computed using Gaussian or variational Bayes approximation. The variational Bayes approximation (VBA) $\tilde{\pi}_{\text{VBA}}(x_i |\pmb \theta,\pmb y)$, proposed by \cite{Niekerk2022ANA}, uses Laplace and variational Bayes (VBA) to correct for the posterior mean of the latent field. It has almost the same accuracy for the mean as the Laplace strategy, but with a lower computational cost.
In INLA the VBA produces 
\begin{equation}
    \pmb x|\pmb\theta \sim N(\pmb x^*, {\pmb Q^*}^{-1}),
    \label{eq:VBA}
\end{equation}
where $\pmb Q^*$ is calculated from the prior precision matrix $\pmb Q_{\text{prior}}$ and the negative second-order partial derivatives of the log-likelihood with respect to the latent field $\pmb x$ around the mode $\pmb x^*$ denoted as $\pmb Q_{\text{like}}$. Then the univariate conditional posterior $\tilde{\pi}_{\text{VBA}}(x_i |\pmb \theta,\pmb y)$ is extracted from the joint conditional posterior \eqref{eq:VBA}. 

All computational operations are performed on the full precision or covariance matrices, which gives these two approximations scalability on the hyperprameter level. The scalability mentioned in the previous point is directly related to the number of evaluations of the distribution $\tilde{\pi}(\pmb \theta|\pmb y)$.  For scenarios characterized by a low dimension (specifically when $t \leq 2$), the INLA/INLA$^+$ approach resorts to a grid methodology, demanding just $2t$ evaluations for effective computation. Conversely, for the CCD method, $p$ evaluations become imperative, with $p$ representing the tally of CCD explored points.\\

It is imperative to mention that the evaluations of $\tilde{\pi}(\pmb \theta|\pmb y)$ are performed on a single computing unit. However, for more computationally intensive tasks or for large-scale evaluations, the framework has been optimized to efficiently distribute the workload across multiple computing units as required, ensuring both flexibility and scalability in our computational approach.

The newly developed framework is implemented in C++ using the Blaze library, which is a publicly available mathematical library designed for efficient computation of dense linear algebra operations \citep{blaze2}. INLA$^+$ framework delivers high computational speed and performance for Bayesian inference.

\subsection{Implementation of INLA\texorpdfstring{$^+$}{+}: OpenMP and MPI parallelization}\label{sec:mpi}

In parallel computing, a node is a discrete computer system with its own processors, memory, and storage. Running code on a single node utilizes only that system's resources. Conversely, multi-node execution leverages the combined resources of interconnected computer systems.

\begin{figure}[hbt!]
\centering
\includegraphics[width=0.8\linewidth]{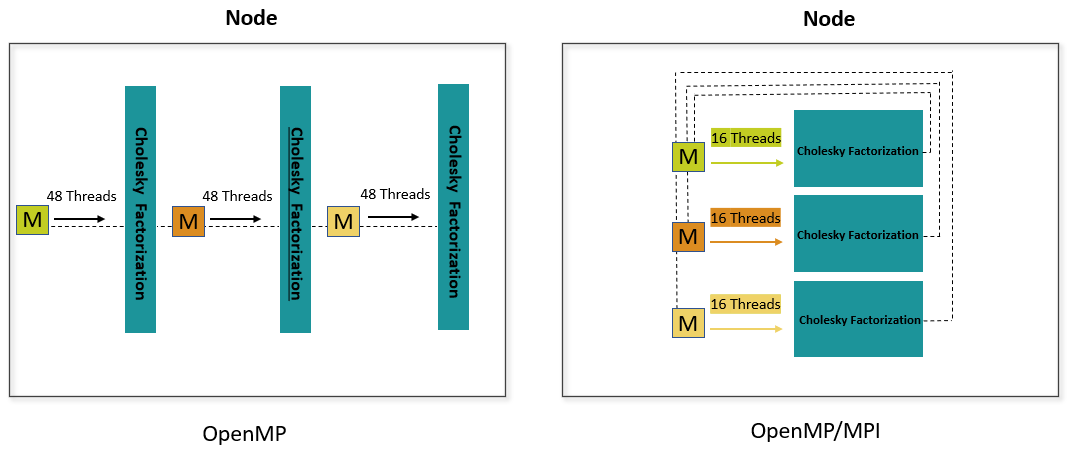}
\caption{Executing Cholesky factorization thrice on one node in two ways. Left: Sequential factorization using OpenMP (48 threads each). Right: Parallel Cholesky factorization using OpenMP (16 threads each) and MPI (three processes). Each “M" denotes a matrix.}
\label{fig:cholesky1node}
\end{figure}

OpenMP is an API enabling shared memory multiprocessing in languages like C, C++, and Fortran. Through directives, it facilitates multi-thread execution on a single node. Alternatively, MPI is a specification for parallel programming across multiple nodes in distributed environments, utilizing message passing for inter-node communication and synchronization.

Figure \ref{fig:cholesky1node} compares two methods of executing three Cholesky factorizations on one node. On the left, OpenMP's 48 threads perform factorizations sequentially. On the right, 16 threads handle each factorization, running them concurrently via MPI. While INLA employs OpenMP for single-node parallelization, INLA$^+$ integrates both OpenMP and MPI for both single and multi-node operations. Both INLA versions primarily confront Cholesky and eigen-decompositions as computational bottlenecks.

\begin{figure}
\centering
\begin{subfigure}{0.5\textwidth}
\centering
\includegraphics[width=\linewidth]{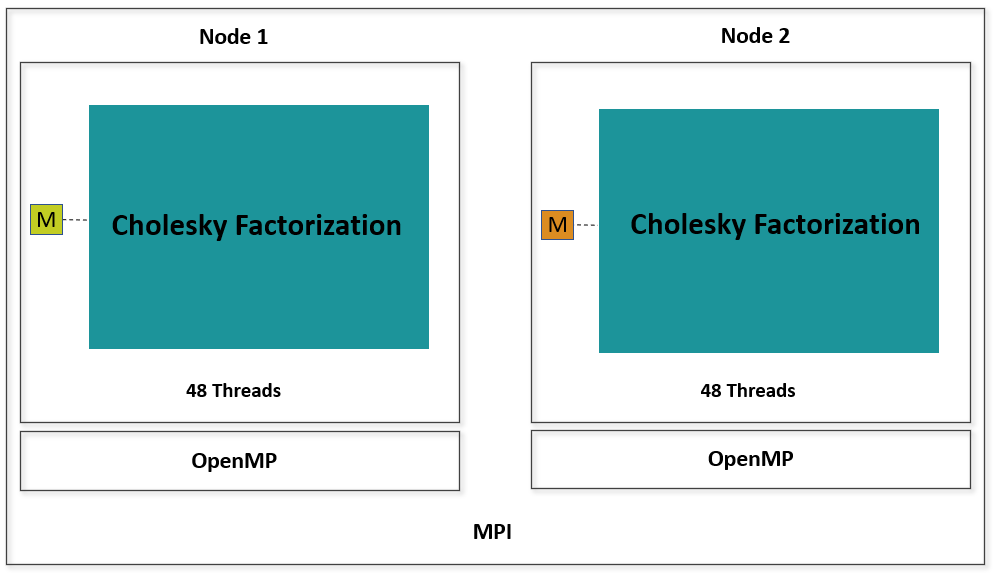}
\caption{Two Cholesky factorizations across 2 nodes.}
\label{fig:node12a}
\end{subfigure}%
\begin{subfigure}{.5\textwidth}
\centering
\includegraphics[width=\linewidth]{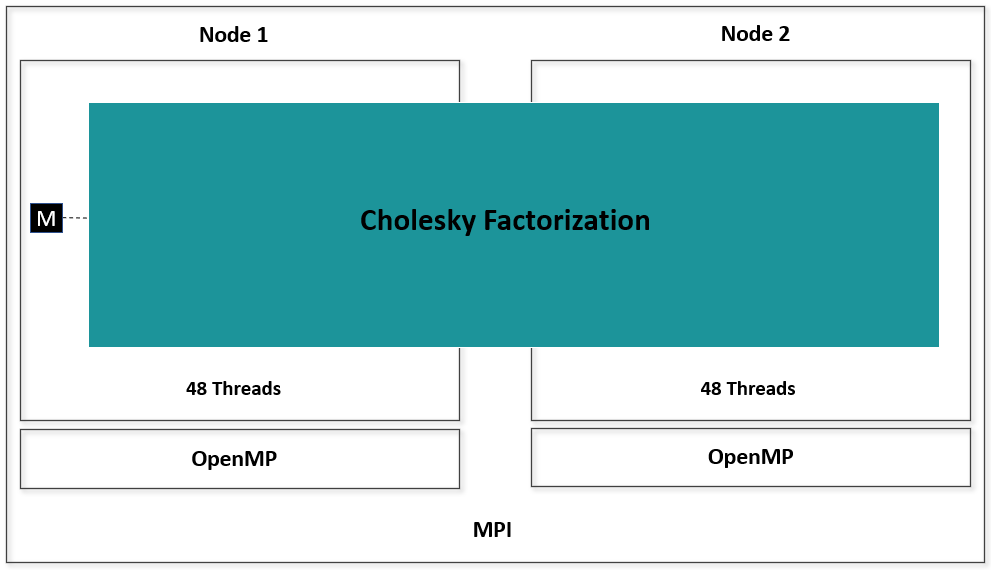}
\caption{Single Cholesky factorization distributed over 2 nodes.}
\label{fig:node12b}
\end{subfigure}
\caption{Cholesky factorization configurations on two nodes. Each “M" symbolizes a matrix.}
\label{fig:node12}
\end{figure}

Given a node, matrix decomposition can be parallelized via OpenMP and MPI. When employing multiple nodes, there are two primary decomposition approaches. The first processes each matrix on an individual node, minimizing communication overheads and ensuring scalability. The limitation here is the matrix size, constrained by the node's memory. The second method distributes the matrix computation across nodes, beneficial for large matrices but requiring thoughtful inter-node communication and synchronization strategies. INLA$^+$ currently targets matrix sizes between $10^3$ and $3 \times 10^4$, primarily adopting the first approach.

Figure \ref{fig:node12} presents the process of Cholesky factorizations using two nodes. Figure \ref{fig:node12a} has each factorization on separate nodes, reducing communication overhead. Figure \ref{fig:node12b} distributes a single factorization across both nodes, an approach that is beneficial for larger matrices.

INLA and INLA$^+$ aim for more than just matrix decompositions. Their primary goal is to manage evaluations of $\tilde{\pi}(\pmb \theta|\pmb y)$, crucial for tasks like gradient estimation, hyperparameter exploration, and asymmetric Gaussian interpolation (detailed in Appendix \ref{Appendix A.1}). INLA$^+$ innovates by transitioning from single-node to multi-node evaluations. For example, in gradient estimation, gradients could be evaluated across multiple nodes instead of one. \\

\begin{figure}
\centering
  \centering
  \includegraphics[width=0.8\linewidth]{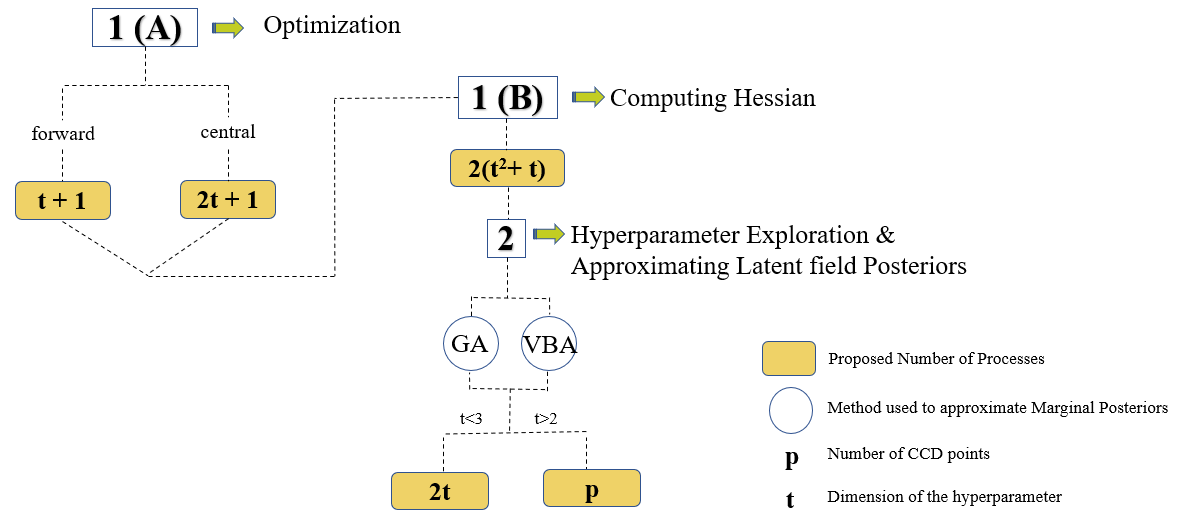}
    \caption{Proposed number of MPI processes in different parallel stages of the INLA$^+$ method. The forward and central terms represent the difference method used to compute the gradient. GA is Gaussian approximation, and VBA is variational Bayes approximation. }
    \label{fig:parallelstagesINLAplus}    
\end{figure}

\textbf{\underline{Gradient estimation:}}

Within INLA$^+$, gradient computation employs both OpenMP and MPI. For hyperparameter $\pmb \theta$ optimization (dimension $t$) as in Stage 1, each iteration requires gradient computation.

In this distributed setup, distinct MPI processes handle different tasks. One process evaluates the function at a specific $\pmb \theta^{(k)}$ during iteration $k$, while others compute gradient components. Upon completion, outputs are sent to a master process. This master node orchestrates processes or threads, assigns tasks, and compiles outputs. Gradient evaluations are then aggregated for further action.

The smart gradient technique \citep{fattah2022smart} adds another layer. After gradient computation at $\pmb \theta^{(k)} = \pmb 0$, bases are established and updated. Additionally, computing the Hessian for the hyperparameter's marginal posterior at its optimal point, $\pmb \theta^*$, is further parallelized, potentially leading to up to $2(t^2 + t)$ evaluations of $\tilde{\pi}(\pmb \theta|\pmb y)$. \\

\textbf{\underline{Hyperparameter exploration:}}

In the numerical integration in Stage 2, for the hyperparameter exploration where 45 CCD points are required, 45 evaluations of $\tilde{\pi}(\pmb \theta|\pmb y)$ ensue. Each evaluation involves matrix factorizations. Ideally, the process count should match or be less than the number of CCD points.

If the number of processes or nodes is insufficient, tasks are allocated proportionally. For example, with six CCD points requiring six evaluations and only five processes available, one process would undertake two evaluations of $\tilde{\pi}(\pmb \theta|\pmb y)$. In Figure \ref{fig:parallelstagesINLAplus}, we illustrate the methodology of INLA$^+$ based on the details presented for Stages 1 and 2. 

Next, we present two examples where dense precision or covariance matrices naturally occur and these are the situations INLA$^+$ is designed for.

\section{Examples}\label{sec:examples}

\subsection{Bypassing constraints complexity in disease mapping models using dense matrices} \label{Identfiability-constraints}

Spatial and temporal counts data require flexible models to uncover the underlying geographical patterns and their temporal changes. However, incorporating additional terms into the model leads to identifiability issues. The literature on spatiotemporal disease mapping is extensive, with a notable research paper by \cite{KnorrHeld2000BayesianMO} addressing spatio-temporal models that involve four different types of interactions. In these models, identifiability problems arise because the overall level can be absorbed by either the spatial or temporal main effects, and the interaction terms become confounded with the main effects \citep{goicoa2018spatio}. When dealing with a large number of areas and/or time points, the computational challenges of inference using the INLA method, that is based on sparse matrices, become pronounced \citep{Schrdle2011SpatiotemporalDM}. To tackle the computational complexity, \cite{OrozcoAcosta2021ScalableBM} propose a ``divide and conquer" approach. They partition the spatial region into sub-regions consisting of contiguous small areas and compute multiple fits, which are then averaged to obtain the final model. Another recent study by \cite{aanes2023faster} suggests dividing the constraints into two separate sets. One set is addressed using a mixed-effects approach, while the other set is handled through the standard conditioning by kriging method. In both studies, the posterior mean of the latent field is initially computed using the improper posterior precision matrix with a small noise added to the diagonal. Subsequently, the posterior mean is corrected using kriging technique by conditioning on the constraints. However, the results are still approximate and not exact, and the time complexity increases quadratically with the number of constraints. We show next an example that illustrates how constraints are auto-imposed using the prior covariance matrix instead of the prior precision matrix.  \\ \\
Assume the number of deaths or incident cases $\pmb y$ follows a Poisson distribution,
\begin{equation}
    \pmb y| \pmb \eta \sim \text{Poisson}(\pmb \phi e^{\pmb \eta}),
\end{equation}
where $\pmb \phi$ is the number of expected cases, and $\pmb \eta = \pmb A \pmb x$ is the linear predictor. The latent field $\pmb x$ represents the effects in the model: an overall risk (intercept), spatial effect (Besag model), linear effect (time), temporal effect (random walk of order 1), spatiotemporal effects (interaction), and other relevant factors. The matrix $\pmb A$ serves as the mapping matrix that relates these effects to the linear predictor $\pmb \eta$. In order to deal with the collinearity issue that comes from the prior precision matrix of the fixed and random effects \citep{goicoa2018spatio}, we use the Moore-Penrose inverse, $\pmb Q^+_{\text{prior}}$, of the prior precision matrix of $\pmb x$. Instead of using the improper precision matrix $\pmb Q_{\text{prior}}$, and then finding the corrected precision matrix of $\pmb x$ by conditioning on some linear constraints $\pmb C$, such that $\pmb Q_{\text{prior}} | \pmb C \pmb x = \pmb 0$, we directly work with the Moore-Penrose inverse, $\pmb Q^+_{\text{prior}}$, that auto-identifies the effects due to the correspondence between the null space of columns $\pmb N$ of $\pmb Q_{\text{prior}}$, and the set of constraints $\pmb C$, since
\begin{equation}
    \pmb N \pmb x = \pmb 0 \quad \text{and} \quad \pmb C \pmb x = \pmb 0.
\end{equation}
The rank of $\pmb Q_{\text{prior}}$ plus the number of constraints (dimension of its null space) equals the size of the latent field $\pmb x$.

We compute the posterior covariance matrix of the latent field $\pmb x|\pmb\theta$ using one of the Woodbury identity variants \citep{Riedel1992ASI} as follows,
\begin{equation}
  \pmb \Sigma^* = \pmb Q^+_{\text{prior}} - \pmb Q^+_{\text{prior}} (\pmb I + \pmb Q_{\text{like}} \pmb Q^+_{\text{prior}})^{-1} \pmb Q_{\text{like}} \pmb Q^+_{\text{prior}},
  \label{woodeqn}
\end{equation}
where $\pmb Q_{\text{like}}$ is the negative second-order partial derivatives of the likelihood function with respect to the latent field. Equation \eqref{woodeqn} is based on dense matrices and requires dense solvers to compute the result. The posterior covariance matrix of the effects is updated using the pseudo-inverse of the prior precision matrix, which makes fitting any spatiotemporal model independent of the complexity of the interactions and the number of linear constraints. This opens the stage to test models with complex interactions and may allow epidemiologists, policy makers, and health researchers to obtain inferences for more complex problems.

\subsection{Crossed effects regression}
Random effects are considered to be crossed (as opposed to nested) when each unit of a group cannot be contained in a single unit of another group. As an example we consider the  Penicillin dataset in the \emph{lme4} R library. The dataset contains the diameter of the zone of inhibited growth of the organism on each of 24 plates on which six samples of penicillin were tested. The aim was to investigate the variability between samples using the B. subtilis method whereby a bulk-inoculated nutrient agar medium is poured onto a petri dish until a diameter of 90mm is reached, this filled petri dish is known as a plate. The medium then sets and six small cylinders are placed equally spaced in the medium, wherein the six different penicillin samples are then administered. After a while, the penicillin inhibits the growth of the organism and a clear circle of inhibited growth around the cylinder can be observed. The diameter of this inhibited growth circle is measured as a way to explain the concentration of the penicillin sample.  
The model can thus be graphical presented as 

\begin{figure}[ht]
    \centering
    \includegraphics[width = 12cm]{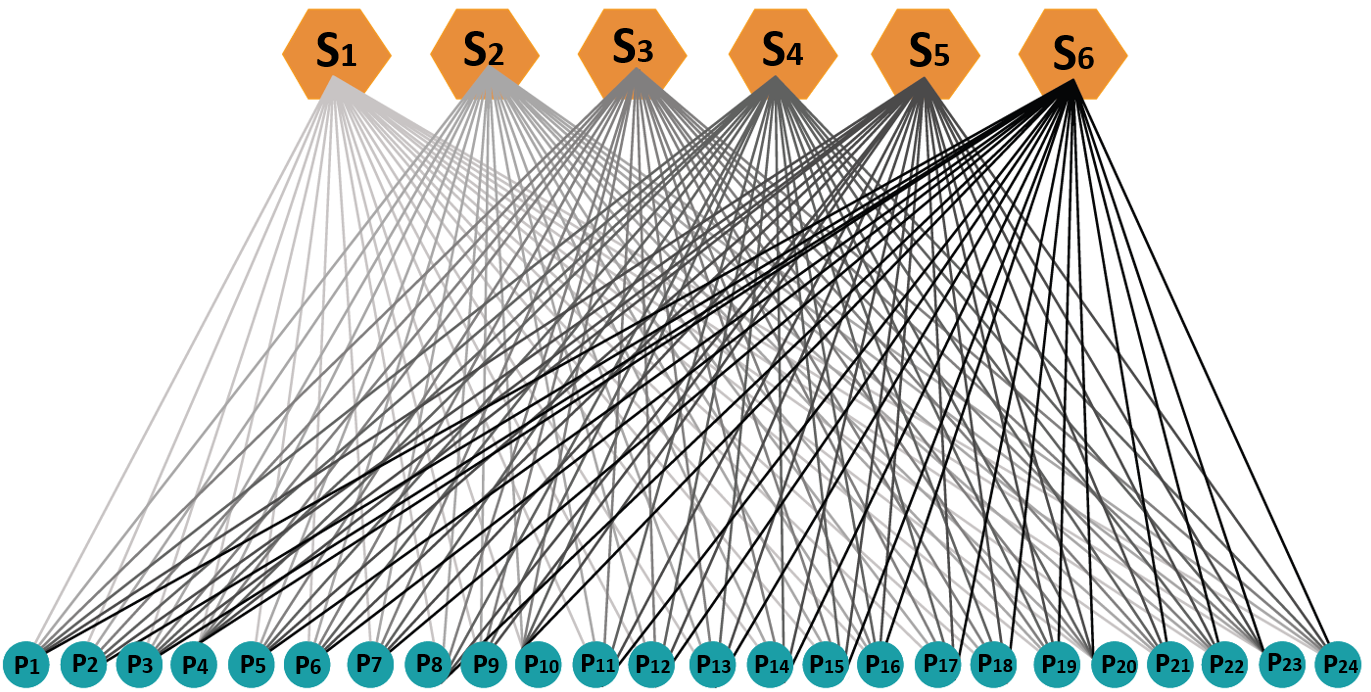}
\label{fig:model-crossed}
\caption{Graphical representation of variability analysis in Penicillin inhibition}
\end{figure}

\begin{figure}[h]
    \includegraphics[width = 8cm]{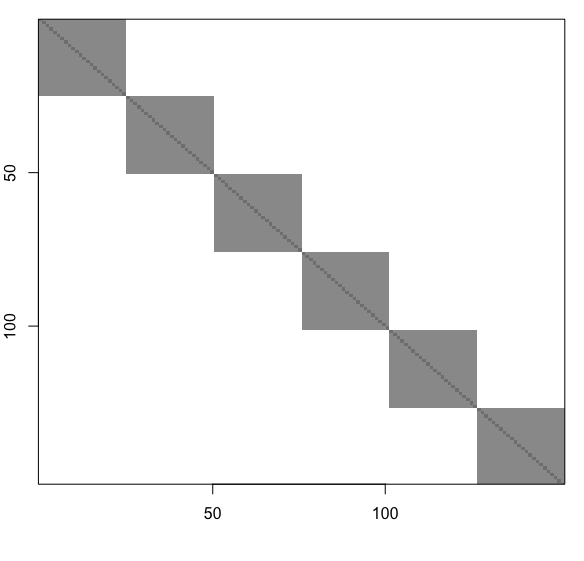}
     \includegraphics[width = 8cm]{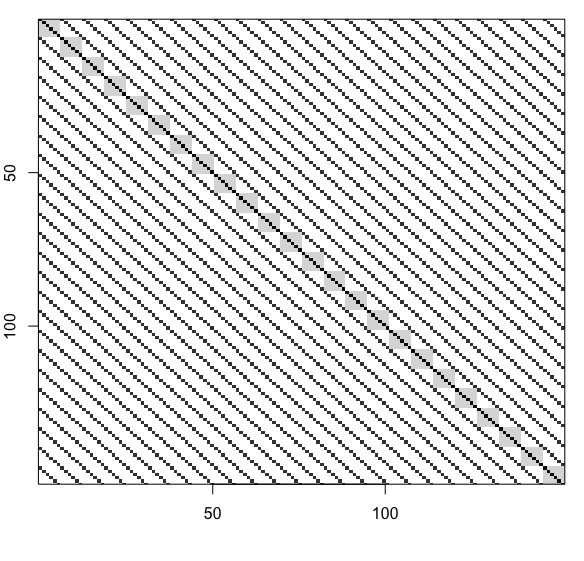}
\caption{Prior covariance matrices of a nested (left) and crossed (right) effects design for the Penicillin study}
\label{fig::crossed}
\end{figure}

It is clear that the effects are not nested. If a nested effects design is desired, then each sample should be used on 24 unique plates and the resulting covariance matrix is presented in Figure \ref{fig::crossed} (left). In this case there are 144 different plates and each sample is administered to a unique set of 24, so that there is correlation within each sample but not between plates with different samples. On the contrary, since the experiment considers the same 24 plates, the design is crossed and the resulting covariance matrix is presented in Figure \ref{fig::crossed} (right). Now we can see that there are correlations within and between samples and this results in a much denser matrix than the nested effects model.

\section{Simulation study of performance}\label{sec:sim}

The hybrid approach used in INLA$^+$ scales well compared to the current implementation of INLA, and makes efficient use of the shared and distributed memory on the multiprocessing nodes. As the size of the application grows, INLA$^+$ has the capability to incorporate additional resources to maintain highly efficient parallelization and outstanding performance across multiple nodes.

In a recent study conducted by \cite{Rustand2022FastAF}, the INLA method was utilized to fit two different multivariate joint models for primary biliary cholangitis. These models involved hyperparameters, denoted as $\pmb \theta$, with dimensions of $t = 50$ and $t = 150$ respectively. However, the study relied solely on the model configuration $\pmb \theta^*$ to integrate over the approximate marginal posterior distribution, $\tilde{\pi}(\pmb \theta | \pmb y)$ (i.e. empirical Bayes). This approach required 50 and 150 evaluations of $\tilde{\pi}(\pmb \theta | \pmb y)$, respectively. While this approach yielded reasonable results, it tended to underestimate the variability.

An alternative method for integrating over $\tilde{\pi}(\pmb \theta | \pmb y)$ is the CCD strategy as mentioned in Section \ref{sec:prop}. However, adopting this strategy would necessitate a large number of evaluations, specifically 4196 evaluations for $t = 50$, and an even greater number for $t = 150$. The computational cost associated with performing these evaluations can be substantial, potentially limiting the practicality of utilizing the INLA method in certain scenarios.

Furthermore, the existing implementation of INLA restricts the execution of all evaluations of $\tilde{\pi}(\pmb \theta | \pmb y)$ to a single computing node. In contrast, the new approach proposed in this paper enables parallel execution of the $\tilde{\pi}(\pmb \theta | \pmb y)$ evaluations by utilizing multiple computing nodes. This parallelization allows for distributing the computational workload and potentially scaling up the evaluations to make use of all available nodes. This enhanced flexibility and scalability make the new approach more efficient and applicable in a wider range of scenarios.

\begin{figure}
        \centering
        \begin{minipage}{0.5\textwidth}
          \centering
          \includegraphics[width=1\linewidth]{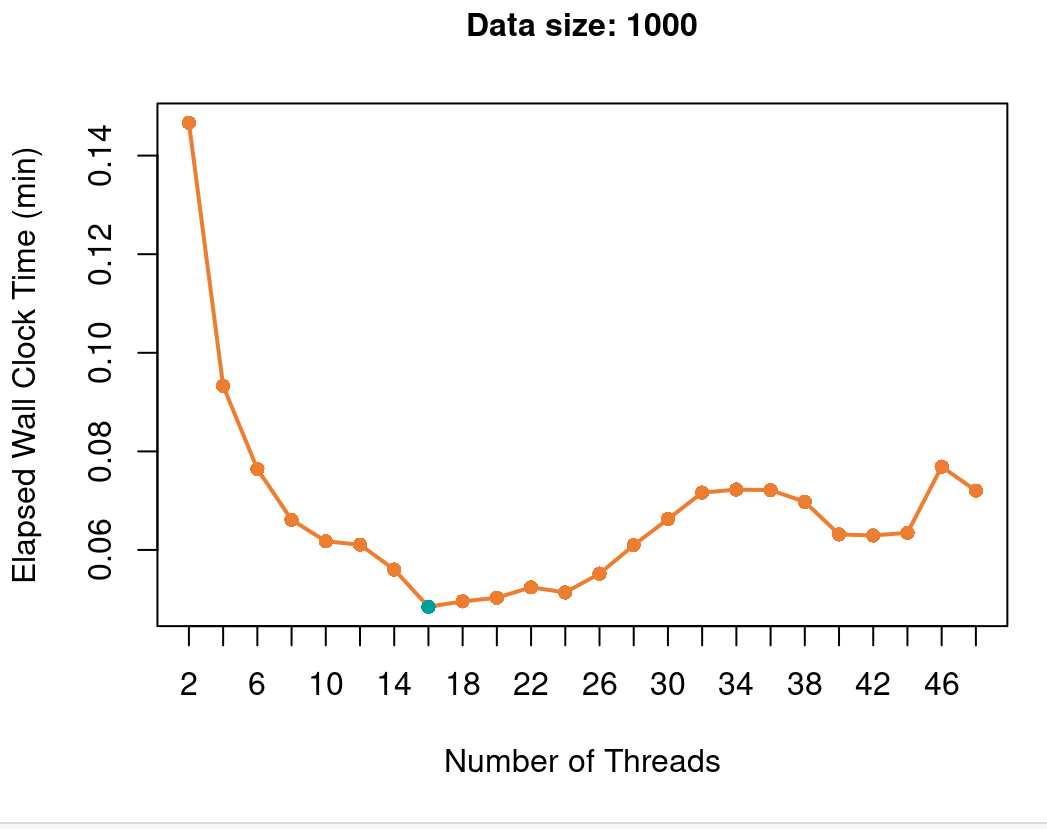}
        \end{minipage}%
        \begin{minipage}{0.5\textwidth}
          \centering
          \includegraphics[width=1\linewidth]{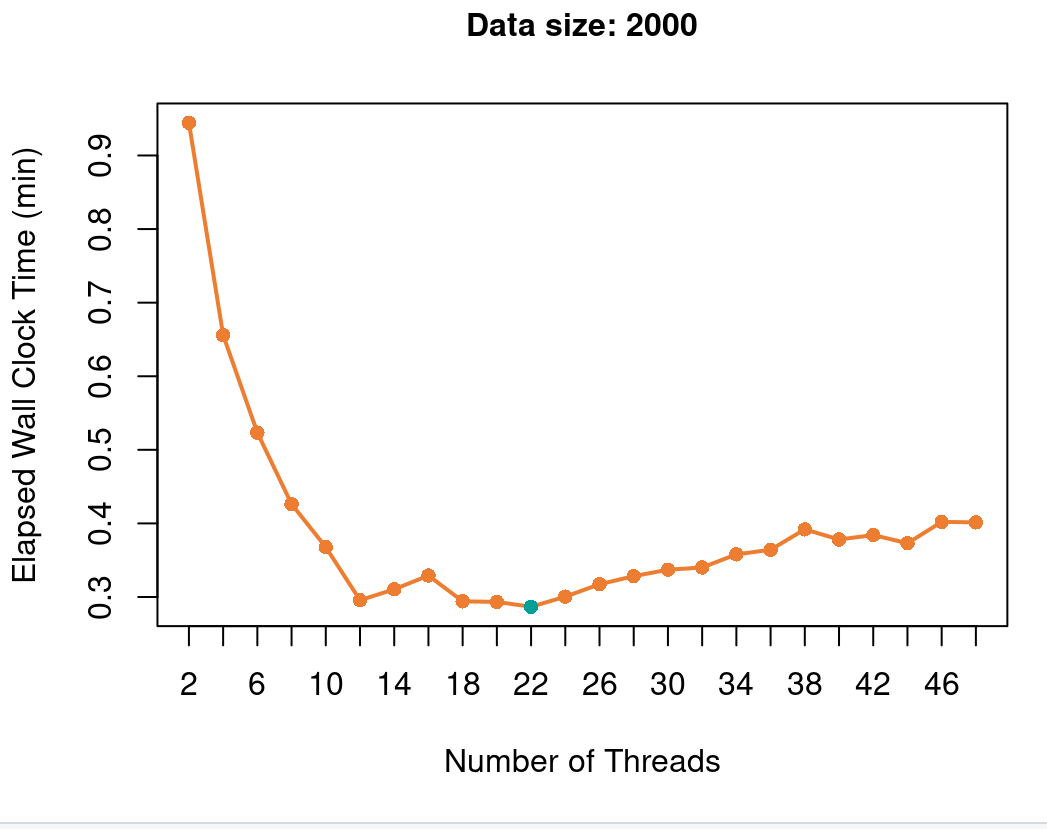}
        \end{minipage}
\begin{minipage}{.5\textwidth}
  \centering
  \includegraphics[width=1\linewidth]{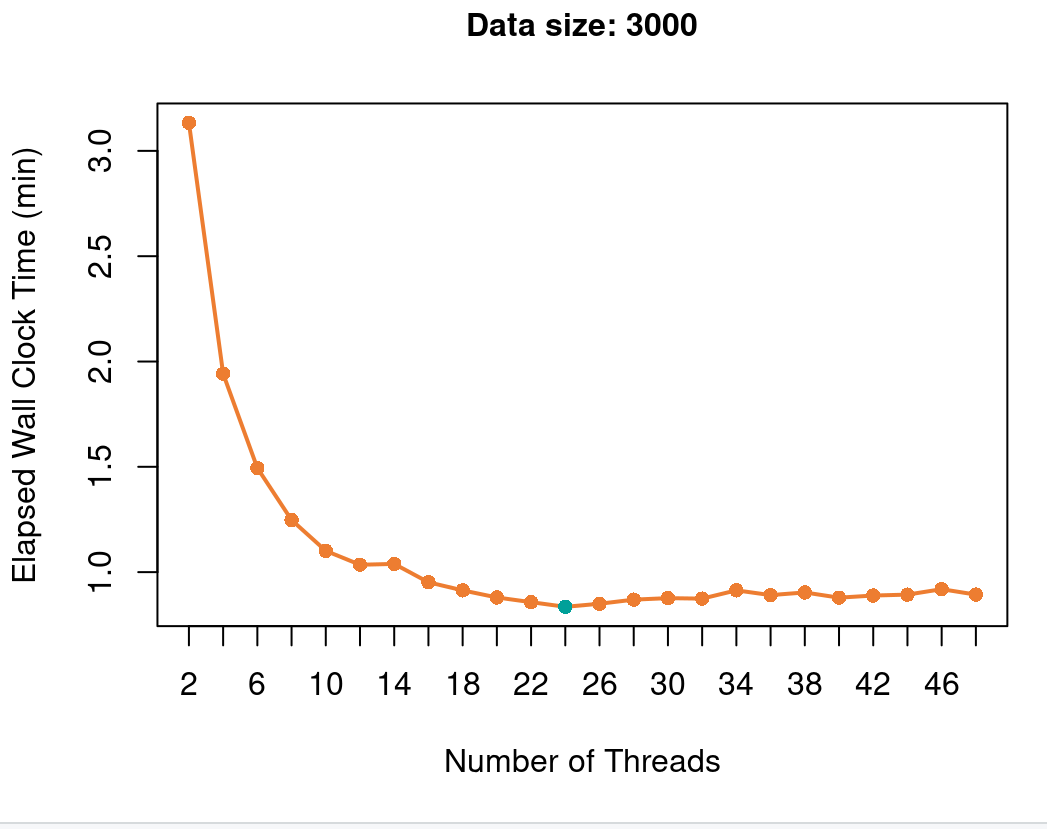}
\end{minipage}%
\begin{minipage}{.5\textwidth}
  \centering
  \includegraphics[width=1\linewidth]{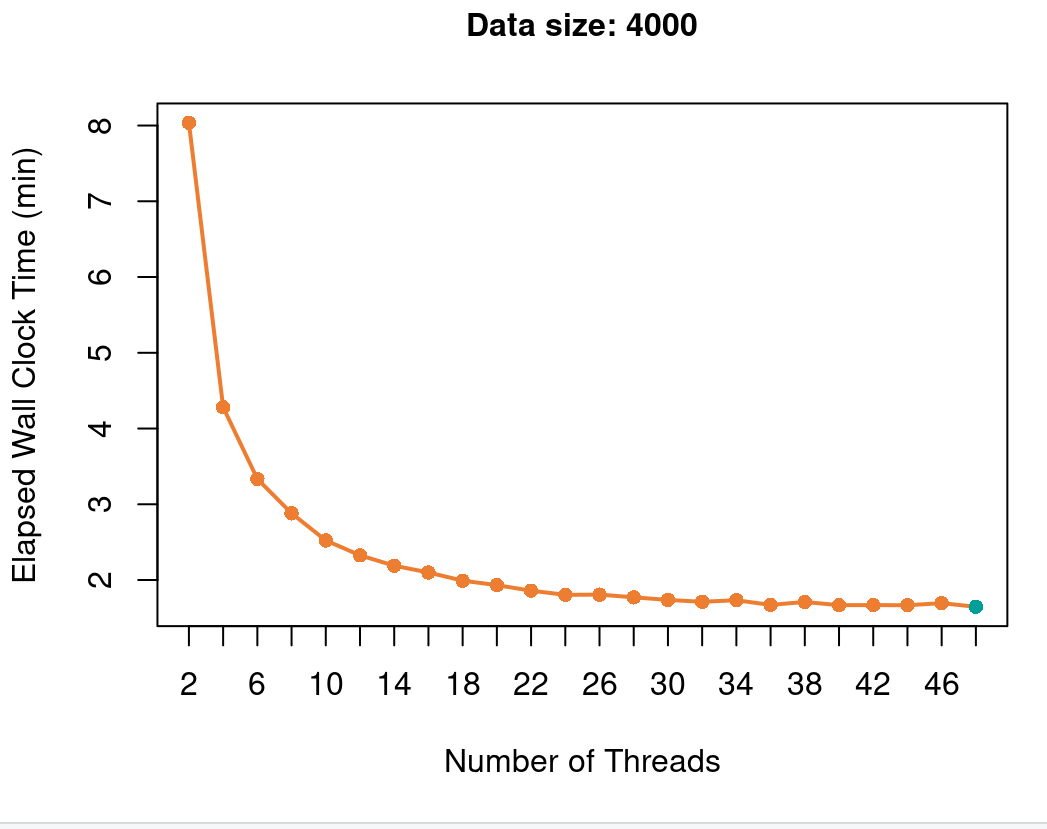}
\end{minipage}
\begin{minipage}{.5\textwidth}
  \centering
  \includegraphics[width=1\linewidth]{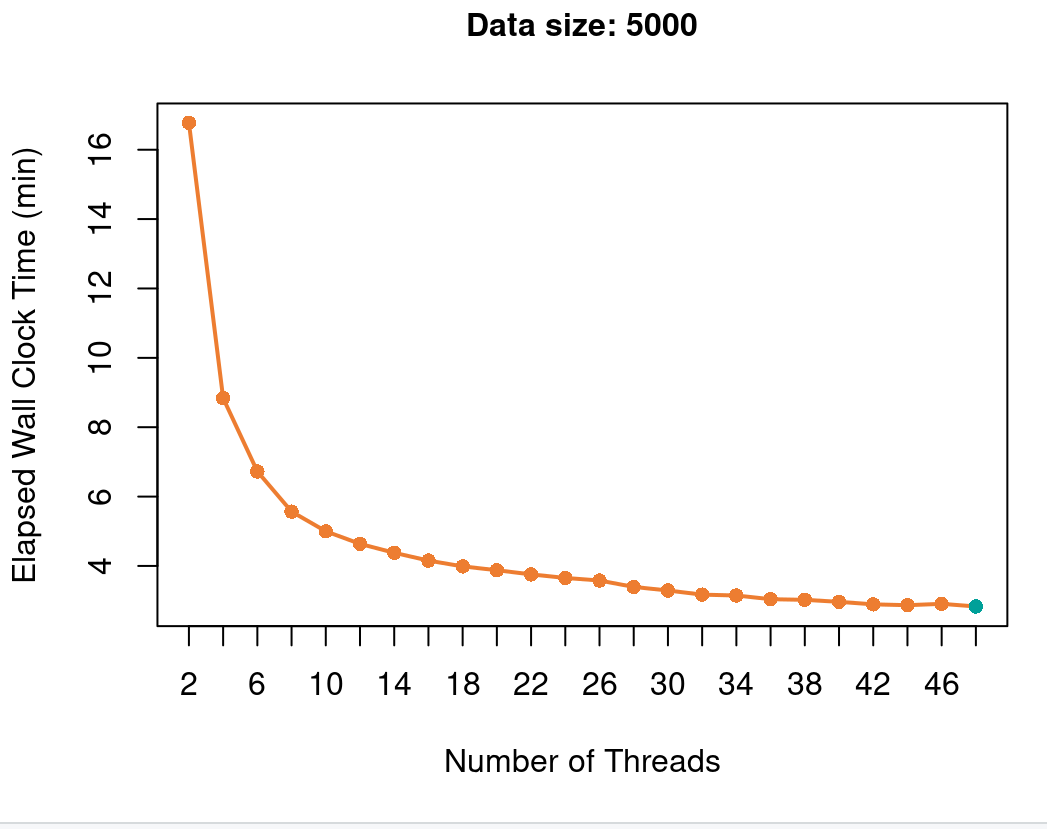}
\end{minipage}%
\begin{minipage}{.5\textwidth}
  \centering
  \includegraphics[width=1\linewidth]{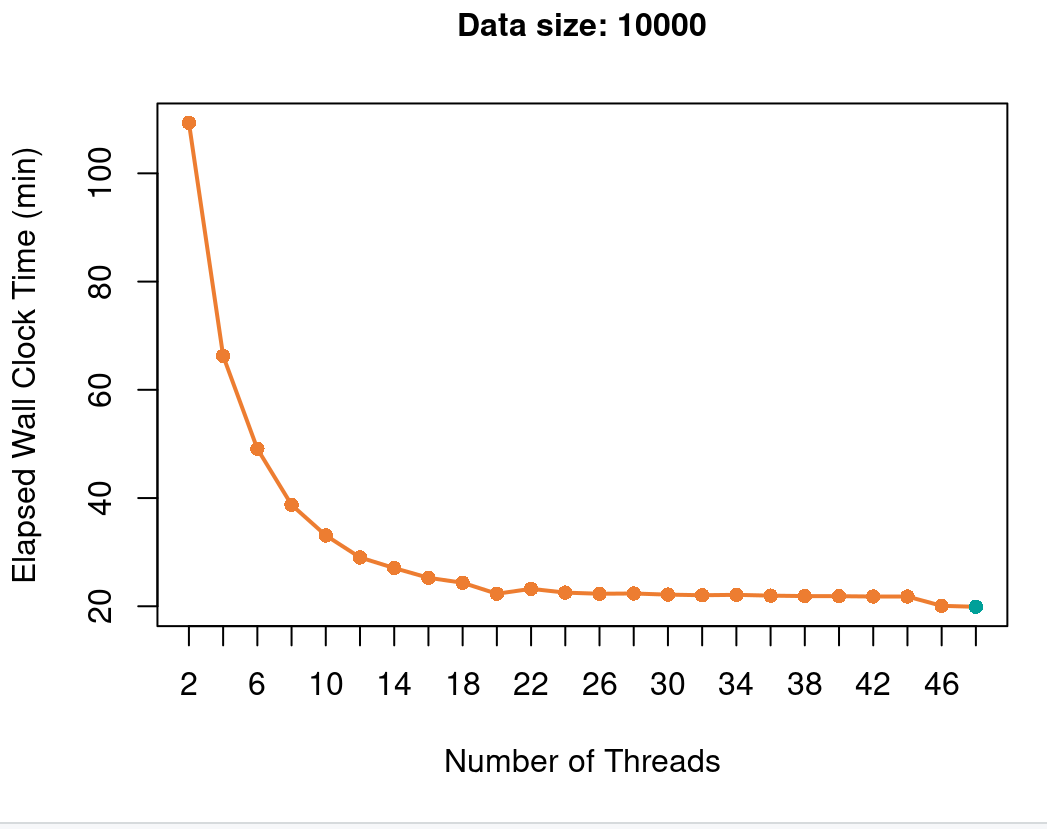}
\end{minipage}
    \caption{The performance of the INLA$^+$ method in computing the marginal posteriors of the latent field and the hyperparameter, marginal likelihood, and deviance information criterion (DIC) using one node as a function of the number of threads for different data sizes. The blue dot represents the minimum elapsed wall clock time in minutes.}
    \label{fig:openmpserver}    
\end{figure}

\begin{table}
    \caption{The performance of the INLA$^+$ method in computing the marginal posteriors of the latent field and the hyperparameter, marginal likelihood, and deviance information criterion (DIC). We compare the execution time and memory needed in different scenarios. The columns indicate the following: data size: the size of the input data, x size: the size of the fixed and random effects, nodes: the number of computing nodes of 128 GB memory, threads per node: the number of threads used per computing node, time (min): the time in minutes taken by the INLA$^+$ method to perform inference, memory per node: the amount of memory used per computing node. Hyperparameter size is 2.}
    \label{ExecusionTimeMemory}   
\centering
\begin{tabular}{cccccc}
\textbf{data size}       & \textbf{x size} & \textbf{nodes} & \textbf{threads per node} & \textbf{time (min)} & \textbf{memory per node (GB)} \\ \hline
1000 & 1009            & 1              & 10                        & 0.041               & \textless 1              \\
1000                     & 1009            & 3              & 32                        & 0.044               & \textless 1              \\
5000                     & 5009            & 1              & 10                        & 1.819               & \textless 1              \\
5000                     & 5009            & 3              & 32                        & 0.846               & \textless 1              \\
10000                    & 10009           & 1              & 10                        & 23.619              & 10.939                   \\
10000                    & 10009           & 3              & 32                        & 11.62               & 11.304                   \\
15000                    & 15009           & 1              & 10                        & 74.499              & 23.683                   \\
15000                    & 15009           & 3              & 32                        & 35.225              & 26.638                   \\
20000                    & 20009           & 1              & 10                        & -                   & out of memory            \\
20000                    & 20009           & 3              & 32                        & 78.441              & 43.323                   \\
30000                    & 30009           & 3              & 32                        & 230.071             & 106.93                   \\
40000                    & 40009           & 3              & 32                        & -                   & out of memory           
\end{tabular}
\end{table}

The computation time in the optimization routine is dominated by the number of iterations needed to find the mode of the hyperparameter. Each evaluation of the objective function (marginal posterior $\tilde{\pi}(\pmb \theta | \pmb y)$) requires one eigen-decomposition (when the prior precision matrix is not full rank) and several (3-5) Cholesky factorizations (1 Cholesky in case of Gaussian likelihood) to solve system of equations and find $\pmb x^*$, see Algorithm \ref{GAALG}. The number of evaluations of the objective function depends on the dimension of the hyperparameter, the numerical differentiation method used (forward or central) for the gradient, and the number of steps used in the line search.

To assess the performance of fitting Poisson count data using INLA$^+$, we generated data under various scenarios. The model utilizes a linear predictor that incorporated both fixed and random effects. To explore the computation effects of using higher-dimensional hyperparameters, we increase the number of random effects that are used to simulate data. The goal of our experiments in this section is to compare the computational efficiency of INLA$^+$ when utilized with different OpenMP thread and MPI process architectures. Specifically, we focus on measuring the runtime required to compute the marginal posteriors of the latent field and the hyperparameter, across the different scenarios. We focus on the Gaussian approximation strategy to compute the marginal posteriors of the latent field. 

\subsection{Running INLA\texorpdfstring{$^+$}{+} on one node}

We conducted inference using INLA$^+$ on a single-node machine with 755 GB of main memory, 52 cores, and 26 dual-socket Intel(R) Xeon(R) Gold 6230R CPU @ 2.10GHz, with varying data sizes. The summarized results are depicted in Figure \ref{fig:openmpserver}. In each experiment, we compute the marginal posteriors of the latent field and the hyperparamater, marginal likelihood, and the deviance information criterion. Our findings indicate that performing Bayesian inference using dense matrices for application size range 1000-10000 on a single node is achievable within a short time frame. By increasing the number of OpenMP threads from 2 to 48, we observe a significant increase in speed. However, adding more OpenMP threads to run the code may not always be efficient, as demonstrated by the data sizes of 1000, 2000, and 3000 shown in Figure \ref{fig:openmpserver}. In such cases, we can utilize MPI processes to take advantage of multiple threads and gain more speed, as discussed in the next section.

\subsection{Running INLA\texorpdfstring{$^+$}{+} on multiple nodes}

Table \ref{ExecusionTimeMemory} presents a comparison of the execution time and memory requirements of the INLA$^+$ method for computing marginal posteriors, marginal likelihood, and deviance information criterion, across different data sizes and computing cluster configurations. Specifically, we compare the performance of the method on data sets of varying sizes, processed on a computing cluster with different numbers of nodes and threads per node. The experiments were conducted using KAUST's Cray X 40 System - Shaheen II \citep{Hadri2015OverviewOT}.

Table \ref{ExecusionTimeMemory} reveals several interesting observations. First, the processing time generally increases with the size of the input data, as evidenced by the comparison of the 1000, 5000, 10000, 15000, and 30000 data size rows. Second, increasing the number of computing nodes tends to reduce the processing time, as seen in the comparison of the 5000, 10000, and 15000 data size rows. Third, the program runs out of memory for input data sizes of 20000 with one node and 40000 with 3 nodes and 32 threads per node. Fourth, the memory usage per node typically increases with the size of the input data and the number of nodes used, as shown in the comparison of the 10000 and 15000 data size rows and the 15000 and 20000 data size rows.


\begin{table}[h]
\caption{Performance comparison of the parallelized INLA$^+$ method on different cluster configurations and dataset sizes in computing the marginal posteriors of the latent field $\pmb x$ and the hyperparameter. The hyperparameter size is 6.}
\label{thetascaling}
\centering
\begin{tabular}{cccccc}
\textbf{data size} & \textbf{x size} & \textbf{nodes} & \textbf{threads per node} & \textbf{processes/nodes} & \textbf{time (min)}                      \\ \hline
10000              & 10422           & 1              & 32                        & 1                        & 104.25                         \\
10000              & 10422           & 1              & 10                        & 3                        & 88.40                            \\
10000              & 10422           & 2              & 16                        & 4                        & 51.90                          \\
10000              & 10422           & 3              & 32                        & 3                        & 34.54                          \\
10000              & 10422           & 7              & 32                        & 7                        & 15.65                          \\
10000              & 10422           & 45             & 32                        & 45                       & 9.93                           \\
30000              & 30422           & 1              & 32                        & 1                        & 2019.93                             \\
30000              & 30422           & 7              & 32                        & 7                        & 328.74 \\
30000              & 30422           & 45             & 32                        & 45                       & 188.31                         
\end{tabular}
\end{table}

With an increase in the hyperparameter size from 2 (Table \ref{ExecusionTimeMemory}) to 6 (Table \ref{thetascaling}), results in Table \ref{thetascaling} show that more nodes can be added to improve the method's performance. For example, for a data size of 10000 with a hyperparameter size of 6, the method can run on one node up to 45 nodes. When the hyperparameter size is 6, the number of CCD points is 45, and using 45 nodes allows 45 matrix decompositions of size 10000 to run in parallel. The processing time decreases from 104.25 minutes for a single node to 9.93 minutes for 45 nodes using 32 threads per node. Additionally, the user can change the configurations on a single node, either using 1 MPI process with 32 threads or 3 MPI processes each with 10 threads, with the latter case providing faster execution time. The scalability of the method is examined when choosing 7 nodes (proposed number of nodes in the optimization routine - stage 1 in Figure \ref{fig:parallelstagesINLAplus}) and 45 nodes (proposed number of nodes to compute the marginal posterior of the latent field - stage 2 in Figure \ref{fig:parallelstagesINLAplus}). For instance, we observe a speedup of 6.14 when using 7 nodes compared to one node in a data of size 30000, and a speedup of 10.72 when using 45 nodes.

\section{Application to disease mapping with a three-dimensional interaction: time x age x space}\label{sec:app}

In the context of disease mapping, conditional autoregressive (CAR) models are often utilized, which typically assume uniform effects across different age groups. While this assumption is reasonable in some scenarios, it fails to account for situations where various age groups are affected differently, especially due to region-specific factors. \cite{goicoa2016age} delve into the potential discrepancies in the effects of age groups and highlight the importance of considering age-specific variations when analyzing disease patterns and their geographical distribution.

This section focuses on analyzing mortality rates across regions, time periods, and age groups to understand spatial variations and temporal trends. The main objective is to identify patterns and interactions between space, time, and age. Initially, additive models with conditional autoregressive (CAR) prior for spatial effects and random walks for time and age effects are considered. However, these models may be overly restrictive as interactions between factors are common. To include interactions between time and space, mainly four interaction types are used as described by \cite{KnorrHeld2000BayesianMO}. In the application we pin INLA$^+$ against the method proposed by \cite{goicoa2016age}, and they used a Type IV interaction. Thus, for illustration and in preparation for the analysis of the real data, we also consider the Type IV interaction model. This model infers that temporal trends differ among distant regions but are similar among adjacent regions. The interaction type should be chosen carefully and with expert knowledge. We use INLA$^+$ for the approximate Bayesian inference to fit the disease mapping model.

\subsection{Model}

We perform approximate Bayesian inference for a generated data model of Poisson counts $\pmb y$ based on a linear predictor $\pmb \eta$ and expected number of cases $\pmb \phi$,
\begin{equation}
    \pmb y| \pmb \eta \sim \text{Poisson}(\pmb \phi e^{\pmb \eta}).
\end{equation}
The linear predictor is formed of an overall risk, temporal, age, and spatial effects and their pairwise interactions,
\begin{equation}
    \pmb \eta = \pmb 1^T \mu + \pmb \alpha_\text{time} + \pmb \alpha_\text{age} + \pmb \alpha_\text{space} + \pmb \zeta_\text{time x age} + \pmb \zeta_\text{time x space} + \pmb \zeta_\text{space x age},
    \label{model8}
\end{equation}
where the random effects are assumed to be centered Gaussian with different precision matrices $\pmb Q_{*} = \tau_{*} \pmb R_{*}$, $* = \{\text{time, space, age, time-space,}\ldots\}$, $\tau_{*}$ is the precision parameter, $\pmb R_{*}$ is the fixed structure of the rank-deficient precision matrix. We assign a PC prior for all precision parameters \citep{Simpson2014PenalisingMC}. Given that $\otimes$ is the Kronecker product, we define the random effects as follows:
\begin{table}[h]
\begin{tabular}{cl}
\multicolumn{1}{c|}{$\mu$}                              & overall mean,                                                                                                                                        \\
\multicolumn{1}{c|}{$\pmb \alpha_\text{time}$}               &  modeled with random walk of order  $o_t$, size $n_t$, and has structure $\pmb R_{t}$,                                             \\
\multicolumn{1}{c|}{$\pmb \alpha_\text{age}$}                & modeled with random walk of order $o_a$, size $n_a$, and has structure $\pmb R_{a}$,                                            \\
\multicolumn{1}{c|}{$\pmb \alpha_\text{space}$}              & modeled with Besag (Intrinsic CAR) model of size $n_s$, and has structure $\pmb R_{s}$,                                               \\
\multicolumn{1}{c|}{$\pmb \zeta_\text{time x age}$}          & models the time x age interaction, and has structure $\pmb R_{t} \otimes \pmb R_{a}$,                       \\
\multicolumn{1}{c|}{$\pmb \zeta_\text{time x space}$}             & models the time x space interaction, and has structure $\pmb R_{t} \otimes \pmb R_{s}$,                       \\
\multicolumn{1}{c|}{$\pmb \zeta_\text{space x age}$}         & models the space x age interaction, and has structure $\pmb R_{s} \otimes \pmb R_{a}$.                       
\end{tabular}
\end{table}

By incorporating pairwise Type IV interactions, we can explore the interplay between time, age, and spatial effects and understand how they collectively impact the outcome. 

\subsection{Identfiability constraints} 

To make the three-way interaction model identifiable, we impose sum-to-zeo constraints . Assume the order of the two random walks can be one or two, then the number of constraints imposed on the effects are:

\begin{itemize}
    \item One constraint on $\pmb \alpha_\text{time}$, one constraint on $\pmb \alpha_\text{age}$ and one constraint on $\pmb \alpha_\text{space}$.
    \item $n_t n_a - (n_t - 1)(n_a - 1)$ constraints on $\pmb \zeta_\text{time x age}$, $n_t n_s - (n_t - 1)(n_s - 1)$ constraints on $\pmb \zeta_\text{time x space}$, and $n_s n_a - (n_s - 1)(n_a - 1)$ constraints on $\pmb \zeta_\text{space x age}$.
\end{itemize}

\noindent If $o_t = 2$ and $o_a = 2$, then we append the linear predictor $\pmb \eta$ as
\begin{equation}
    \pmb \eta^{'} = \pmb \eta + \pmb \beta_t \pmb t + \pmb \beta_a \pmb a,
    \label{addlin}
\end{equation}
where $\pmb \beta_t$ and $\pmb \beta_a$ are fixed effects for the linear trends of time $\pmb t$, and age $\pmb a$ respectively. Subsequently, we need to add more constraints to make this model identifiable \citep{goicoa2018spatio}. The number of constraints in this case becomes:
\begin{itemize}
    \item Two constraints on $\pmb \alpha_\text{time}$, two constraints on $\pmb \alpha_\text{age}$ and one constraint on $\pmb \alpha_\text{space}$.
    \item $n_t n_a - (n_t - o_t)(n_a - o_a)$ constraints on $\pmb \zeta_\text{time x age}$, $n_t n_s - (n_t - o_t)(n_s - 1)$ constraints on $\pmb \zeta_\text{time x space}$, and $n_s n_a - (n_s - 1)(n_a - o_a)$ constraints on $\pmb \zeta_\text{space x age}$.
\end{itemize}

The INLA method corrects for these constraints using the Kriging technique \citep{Schrdle2011SpatiotemporalDM}. The cost when using this technique grows quadratically, and its time complexity is proportional to $\mathcal{O}(sk^2)$, where $s$ is the size of the fixed and random effects and $k$ is the total number of constraints. For high $k$, the cost of this technique dominates the overall cost for approximate inference. Adding more computing resources does not solve the computational issue since INLA does not scale on distributed nodes. 

Conversely, we use the Moore-Penrose inverse of the precision matrix to bypass the complexity of the constraints, as proposed in Section \ref{Identfiability-constraints}. Then we compute the posterior covariance matrix  using \eqref{woodeqn}. In Appendix \ref{Appendix C}, we present a numerical example that opposes the two approaches: the Kriging technique and \eqref{woodeqn}, respectively, to correct for the constraints.

We compare the performance of INLA and the new INLA$^+$ approach through a simulated example. Then we fit a real data to analyze a prostate cancer mortality in Spain using all pairwise interactions: time, space and age.

\subsection{Simulation study} 

To compare the execution time needed to obtain inference using INLA and INLA$^+$, we use the linear predictor:
\begin{equation}
    \pmb \eta^{''} = \pmb 1^T \mu + \beta_t \pmb t + \pmb \alpha_\text{time} + \pmb \alpha_\text{space} + \pmb \zeta_\text{time x space}, 
\end{equation}
for the Poisson counts, $ \pmb y| \pmb \eta \sim \text{Poisson}(e^{\pmb \eta})$. We assign a weakly informative centered Gaussian priors
with low precision for the overall intercept $\mu$ and fixed effect $\beta_t$. The time effect $\pmb \alpha_\text{time}$ is a random walk of order 2, and $\pmb \alpha_\text{space}$ is the Besag model, that is simulated using the code in Appendix \ref{Appendix code}. The number of linear constraints is $n_t$ + 2$n_s$ + 1 (2 constraints on $\pmb \alpha_\text{time}$, 1 constraint on $\pmb \alpha_\text{space}$ and $n_t$ + 2$n_s$ - 2 constraints on $\pmb \zeta_\text{time x space}$). Both approaches, INLA and INLA$^+$ find the mode of the hyperparameter and estimate the marginal posteriors of the hyperparameter and the latent field using a Gaussian approximation. Subsequently, the posterior results are the same. We compare the execution time for inference in Table \ref{CINLAINLA1234}.

\begin{table}[h]
\caption{Comparison between INLA and the new approach.}
\label{CINLAINLA1234}
\centering
\renewcommand{\arraystretch}{1.3}
\begin{tabular}{c|c|c|c|cc|c}
\multirow{2}{*}{\begin{tabular}[c]{@{}c@{}}Space\\ size\end{tabular}} & \multirow{2}{*}{\begin{tabular}[c]{@{}c@{}}Time\\ size\end{tabular}} & \multirow{2}{*}{\begin{tabular}[c]{@{}c@{}}Effects\\ size (s)\end{tabular}} & \multirow{2}{*}{\begin{tabular}[c]{@{}c@{}}Constraints\\ (k)\end{tabular}} & \multicolumn{2}{c|}{Execution Time (s)} & \multirow{2}{*}{\begin{tabular}[c]{@{}c@{}}Speedup\\ Ratio\end{tabular}} \\
\cmidrule{5-6}
 &  &  &  & INLA & INLA$^+$ & \\ \hline
200 & 5 & 1207 & 406 & 53.06 & 2.89 & 18.35 \\
400 & 5 & 2407 & 806 & 366.89 & 14.07 & 26.08 \\
800 & 5 & 4807 & 1606 & 1822.32 & 82.92 & 21.98 \\
\end{tabular}
\end{table}

We run INLA on a single Cascade Lake CPU node, 40 cores, 2.50 GHz, 384 GB/usable 350 GB, and we exploit fully the number of threads present (10:4). However, we run the new approach on 25 nodes Cascade Lake nodes. Table \ref{CINLAINLA1234} shows that the new approach has a significantly faster execution time for computing the marginal posteriors in these types of models. The computations in the new approach is independent of the number of constraints. However, INLA depends on the sparsity of the precision matrices and the number of constraints. In the presence of adequate computational resources, the presented approach outperforms INLA in the presence of high interactions among the latent field's components.

\begin{figure}
\centering
  \centering
  \includegraphics[width=0.8\linewidth]{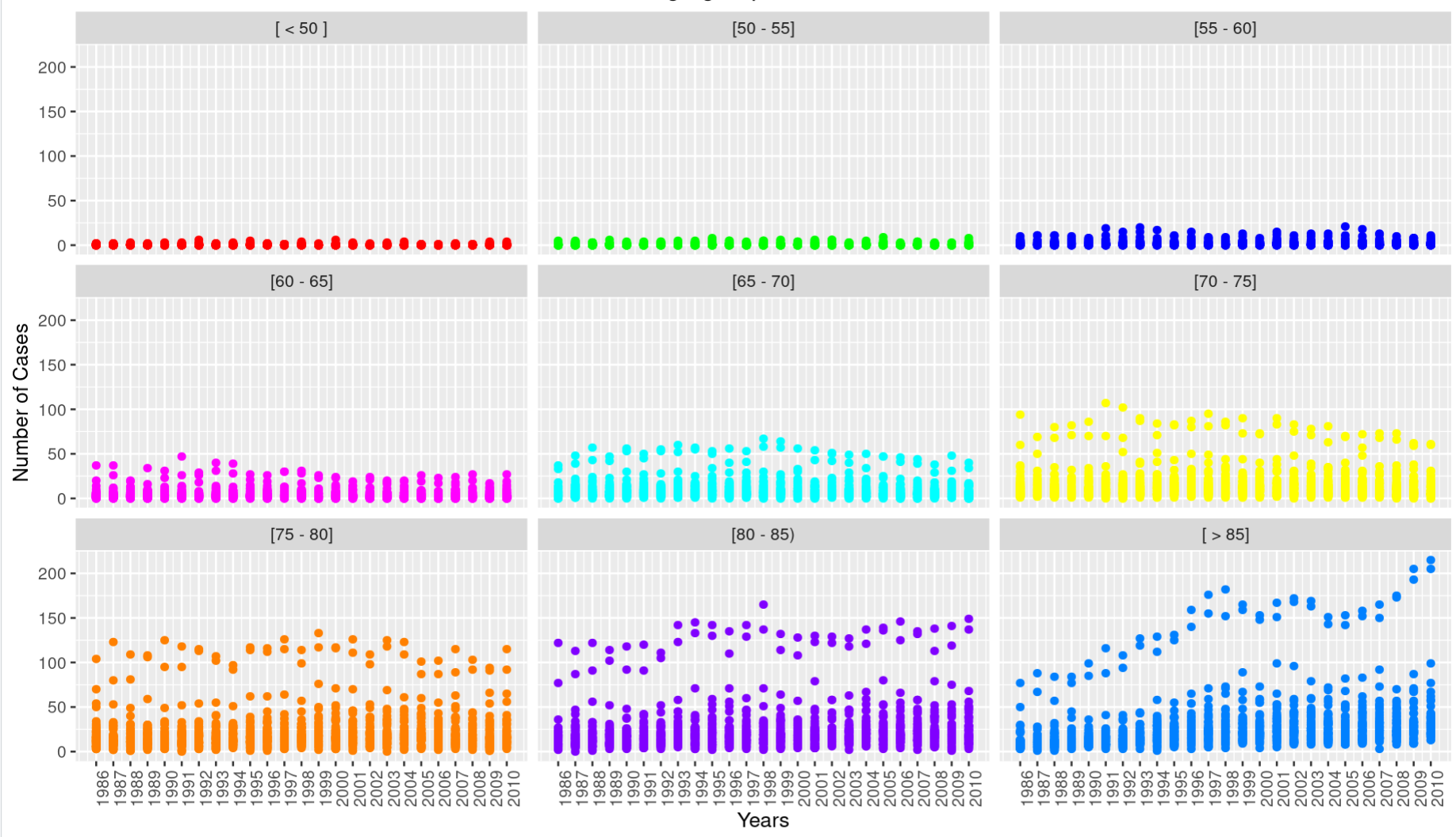}
    \caption{Number of death cases due to prostate cancer in Spain for the period 1986 - 2010 for different age groups. Total number of provinces per year and age group is 50.}
    \label{fig:desc1}    
\end{figure}

\begin{figure}
\centering
\begin{subfigure}{0.33\textwidth}
  \centering
  \includegraphics[width=\linewidth]{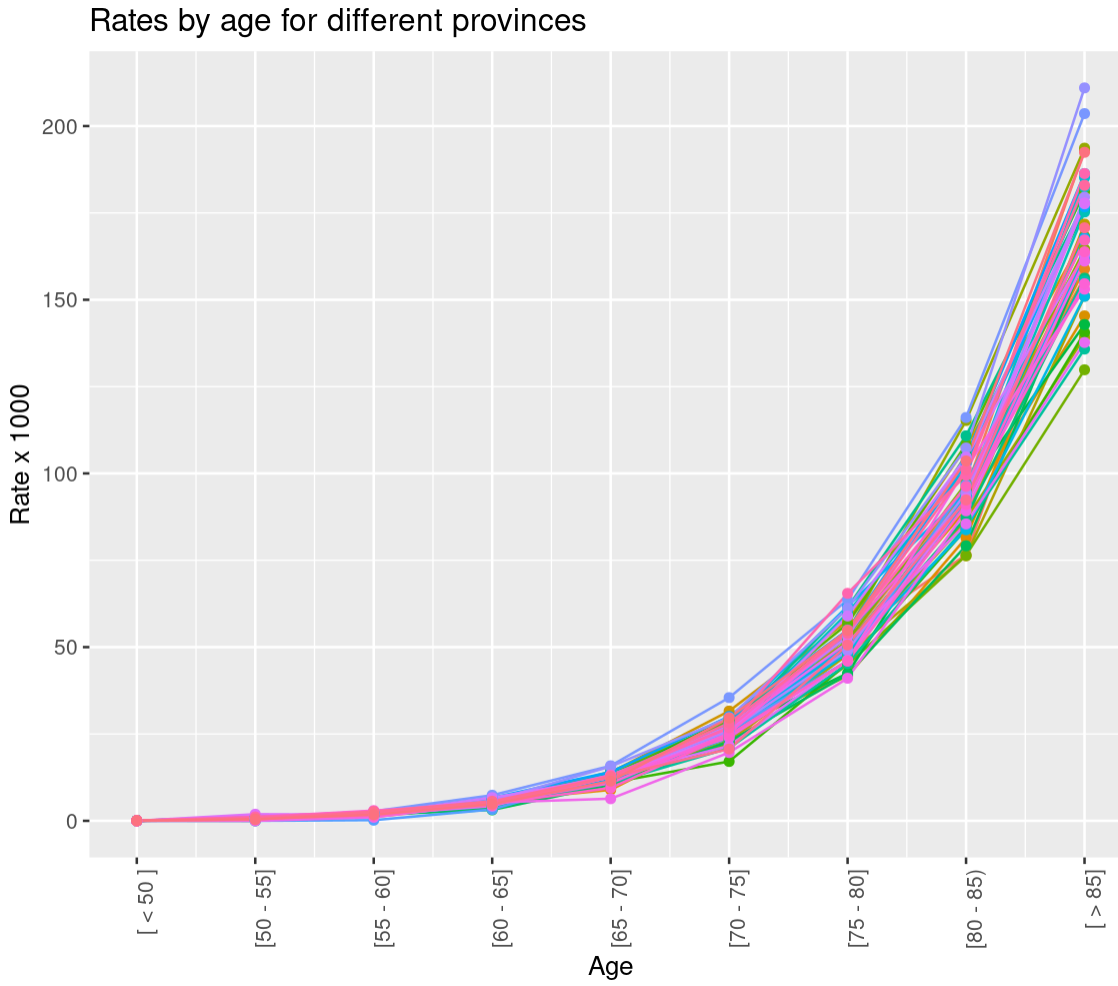}
\end{subfigure}%
\begin{subfigure}{.33\textwidth}
  \centering
  \includegraphics[width=\linewidth]{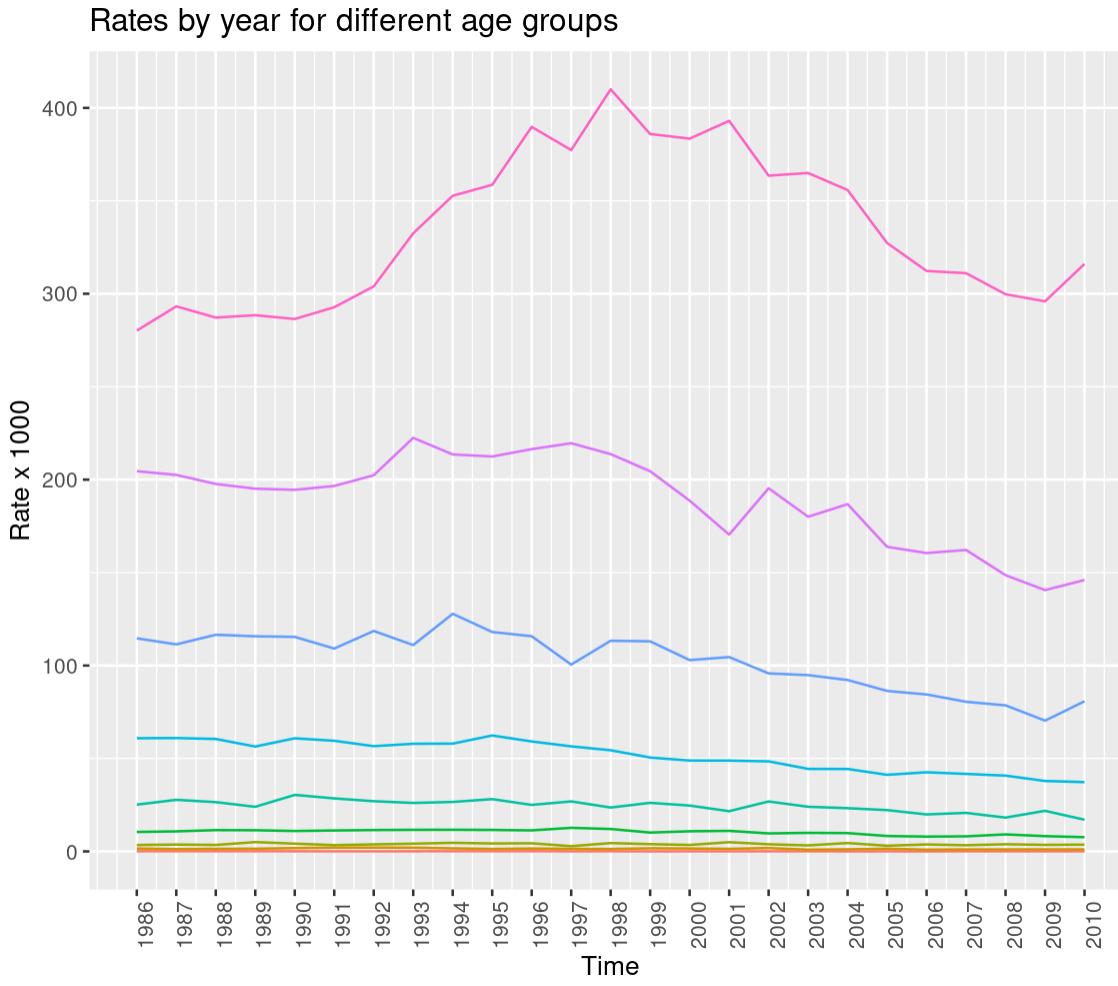}
\end{subfigure}%
\begin{subfigure}{.33\textwidth}
  \centering
  \includegraphics[width=\linewidth]{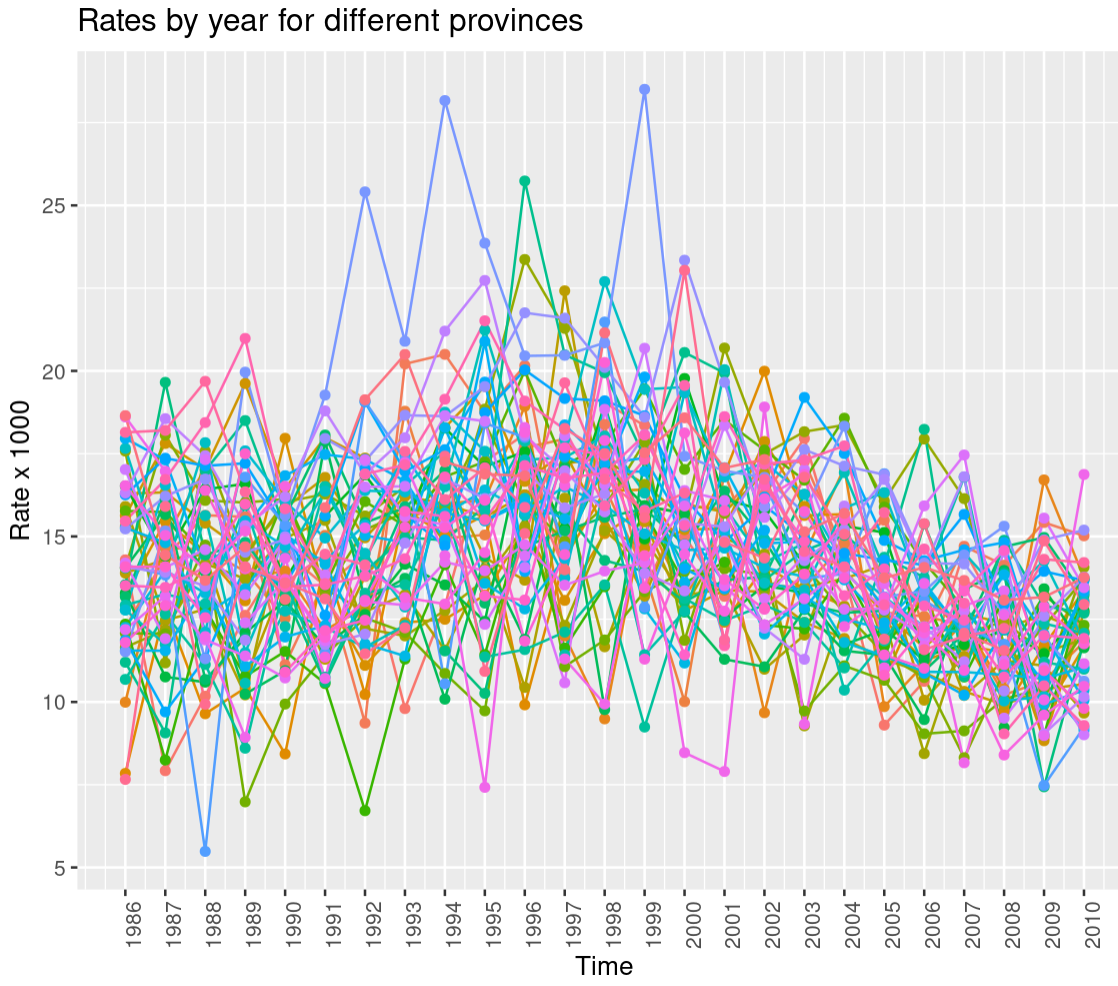}
\end{subfigure}
\caption{Exploring interactions in prostate cancer mortality rates: age x space, year x age, and time x space trends.}     
\label{exploreinteractions}
\end{figure}

\subsection{Cancer mortality data in Spain}

In this section, we replicate the three-way interaction model proposed by \cite{goicoa2016age} in their analysis of prostate cancer mortality data across 50 provinces in Spain from 1986 to 2010. The model incorporates interactions between space, time, and age, and their analysis revealed interesting results using model \eqref{model8}. They then compared this model with a more complex model that includes a space-age-time interaction effect. However, fitting the complex model took a significantly longer execution time (6 days), while \eqref{model8} only took several hours. In contrast, using INLA$^+$, we were able to fit both models in just minutes, as shown in Table \ref{thetascaling}.

\begin{figure}
\centering
\begin{subfigure}{0.3\textwidth}
  \centering
  \includegraphics[width=\linewidth]{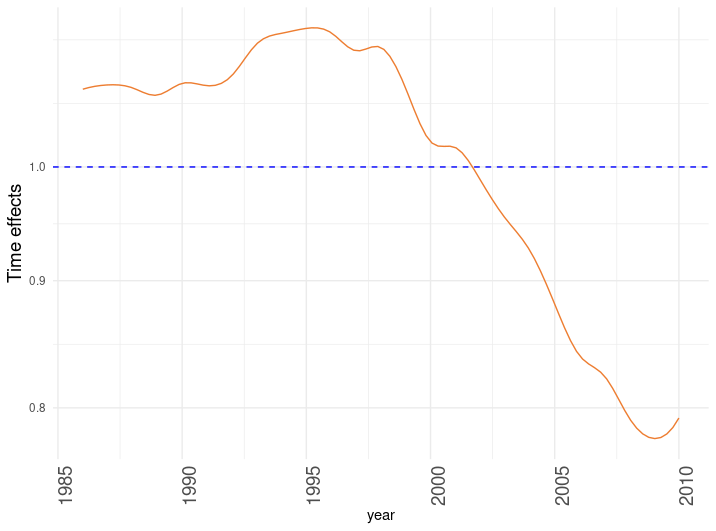}
\end{subfigure}%
\begin{subfigure}{.4\textwidth}
  \centering
  \includegraphics[width=\linewidth]{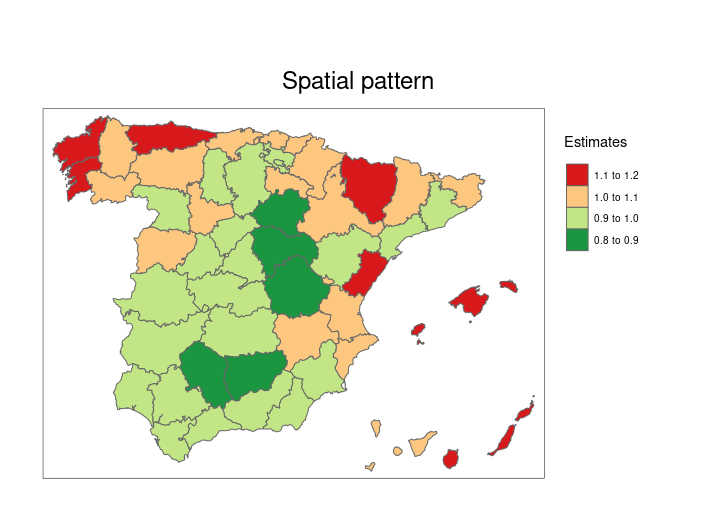}
\end{subfigure}%
\begin{subfigure}{.3\textwidth}
  \centering
  \includegraphics[width=\linewidth]{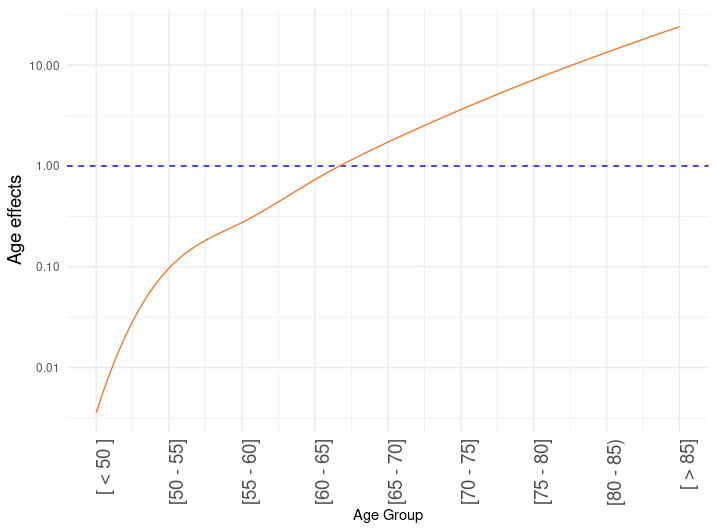}
\end{subfigure}
\caption{Prostate cancer temporal, spatial effect and age effects (in exponential scale) for the period 1986-2010, 50 provinces and nine different age groups.}     
\label{results:1}
\end{figure}

We define a linear predictor as described in \eqref{addlin}, with random walk of order $1$ for temporal effect, random walk of order $1$ for age effect, and Besag model for the spatial effect. \cite{goicoa2016age} categorized age groups for prostate cancer analysis: [$<$ 50], [50 - 55], [55 - 60], [60 - 65], [65 - 70], [70 - 75], [75 - 80], [80 - 85), and [$>$ 85], as illustrated in Figure \ref{fig:desc1}. To ensure model identifiability, we need to add 2010 constraints, which are automatically imposed by using the Moore-Penrose inverse of the prior precision matrix, as shown in \eqref{woodeqn}. 

We assign a Penalized Complexity (PC) prior \citep{Simpson2014PenalisingMC} to each precision parameter $\tau_*$, where $* = \{$time, space, age, time-space, time-age, age-space, space-time-age$\}$. In Figure \ref{exploreinteractions}, we explore the interaction trends between age x space, year x age, and time x space. The figure indicates an exponential trend for space x age interaction, a mixture of linear and non-linear trends for age x time interactions, and complex trends for space and time interactions.

We use the variational Bayes approximation \cite{Niekerk2022ANA} to estimate the marginal posteriors of the latent field, and the resulting posterior mean estimates are depicted in Figures \ref{results:1}, \ref{results:2}, \ref{results:3}, and \ref{results:4}. Analysis of these figures reveals several noteworthy interpretations. Firstly, Figure \ref{results:1} shows a significant influence of the temporal factor on the escalation of cancer rates from 1986 to 2002. Additionally, in various northern provinces such as La Coruna, Huesca, Navarra, and Asturias, the spatial effect further augments the overall cancer rates. Notably, the global cancer rate demonstrates a distinct increase specifically among individuals aged 65 and above.

Furthermore, the impact of the area varies across different age groups, as indicated by Figure \ref{results:2}. For the age group [65 - 70] and older, there was an upward trend observed in Gerona and Zamora, while Almeria and Badajoz displayed a downward trend. Moreover, a notable contribution to the rate increase was observed among the [+85] age group across several spatial areas, as depicted in Figure \ref{results:3}. Analyzing the interaction between time and age, we observe that the global cancer rate increased for higher age groups over a span of 25 years, while it decreased for lower age groups. However, regarding the interaction between space and time, no clear trend was discernible.

It is important to highlight that the interpretations mentioned above offer only a limited glimpse into the insights provided by the figures. Conducting further analyses can uncover additional valuable interpretations. For a more in-depth understanding of the data and results, we recommend referring to \cite{goicoa2016age}.

\section{Conclusion}

We introduce a novel framework that utilizes dense matrices for approximate Bayesian inference using the INLA approach, across multiple computing nodes. The framework is specifically designed to handle complex models and enhance the method's capabilities by incorporating additional features. Notably, it exhibits superior scalability compared to the existing INLA method when dealing with dense precision or covariance matrices. Leveraging the computational power of multiprocessors in shared and distributed memory architectures, it effectively utilizes modern computational resources.

The efficacy of the proposed approach is validated through a simulation study, offering substantial potential for researchers and practitioners employing INLA$^+$ in various fields of application where dense precision matrices are apparent. Additionally, insights into the performance of the INLA$^+$ method on large datasets and different computing configurations are provided. As an example, a computational task involving a dataset of size 30,000, employing dense solvers requires approximately 188 minutes to complete. Furthermore, the trade-offs between processing time and memory requirements are highlighted, enabling researchers to optimize their computational resources for statistical inference. 

An application of the new approach is demonstrated through approximate Bayesian inference for a three-way spatio-temporal disease mapping interaction model, where numerous constraints are necessary to ensure model identifiability. By utilizing a dense prior covariance matrix for the latent field, the complexity associated with imposing a high number of linear constraints is effectively addressed. The INLA$^+$ method is applied to analyze cancer mortality data in Spain using a space-time-age interaction model.

To further improve the performance of the current INLA$^+$ framework, several avenues can be explored. One potential extension is the utilization of GPUs instead of CPUs to accelerate computation, particularly for large-scale matrix decomposition or factorization problems. This would significantly enhance the computation speed. Additionally, implementing matrix decomposition on multiple nodes, as opposed to a single node, could overcome memory limitations and facilitate processing of larger datasets, leading to improved outcomes. These extensions hold significant promise in enhancing the efficiency and scalability of the current framework and warrant further exploration in future research.

\newpage
\appendix
\section*{Appendices}
\addcontentsline{toc}{section}{Appendices}
\renewcommand{\thesubsection}{\Alph{subsection}}

\subsection{Computing the marginal posterior distributions of the hyperparameters} \label{Appendix A.1}

\subsubsection{Mode of the hyperparameters} \label{modelhyper}

The INLA method constructs $\tilde{\pi}(\pmb \theta|\pmb y)$ by replacing the full conditional of $\pmb x$: $\pi(\pmb x|\pmb y,\pmb \theta)$ with its Gaussian approximation: $\tilde{\pi}_G(\pmb x|\pmb y,\pmb \theta) \sim \mathcal{N}(\pmb x^*, \pmb \Sigma^*)$, then it writes the Laplace approximation of $\pi(\pmb \theta|\pmb y)$, and evaluate it at the mode $\pmb x^*$,
\begin{equation}
   \tilde{\pi}(\pmb \theta|\pmb y) \propto \displaystyle  \frac{\pi(\pmb \theta) \pi(\pmb x|\pmb \theta) \pi(\pmb y|\pmb x,\pmb \theta)}{\tilde{\pi}_G(\pmb x|\pmb y,\pmb \theta)}\Bigg|_{\pmb x^*}.  
   \tag{41}
   \label{mode-approx}
\end{equation}
We get the model configuration $\pmb \theta^*$ using BFGS algorithm \citep{nocedal2006numerical}, an iterative method for unconstrained nonlinear minimization problems,
\begin{equation}
    \pmb \theta^*  = \underset{\pmb \theta}{\mathrm{argmin}} \hspace{3mm} -\tilde{\pi}(\pmb \theta|\pmb y).
\end{equation}
The estimated gradient $\nabla \tilde{\pi}(\pmb \theta|\pmb y)$ at each iteration for this algorithm and the estimated Hessian $\nabla^2 \tilde{\pi}(\pmb \theta|\pmb y)$ of $\tilde{\pi}(\pmb \theta|\pmb y)$ at $\pmb \theta^*$ are approximated using numerical differentiation methods boosted by the Smart Gradient and Hessian Techniques \cite{fattah2022smart}. 

\subsubsection{Asymmetric Gaussian interpolation} \label{AGI}

First, we estimate the negative hessian matrix $-\nabla^2 \tilde{\pi}(\pmb \theta|\pmb y) = (\pmb \Sigma^*_\theta)^{-1}$ of $\tilde{\pi}(\pmb \theta|\pmb y)$ at the mode $\pmb \theta^*$, and we use the eigen-decomposition of this covariance matrix $\pmb \Sigma^*_{\theta^*} = \pmb V \pmb \Lambda^{1/2} \pmb V^T$ to define a new variable $\pmb z$ as a new parameterization for $\pmb \theta$ to make the densities more regular, which corrects for scale and rotation, 
\begin{equation}
    \pmb \theta (\pmb z) = \pmb \theta^* + \pmb V \pmb \Lambda^{1/2} \pmb z.
    \label{thetatozscale} 
\end{equation}
The marginal distribution $\pi(\theta_i|\pmb y)$ is not necessarily Gaussian. With the information we have from the Hessian, we can approximate the joint distribution $\tilde{\pi}(\pmb \theta|\pmb y)$ with a multivariate Gaussian distribution,
$$\pmb \theta - \pmb \theta^* \sim N(\pmb 0,\pmb \Sigma^*_{\theta^*}),$$
then we correct for skewness for the marginals using the scaling parameters $\sigma^2_{i*}$, $i = 1, \ldots m$, as defined in the $\pmb z$ scale with the marginal densities,
\begin{equation}
 \begin{split}
     \sum_{i=1}^{m} \log \tilde{\pi}(z_i(\pmb \theta)|\pmb y) \approx -\dfrac{1}{2} \dfrac{z_1^2}{\sigma^2_{1*}} -\dfrac{1}{2} \dfrac{z_2^2}{\sigma^2_{2*}} \ldots -\dfrac{1}{2} \dfrac{z_m^2}{\sigma^2_{m*}},
 \end{split}  
 \label{marghyper}
\end{equation}
where \
\[ \sigma^2_{i*}  = \begin{cases} 
\sigma^2_{i+} & z_i \geq 0\\
\sigma^2_{i-} & z_i < 0
\end{cases}.
\]
These scaling parameters $\sigma^2_{i*}$ differ for each axis of vector $\pmb z$ and depend on whether the axis is positive or negative. When these are known, it becomes easy to approximate $\pi(\theta_i|\pmb y)$. The Gaussian approximation for the joint distribution has no extra computation burden since the hessian and the mode are already computed previously in the optimization routine, see Section \ref{modelhyper}. 

The multi-dimensional numerical integration available to integrate the hyperparameters in Equation (\ref{marghyper}) is unstable, and we follow the numerical integration free algorithm \citep{Martins2013BayesianCW} as an alternative, accurate and fast approximation to calculate the scaling parameters.

\subsubsection{Marginal posteriors of the hyperparameters}

After getting the scaling parameters using Equation (\ref{marghyper}), we compute the marginals of $\pmb \theta$ using the following \textit{Lemma},

\noindent \textit{\textbf{Lemma 3}},  

\noindent \textit{If $\pmb x = (x_1,x_2, \ldots, x_n)^T \sim \text{Gaussian}(\pmb \mu, \pmb \Sigma) $, then for all $ x_i $}
$$- \dfrac{1}{2} (x_i - \mu_i, E[\pmb x_{-i}|x_i] -\pmb \mu_{-i}) 
\pmb \Sigma^{-1}_{-i} (x_i - \mu_i, E[\pmb x_{-i}|x_i] - \pmb \mu_{-i})^T = - \dfrac{1}{2} \dfrac{(x_i-\mu_i)^2}{\Sigma_{ii}}$$
\textit{where $x_i$ is the $i^{th}$ position of $\pmb x$.} 

\vspace{0.2cm}
\noindent and assuming that $\log \tilde{\pi}(\pmb z|\pmb y)$ has a Gaussian kernel, the marginal distribution $\pi(z_i|\pmb y)$ is approximated by evaluating the joint distribution $\log \tilde{\pi}(\pmb \theta(\pmb z)|\pmb y)$ at $z_i$ and the conditional mean $E(\pmb z_{-i}| z_i)$. Then we transform the z-scale to $\theta$-scale to get the marginals $\pi(\theta_i|\pmb y)$ using Equation (\ref{thetatozscale}).

\subsection{Computing the marginal posterior approximations of the latent field} \label{Appendix A.2}

We explore two approximations for the marginal posterior of the latent field $\pmb x$. We begin our analysis with the most basic approximation, the Gaussian $\tilde{\pi}_G(x_i|\pmb \theta,\pmb y)$. Then, we present a new hybrid approach, the variational Bayes approximation $\tilde{\pi}_{\text{VBA}}(x_i |\pmb \theta,\pmb y)$, proposed by \cite{Niekerk2022ANA}, that uses Laplace and variational Bayes to correct for the posterior mean. It has almost the same accuracy as Laplace but with a lower computational cost.

\subsubsection{Gaussian approximation strategy} \label{gastr}

\noindent The marginal posterior or the full conditional distribution of the latent field $\pmb x$, given $\pmb \theta$, can be written as,
\begin{equation}
     \pi(\pmb x|\pmb \theta,\pmb y) \propto \pi(\pmb y|\pmb x,\pmb \theta) \pi(\pmb x|\pmb \theta).
\end{equation}
Usually $\pi(\pmb y|\pmb x,\pmb \theta)$ is not Gaussian and $\pi(\pmb x|\pmb \theta)$ is the latent field with zero mean and precision matrix $\pmb Q_x$. To approximate this conditional posterior by Gaussian, we write it in this form,
\begin{equation} 
\begin{split}
\tilde{\pi}_{G}(\pmb x|\pmb \theta,\pmb y) & \propto \exp\Big(-\displaystyle\frac{1}{2}(\pmb x - \pmb x^*)^{T} \pmb Q^{*}(\pmb x - \pmb x^*)\Big).
\end{split}
\end{equation}
After some expansion, the approximation becomes
\begin{equation} 
\begin{split}
\tilde{\pi}_{G}(\pmb x|\pmb \theta, \pmb y) & \propto \exp\Big(-\displaystyle\frac{1}{2}x^{T} \pmb Q^{*}x + {\pmb x^*}^T \pmb Q^* \pmb x\Big) = \exp\Big(-\displaystyle\frac{1}{2}\pmb x^{T} \pmb Q^{*}x + \pmb b^T \pmb x\Big),
\end{split}
\end{equation}
where $ \pmb Q^{*} = \pmb Q_{x} +  \pmb Q_{l}$, $\pmb Q_{l}$ is the second-order partial derivatives of the likelihood $\pi(\pmb y|\pmb x,\pmb \theta)$ evaluated at $\mu_l$ and $\pmb b$ is 
\begin{equation}
    \pmb b = {\pmb x^*}^T \pmb Q^* = \pmb b_{x} + \pmb b_{l} = \nabla g(\pmb x) - \nabla^2 g(\pmb x) \pmb \mu_{l},
\end{equation}
where $\pmb \mu_{l}$ is what maximizes the likelihood. Since the latent field $\pmb x$ is maximized at zero, then it is also $\pmb x^*$ maximizes the full conditional latent field and it becomes,
\begin{equation}
   \pmb b = {\pmb x^*}^T \pmb Q^* = \nabla g (\pmb x) - \nabla^2 g(\pmb x)\pmb x^*.
   \label{bbecomes}
\end{equation} 
Equation \ref{bbecomes} can be solved after some iterations using Algorithm \ref{GAALG}. Then, we get
\begin{equation}
    \pi(x_i|\pmb \theta,\pmb y) \sim N(\mu_{i}^*,\sigma_{i}^*),
\end{equation}
where $\mu_{i}^* = x^*_i$ and $\sigma_{i}^* = {Q^*_{ii}}^{-1}$. 

\begin{algorithm}
	$\pmb x^{(0)} = \pmb x^* = 0$\\
	\Do{$|\pmb x^* - \pmb x^{(0)}| \geq$ \text{some tolerance}}{
		calculate $\pmb Q^* =\pmb Q_x + \pmb Q_{l}(\pmb x^{(0)})$\\
		$\pmb b$ =  $\nabla g(\pmb x) + \pmb Q_{l} \pmb x^{(0)}$\\
		$\pmb x^{(0)} = \pmb x^*$\\
		solve $\pmb Q^* \pmb x^* = \pmb b$
	}
	\caption{Gaussian Approximation}
        \label{GAALG}
\end{algorithm}

\subsubsection{Variational Bayes approximation strategy}

We use the Kullback-Leibler divergence measure (or relative entropy) to correct the posterior mean approximation of the Laplace method. We derive the approximations based on dense matrices, and we follow the steps presented in \cite{Niekerk2022ANA}. We assume a Gaussian posterior distribution for $\pmb x$ with corrected mean $\pmb \mu^* + \pmb \lambda$,
\begin{equation}
    \pmb x| \pmb y, \pmb \theta \sim \mathcal{N}{(\pmb \mu^* + \pmb \lambda, {\pmb Q^{*}}^{-1})},
\end{equation}
where $\pmb \mu^*$ and $\pmb Q^{*}$ are obtained from the Laplace method, and they are fixed. We can estimate $\pmb \lambda$ using the Kullback-Leibler divergence (KLD) measure (or relative entropy),
\begin{equation}
\begin{split}
     \pmb \lambda^* &= \underset{\pmb \lambda}{\mathrm{argmin}}  \Big( \text{E}_{\pmb x| \pmb y, \pmb \theta}(- \log \pi(\pmb y| \pmb x, \pmb \theta)) + \text{KLD} (\pi(\pmb x| \pmb y, \pmb \theta) || \pi(\pmb x| \pmb \theta)) \Big) 
\end{split}
\label{relativeKLDVBA}
\end{equation}
where the relative entropy,
\begin{equation}
\begin{split}
    \text{KLD} (\pi(\pmb x| \pmb y, \pmb \theta) || \pi(\pmb x| \pmb \theta))  &= \dfrac{1}{2}\text{tr} ( \pmb Q {\pmb Q^{*}}^{-1} ) \\
    &+ ( \pmb \mu^* + \pmb \lambda)^T  \pmb Q ( \pmb \mu^* + \pmb \lambda) \\
    & - s - \log \text{det}( \pmb Q) + \log \text{det}( \pmb Q^* )
\end{split}
\end{equation}
and the expected value is approximated with Gauss-Hermite quadrature using $d^w$ points: $\pmb r^w$ and their respective quadrature weights $\pmb w$, but with linear predictor parameterization.

The expected value for linear predictor $\pmb \eta$ with corrected mean $\pmb A \pmb \mu^* + \pmb A \pmb \lambda = \pmb \nu + \pmb \delta$ is
\begin{equation}
    \text{E}_{\pmb \eta|\pmb y, \pmb \theta}(- \log \pi(\pmb y|\pmb \eta, \pmb \theta)) = \int \sum_{j}^d - \log \pi(y_i|\eta_j,\pmb \theta) \dfrac{1}{\sigma_j\sqrt{2\pi}} \exp\Big(-\dfrac{(\eta_j - \nu_j - \delta_j)^2}{2\sigma_j^2}\Big) d\eta_j. 
\end{equation}
We want the Taylor expansion around $\delta_i = 0$ of
\begin{equation}
    I(\delta_i) = \text{E}_{\eta_i | y_i,\pmb \theta \sim \mathcal{N}(\nu_i + \delta_i, \sigma_i)}(-\log\pi(y_i|\eta_i, \pmb \theta)) = \int f(\eta_i) \dfrac{1}{\sigma_i\sqrt{2\pi}} \exp\Big(-\dfrac{(\eta_i - \mu_i - \delta_i)^2}{2\sigma_i^2}\Big) d\eta_i,
\end{equation}
where $f(\eta_i) = - \log \pi(y_i|\eta_j,\pmb \theta)$. By doing the change of variable of $\zeta_i = \dfrac{\eta_i - \nu_i - \delta_i}{\sqrt{2} \sigma_i}$, we get
\begin{equation}
    I(\delta_i) = \int l(\sqrt{2}\sigma_i \zeta_i + \nu_i + \delta_i) \dfrac{1}{\sqrt{\pi}} e^{-\zeta_i^2} d\zeta_i.
\end{equation}
\noindent Then the expansion,
\begin{equation}
    \begin{split}
        I(\delta_i) \approx I(0) + I^{'}(0)\delta_i + \dfrac{1}{2}  I^{''}(0) \delta_i^2
    \end{split}
\end{equation}
where $I^{'}(\delta_i)$ is computed using Gauss-Hermite quadrature,
\begin{equation}
\begin{split}
    I^{'}(\delta_i) &= \int l^{'}(\sqrt{2}\sigma \zeta_i + \nu_i + \delta) \dfrac{1}{\sqrt{\pi}} e^{-\zeta_i^2} d\zeta_i\\
    &\approx \dfrac{1}{\sqrt{\pi} }\sum_j^{d^w} w_j l^{'}(\sqrt{2}\sigma r^w_j + \nu_i + \delta_i) d\zeta_i
\end{split}
\end{equation}
and
\begin{equation}
\begin{split}
    I^{''}(\delta_i) &= \int l^{''}(\sqrt{2}\sigma_i \zeta_i + \nu_i + \delta_i) \dfrac{1}{\sqrt{\pi}} e^{-\zeta_i^2} d\zeta_i\\
    &\approx \dfrac{1}{\sqrt{\pi} }\sum_j^{d^w} w_j l^{''}(\sqrt{2}\sigma_i r^w_j + \nu_i + \delta_i) d\zeta_i.
\end{split}
\end{equation}
For correcting the mean of the whole linear predictor, we consider
\begin{equation}
    \begin{split}
        \pmb I(\pmb \delta) \approx \pmb I(\pmb 0) + \pmb I^{'}(\pmb 0)^T\pmb \delta + \dfrac{1}{2}  \pmb \delta^T \text{diag}(\pmb I^{''}(\pmb 0) \pmb \delta,
    \end{split}
\end{equation}
and that for the correction $\pmb \lambda$,
\begin{equation}
    \begin{split}
        \pmb I(\pmb \lambda) \approx  \pmb I(\pmb 0) +  (\pmb I^{'}(\pmb 0)  \pmb A)^T \pmb \lambda + \dfrac{1}{2}  \pmb \lambda^T \pmb A^T \text{diag}(\pmb I^{''}(\pmb 0)) \pmb A \pmb \lambda.
    \end{split}
    \label{KLDVBC}
\end{equation}

\noindent Using these together with equation (\ref{relativeKLDVBA}) and (\ref{KLDVBC}), we get 
\begin{equation}
    \pmb \lambda^* = \underset{\pmb \lambda}{\mathrm{argmin}} \Big((\pmb I^{'}(\pmb 0)  \pmb A)^T \pmb \lambda + \dfrac{1}{2}\pmb \lambda^T \pmb A^T \text{diag}(\pmb I^{''}(\pmb 0)) \pmb A \pmb \lambda + \dfrac{1}{2} ( \pmb \mu^* + \pmb \lambda)^T\pmb Q ( \pmb \mu^* + \pmb \lambda) \Big),
\end{equation}
and we solve this iteratively by using Algorithm \ref{alg::vb} and this form,
\begin{equation}
     \pmb c \pmb \lambda^T + \dfrac{1}{2} \pmb \lambda^T \pmb Q^{\pmb c^*} \pmb \lambda,
\end{equation}
where 
\begin{equation}
    \pmb c = (\pmb I^{'}(\pmb 0)  \pmb A)^T + \pmb Q  \pmb \mu^* ~~~~\text{and} ~~~~\pmb Q^{\pmb c} = \pmb Q + \pmb A^T \text{diag}(\pmb I^{''}(\pmb 0)) \pmb A.
\end{equation}

\vspace{0.3cm}
\begin{algorithm}[H]
	$\pmb \mu^{*(0)} = \pmb \mu^{*}, t = 1$\\
	\Do{$\norm{\pmb \mu^{*(t)} - \pmb \mu^{*(t-1)}} \geq$ \text{some tolerance}}{
	    Compute $\pmb Q^{\pmb c}$ and $\pmb c$\\
		solve for $\pmb \lambda^{(t-1)} $: $\pmb Q^{\pmb c} \pmb \lambda^{(t-1)} = \pmb c$\\
		$ \pmb \mu^{*(t)} = \pmb \mu^{*(t-1)} + \pmb \lambda^{(t-1)}$\\
	}
	$ \pmb \mu^{*(T)} =  \pmb \mu^{*(t)}$\\
	\caption{Correction of Posterior Mean using Variational Bayes}
	\label{alg::vb}
\end{algorithm}
\vspace{0.3cm}
\noindent The corrected mean is $\pmb \mu^{*(T)}$. 

\subsection{Simulating a random Besag graph} \label{Appendix code}

\begin{lstlisting}
get_random_besag_graph <- function(n){
  while(TRUE)
  {
    p <- 0.3
    A <- matrix(0, n, n)
    A[] <- rbinom(n^2, size = 1, prob = p)
    Q <- A %*% t(A)
    Q[Q != 0] <- -1
    diag(Q) <- 0
    diag(Q) <- -rowSums(Q)
    g <- inla.read.graph(Q)
    # until number of connected components is 1
    if (g$cc$n == 1) break
  }

  return(Q)
}
\end{lstlisting}

\subsection{Full conditional distribution of an IGMRF under linear constraints} \label{Appendix C}

\noindent The full conditional distribution of the IGMRF $\pmb x$ under linear constraints $\pmb C \pmb x = \pmb 0$, given $\pmb \theta$, is
\begin{equation}
     \pi(\pmb x|\pmb \theta,\pmb y, \pmb C \pmb x = \pmb 0) = \pi(\pmb y|\pmb x,\pmb \theta) \pi(\pmb x|\pmb \theta, \pmb C \pmb x = \pmb 0).
\end{equation}
Usually $\pi(\pmb y|\pmb x,\pmb \theta)$ is not Gaussian and $\pi(\pmb x|\pmb \theta, \pmb C \pmb x = \pmb 0)$ is a Gaussian prior distribution of the latent field with zero mean and generalized inverse $\pmb {\tilde{\Sigma}_x}$ of its precision matrix. To approximate this marginal posterior by Gaussian distribution of covariance matrix $\pmb \Sigma^{*}$, we write it in this form,
\begin{equation} 
\begin{split}
\tilde{\pi}_{G}(\pmb x|\pmb \theta,\pmb y,\pmb C \pmb x = \pmb 0) & \propto \exp\Big(-\displaystyle\frac{1}{2}(\pmb x - \pmb x^*)^{T} \pmb {\Sigma}^{*+}(\pmb x - \pmb x^*)\Big)
\end{split}
\end{equation}
and after some expansion,
\begin{equation} 
\begin{split}
\tilde{\pi}_{G}(\pmb x|\pmb \theta, \pmb y,\pmb C \pmb x = \pmb 0) & \propto \exp\Big(-\displaystyle\frac{1}{2}\pmb x^{T} \pmb \Sigma^{*+}\pmb x + \pmb x^{*T} \pmb \Sigma^{*+} \pmb x\Big) = \exp\Big(-\displaystyle\frac{1}{2}\pmb x^{T} \pmb \Sigma^{*+} \pmb x + \pmb b^T \pmb x \Big)
\end{split}
\end{equation}
where,
\begin{equation}
    \pmb b = {\pmb x^*}^T \pmb \Sigma^{*+}  = \nabla g(\pmb x) - \nabla^2 g(\pmb x) \pmb \mu_{l},
    \label{GAequation}
\end{equation}
and $\pmb \mu_{l}$ is what maximizes the likelihood and since the latent field $\pmb x$ is maximized at zero, then ${\pmb x^*}$ also maximizes the full conditional latent field and it becomes,

\begin{equation}
    \pmb b = \nabla g (\pmb x) - \nabla^2 g(\pmb x) \pmb x^*.
\end{equation}

The covariance $\pmb \Sigma^{*}$ is updated using \eqref{woodeqn}. Equation (\ref{GAequation}) is solved after some iterations before getting the Gaussian approximation,

\begin{equation}
    \pmb x|\pmb \theta,\pmb y,\pmb C \pmb x = \pmb 0 \sim \mathcal{N}(\pmb x^*,\pmb {\Sigma}^*).
\end{equation}

\textbf{Example:} We compute here the precision matrix $\pmb {\Sigma}^*$ in two ways: kriging technique and our proposed approach. Given the Poisson model,
\begin{equation}
    \pmb y \sim \text{Poisson}(\pmb \eta = \pmb A \pmb e),
\end{equation}
where $\pmb y$ is the observed count points, $\pmb \eta$ is the linear predictor, $\pmb e$ are the effects and $\pmb A \in \mathbb{R}^{4x3}$ is the mapping matrix, $\{( (1, 1, 0, 0), (1, 0, 1, 0), (0, 0, 0, 1) \}$. We assume the precision matrices,

\begin{equation}
\pmb Q_{l}(\pmb \eta) = - \dfrac{\partial^2 \pi(\pmb y| \pmb \eta^2)}{\partial \pmb \eta^2} = \text{diag}(1.796,2.033,0.896)
\text{ and }  
\pmb Q_{\pmb e} = \begin{pmatrix}
~~~1 & -1 & ~~~0 & ~~~0\\
-1 & ~~~2 & -1 & ~~~0\\
~~~0 & -1 & ~~~2 & -1\\
~~~0 & ~~~0 & -1 & ~~~1
\end{pmatrix}.
\end{equation}

The imposed constraint is $\pmb C \pmb e = (1~~1~~1~~1) ~ \pmb e  = 0$. We compute the uncorrected precision matrix,

\begin{equation}
    \pmb \Sigma^*_{un} = (\pmb A^T \pmb Q_l \pmb A + \pmb Q_{\pmb e} + \varepsilon \pmb I)^{-1} = \begin{pmatrix}
~~~0.350 & -0.150 & -0.293 & -0.320\\
-0.150 & ~~~0.350 & ~~~0.207 & ~~~0.180\\
-0.293 & ~~~0.207 & ~~~0.554 & ~~~0.430\\
-0.320 & ~~~0.180 & ~~~0.430 & ~~~0.905
\end{pmatrix},
\end{equation}

where $\varepsilon$ is a tiny noise (say $1e^{-4}$). Using equation \eqref{woodeqn}, we correct for the constraints to get, 

\begin{equation}
    \pmb \Sigma^* = \begin{pmatrix}  ~~~0.274 & -0.044 &-0.129 &-0.102 \\
 -0.044 &~~~0.198 &-0.025 &-0.129 \\
 -0.129 &-0.025 &~~~0.198 &-0.044 \\
 -0.102 &-0.129 &-0.044 &~~~0.274 \end{pmatrix}.
\end{equation}
Equivalently, we get the same results using the Woodbury formula, 

\begin{equation}
  \pmb \Sigma^* = \pmb Q^{+}_{\pmb e} - \pmb Q^{+}_{\pmb e} (\pmb I + \pmb A^T \pmb Q_l \pmb A \pmb Q^{+}_{\pmb e})^{-1} \pmb A^T \pmb Q_l \pmb A \pmb Q^{+},
\end{equation}

where $\pmb Q^{+}_{\pmb e}$ is the pseudo-inverse of $\pmb Q_{\pmb e}$,
\begin{equation}
    \pmb Q^{+}_{\pmb e} = \begin{pmatrix}  ~~~0.875 & ~~~0.125 & -0.375 & -0.625\\
~~~0.125 & ~~~0.375 & -0.125 & -0.375\\
-0.375 & -0.125 & ~~~0.375 & ~~~0.125\\
-0.625 & -0.375 & ~~~0.125 & ~~~0.875 \end{pmatrix}.
\end{equation}

\subsection{Additional plots for the three-way interaction model} \label{3wayplots}

\begin{figure}[ht]
\centering
\begin{subfigure}{0.4\textwidth}
  \centering
    \includegraphics[width=\linewidth, height=0.25\textheight]{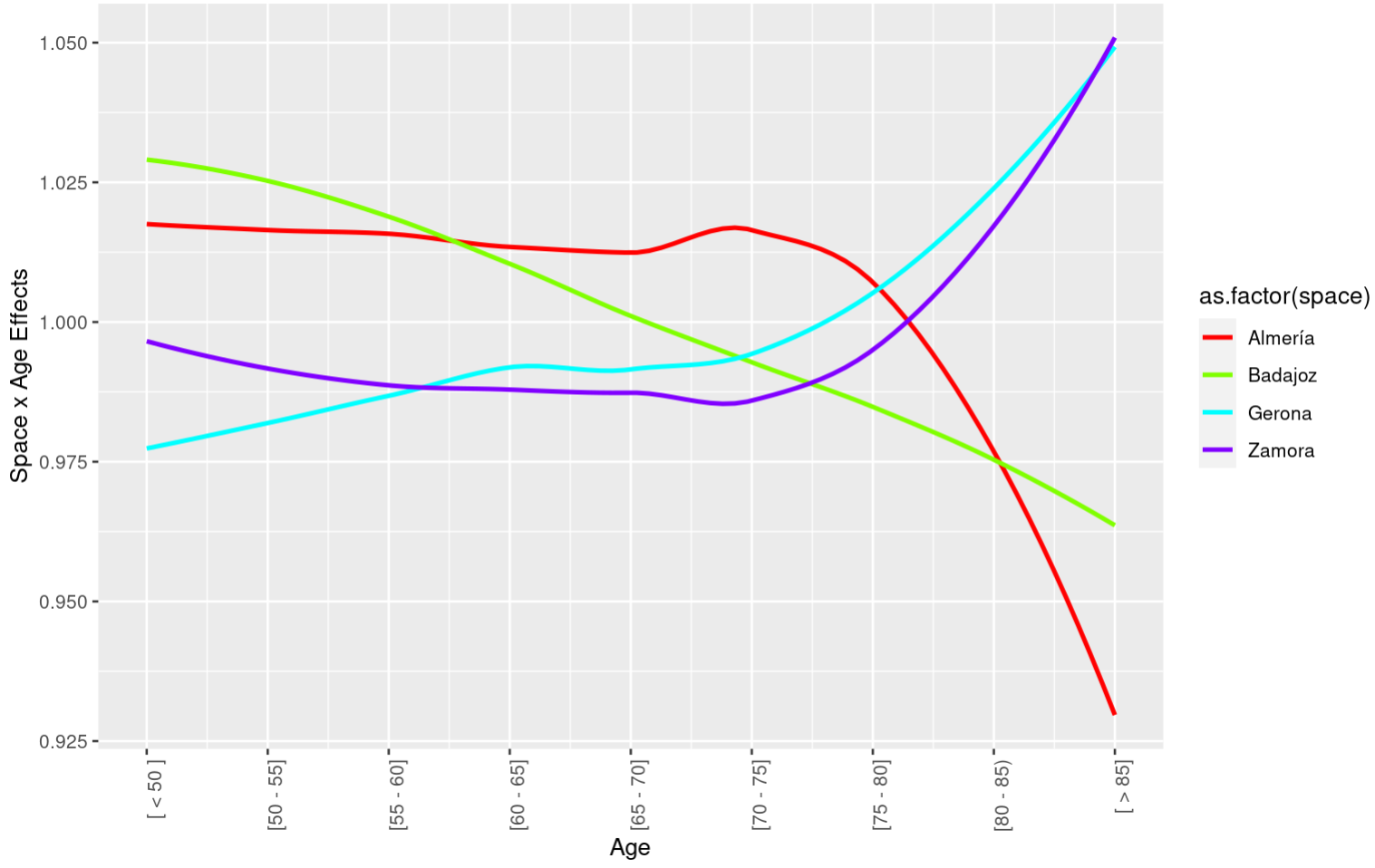}
\end{subfigure}%
\begin{subfigure}{0.6\textwidth}
  \centering
    \includegraphics[width=\linewidth, height=0.25\textheight]{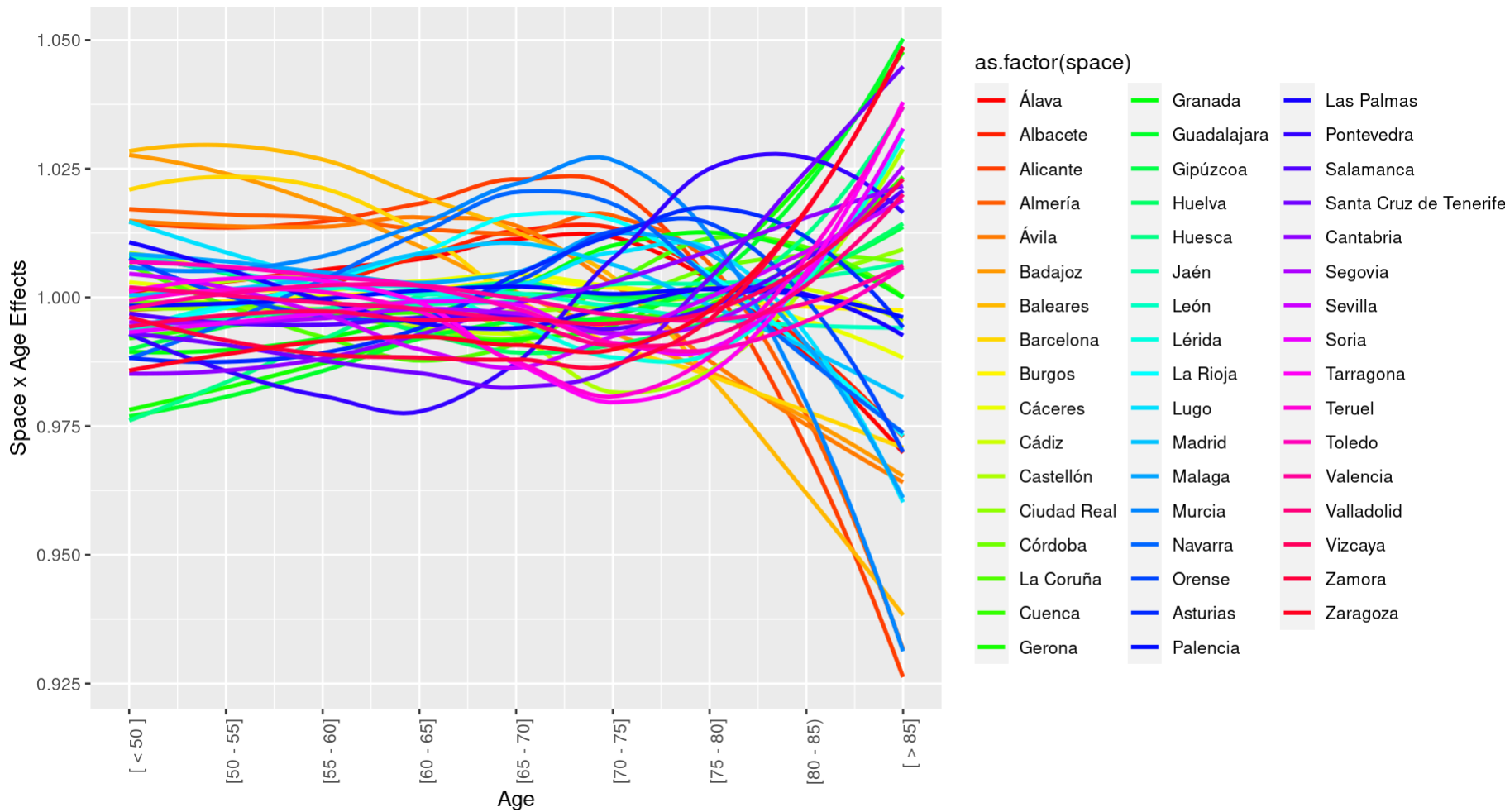}
\end{subfigure}\\
\begin{subfigure}{0.7\textwidth}
  \centering
  \includegraphics[width=\linewidth]{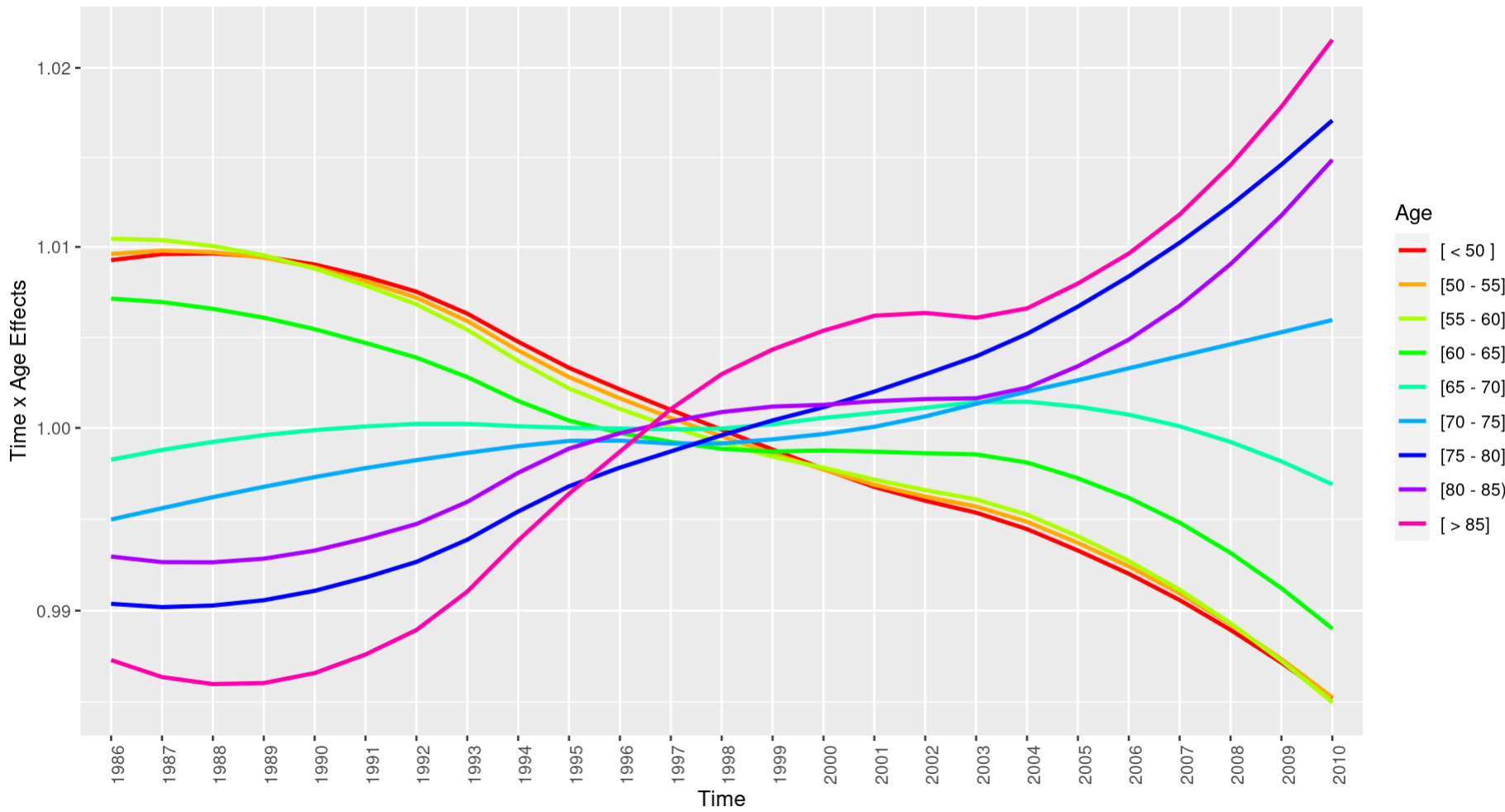}
\end{subfigure}\\
\caption{Space x age interaction, and time x age interaction estimates(exponential scale) in prostate cancer mortality.}     
\label{results:2}
\end{figure}

\begin{figure}
\centering
\begin{subfigure}{0.3\textwidth}
  \centering
  \includegraphics[width=\linewidth]{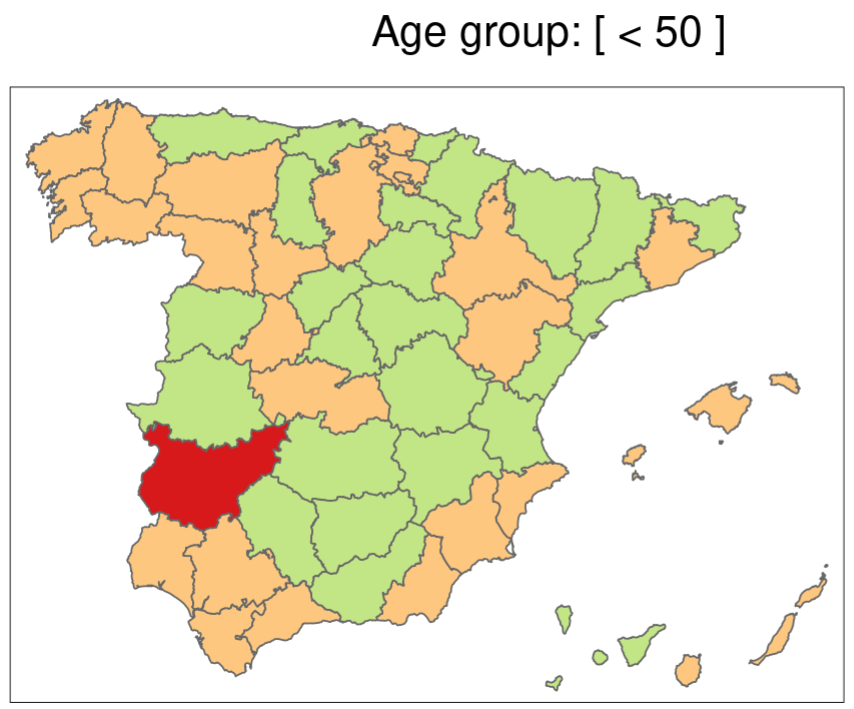}
\end{subfigure}%
\begin{subfigure}{.3\textwidth}
  \centering
  \includegraphics[width=\linewidth]{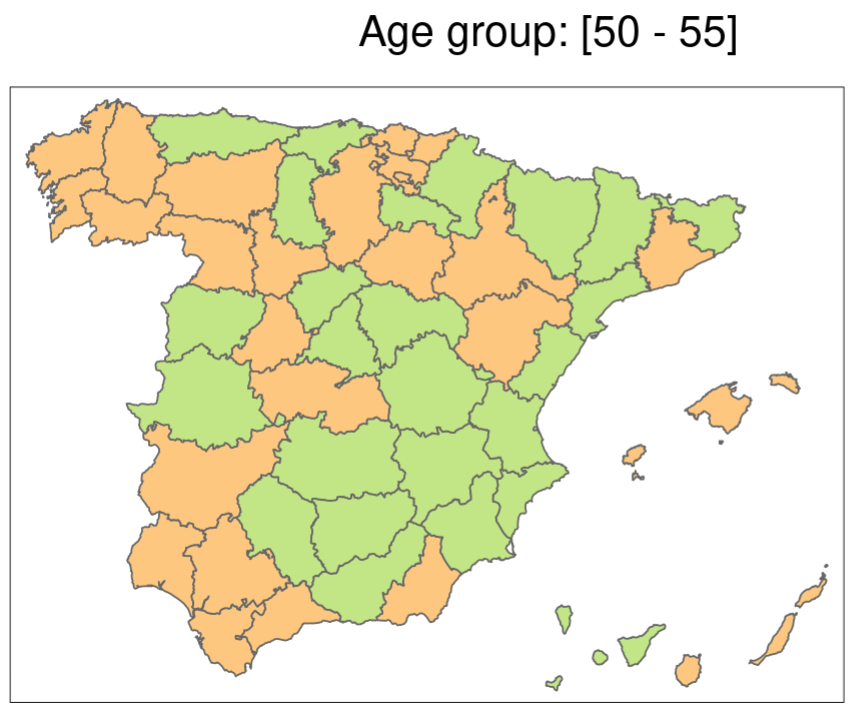}
\end{subfigure}\\
\begin{subfigure}{.3\textwidth}
  \centering
  \includegraphics[width=\linewidth]{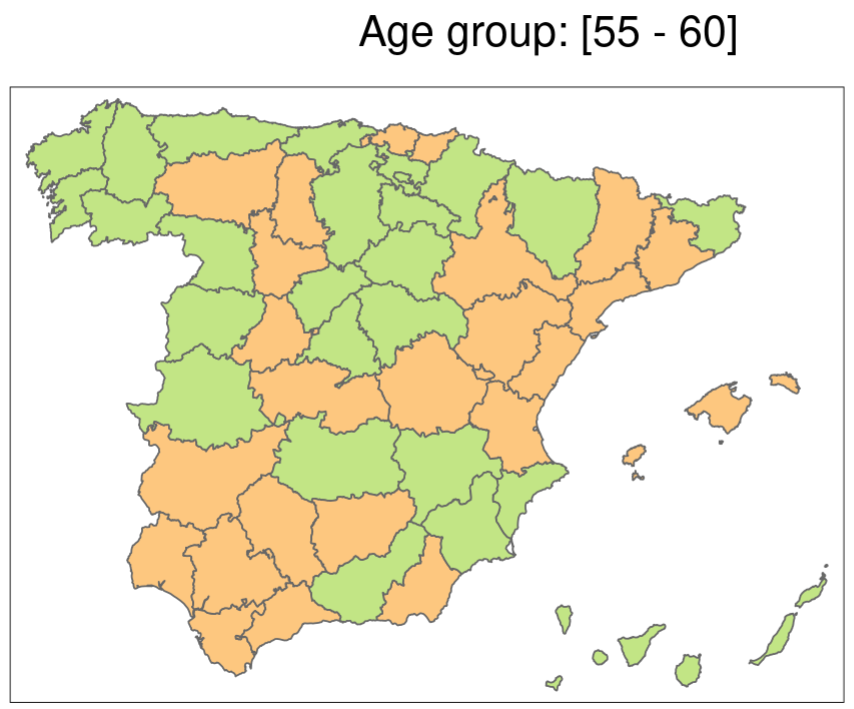}
\end{subfigure}%
\begin{subfigure}{0.3\textwidth}
  \centering
  \includegraphics[width=\linewidth]{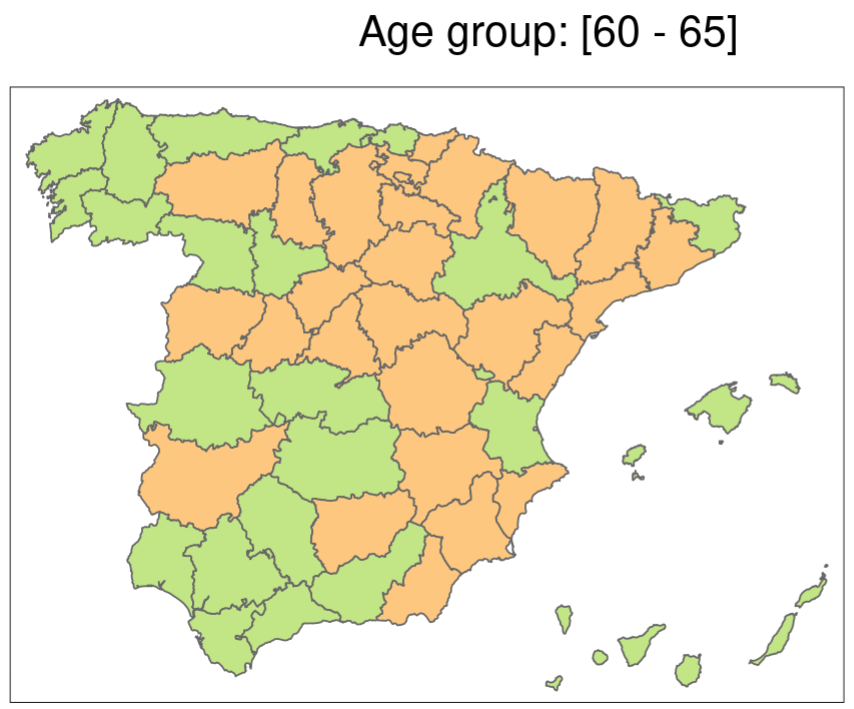}
\end{subfigure}\\
\begin{subfigure}{.3\textwidth}
  \centering
  \includegraphics[width=\linewidth]{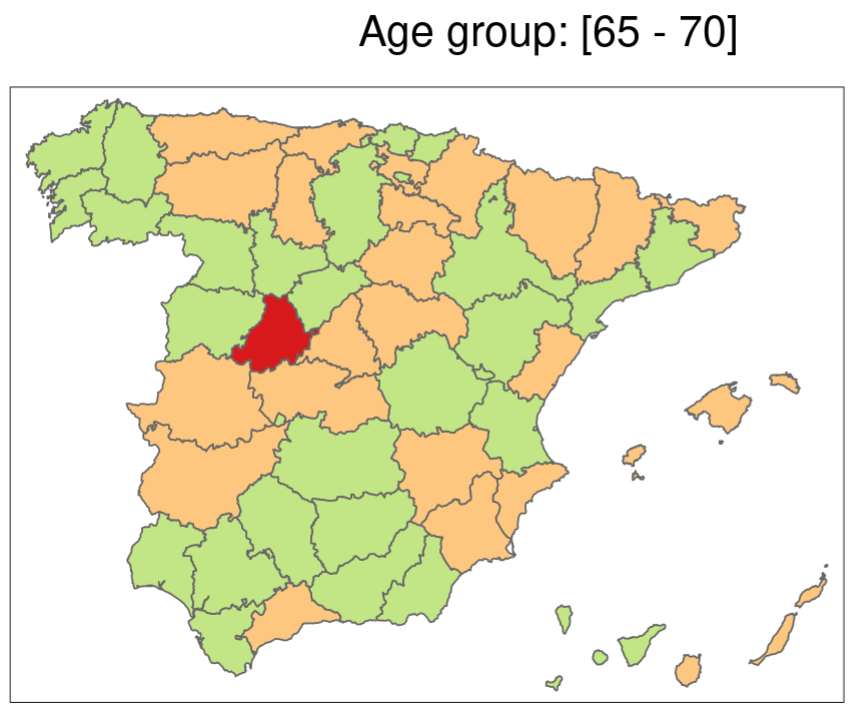}
\end{subfigure}%
\begin{subfigure}{.3\textwidth}
  \centering
  \includegraphics[width=\linewidth]{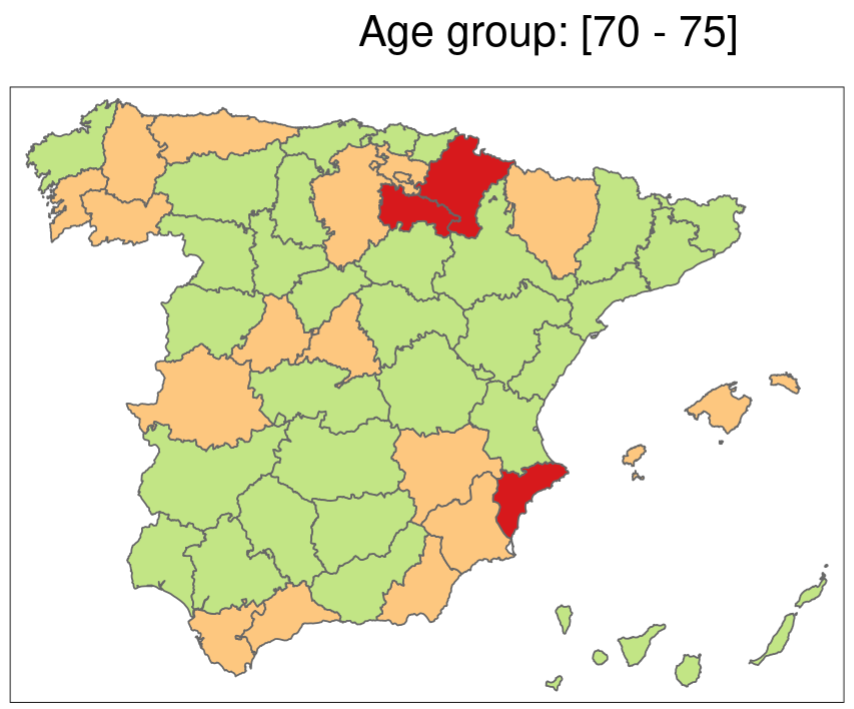}
\end{subfigure}\\
\begin{subfigure}{0.3\textwidth}
  \centering
  \includegraphics[width=\linewidth]{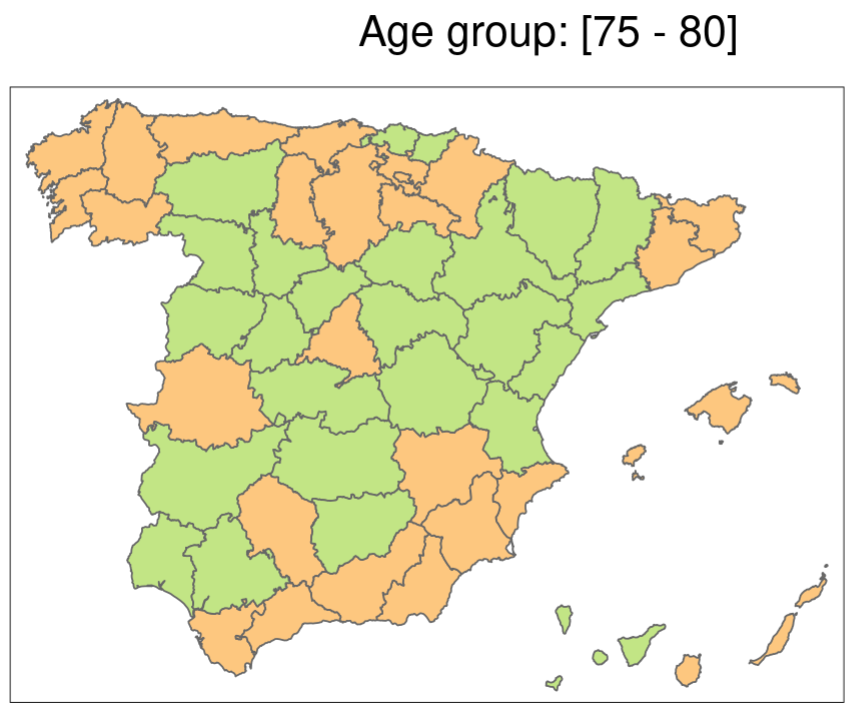}
\end{subfigure}%
\begin{subfigure}{.3\textwidth}
  \centering
  \includegraphics[width=\linewidth]{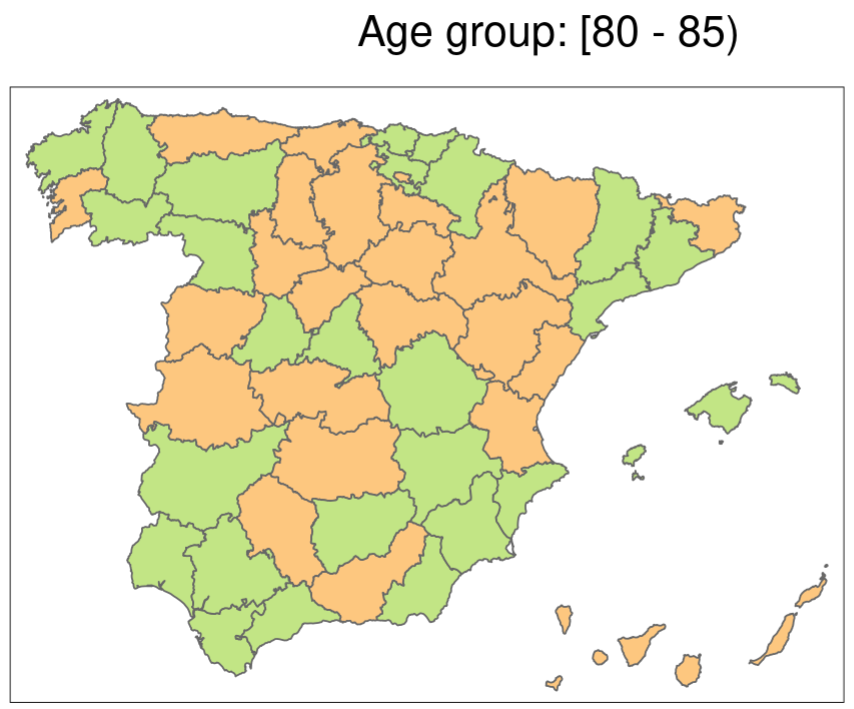}
\end{subfigure}\\
\begin{subfigure}{.3\textwidth}
  \centering
  \includegraphics[width=\linewidth]{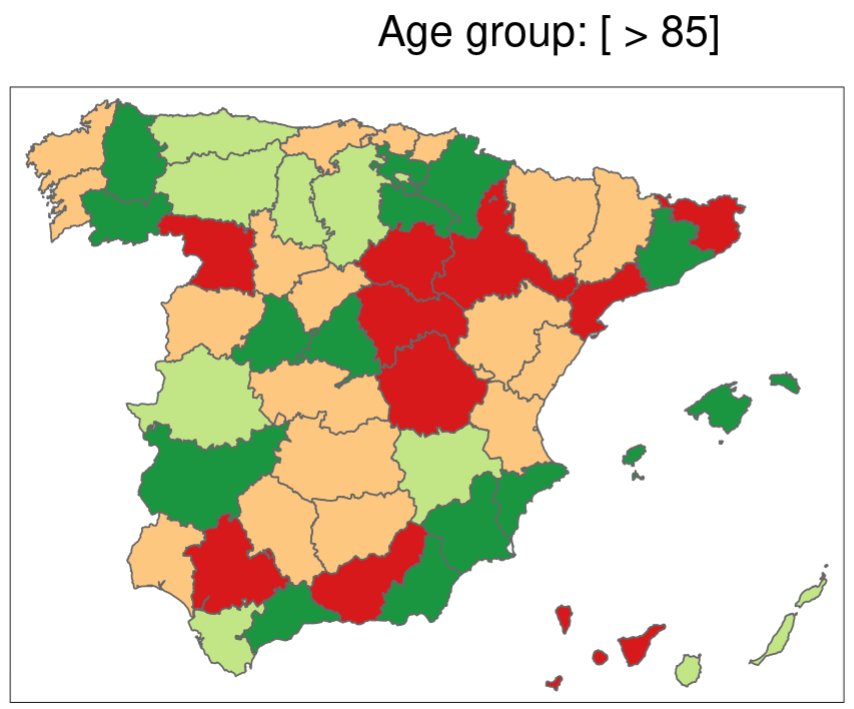}
\end{subfigure}%
\begin{subfigure}{.1\textwidth}
  \centering
  \includegraphics[width=\linewidth]{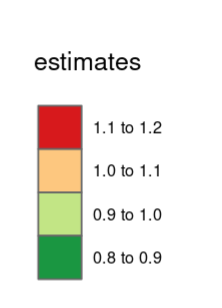}
\end{subfigure}
\caption{Space x age interaction (exponential scale) in prostate cancer mortality (regional effects for different age groups).}
\label{results:3}
\end{figure}

\begin{figure}
\centering
\begin{subfigure}{0.20\textwidth}
  \centering
  \includegraphics[width=\linewidth]{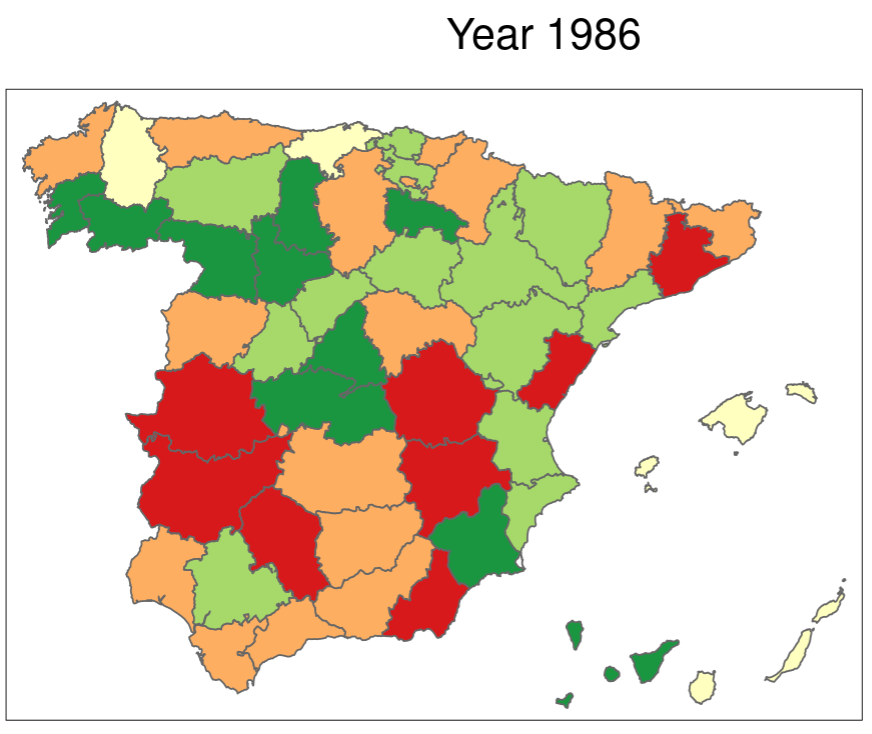}
\end{subfigure}%
\begin{subfigure}{.20\textwidth}
  \centering
  \includegraphics[width=\linewidth]{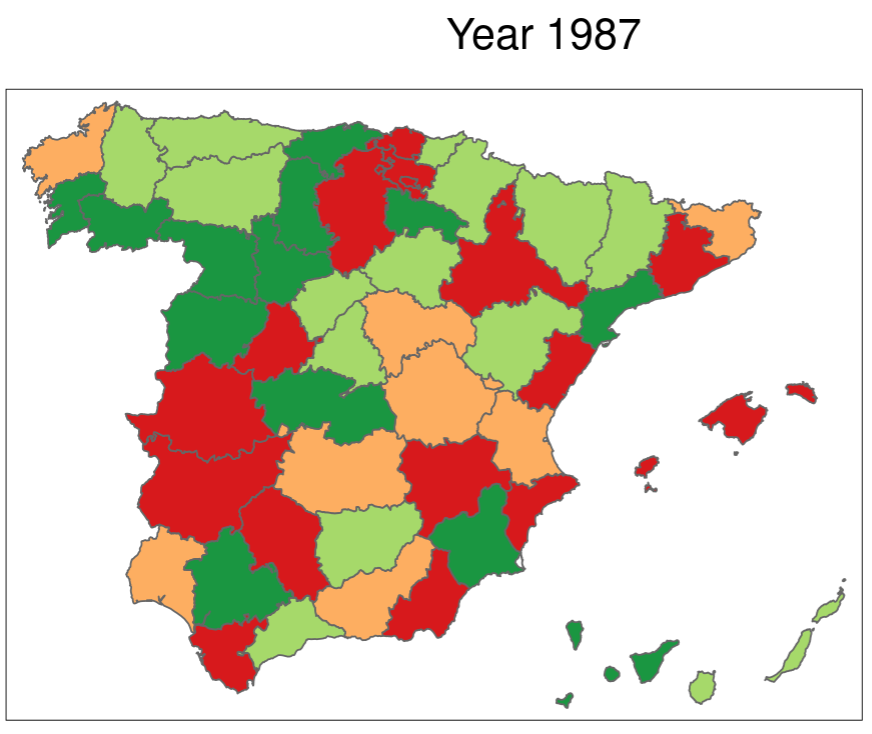}
\end{subfigure}%
\begin{subfigure}{.20\textwidth}
  \centering
  \includegraphics[width=\linewidth]{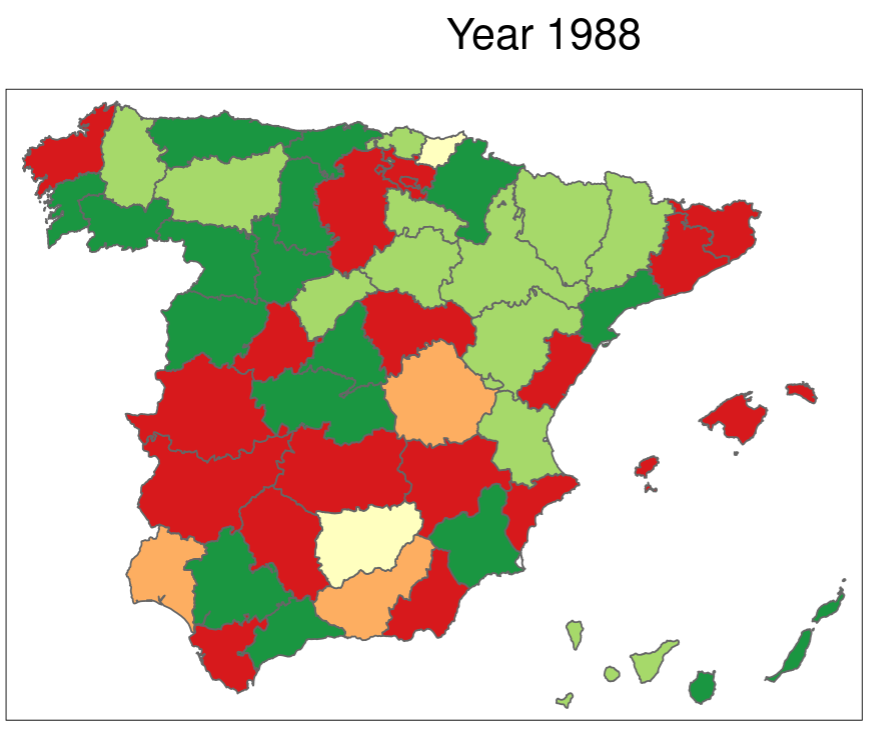}
\end{subfigure}%
\begin{subfigure}{0.20\textwidth}
  \centering
  \includegraphics[width=\linewidth]{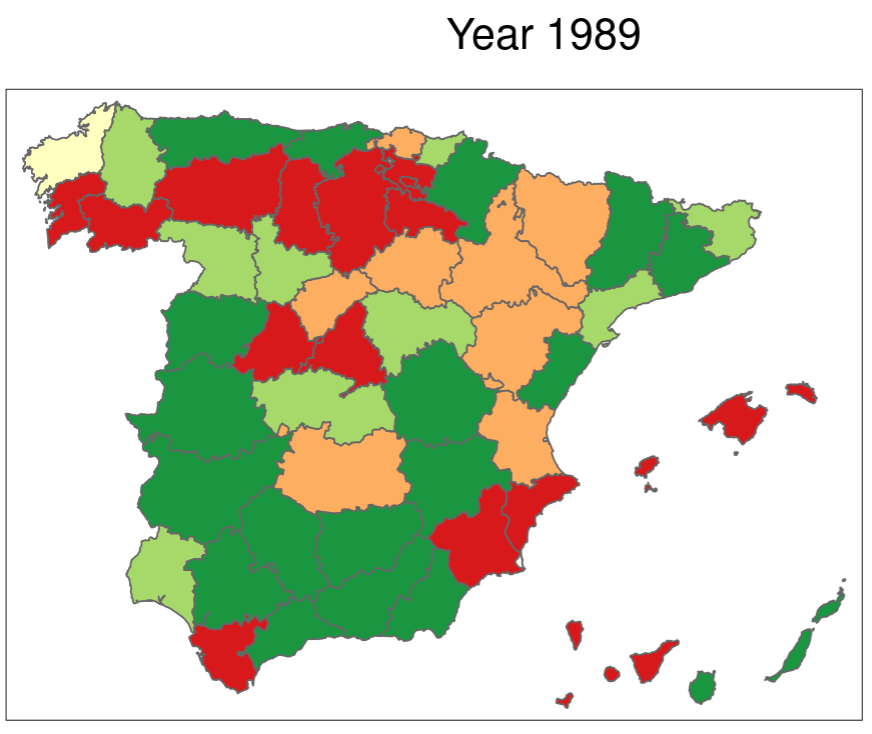}
\end{subfigure}\\
\begin{subfigure}{.20\textwidth}
  \centering
  \includegraphics[width=\linewidth]{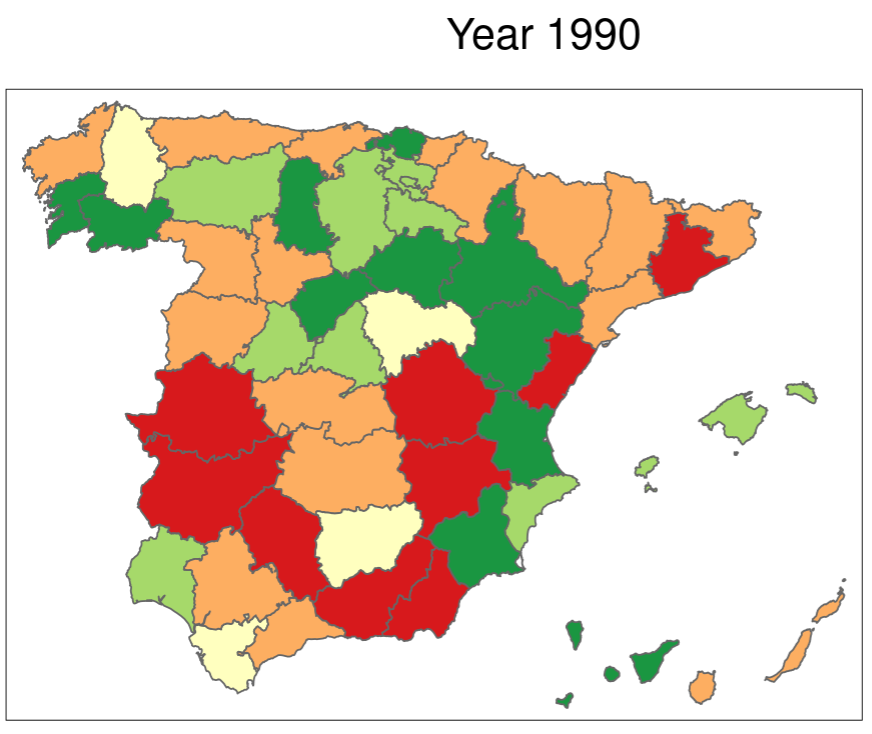}
\end{subfigure}%
\begin{subfigure}{.20\textwidth}
  \centering
  \includegraphics[width=\linewidth]{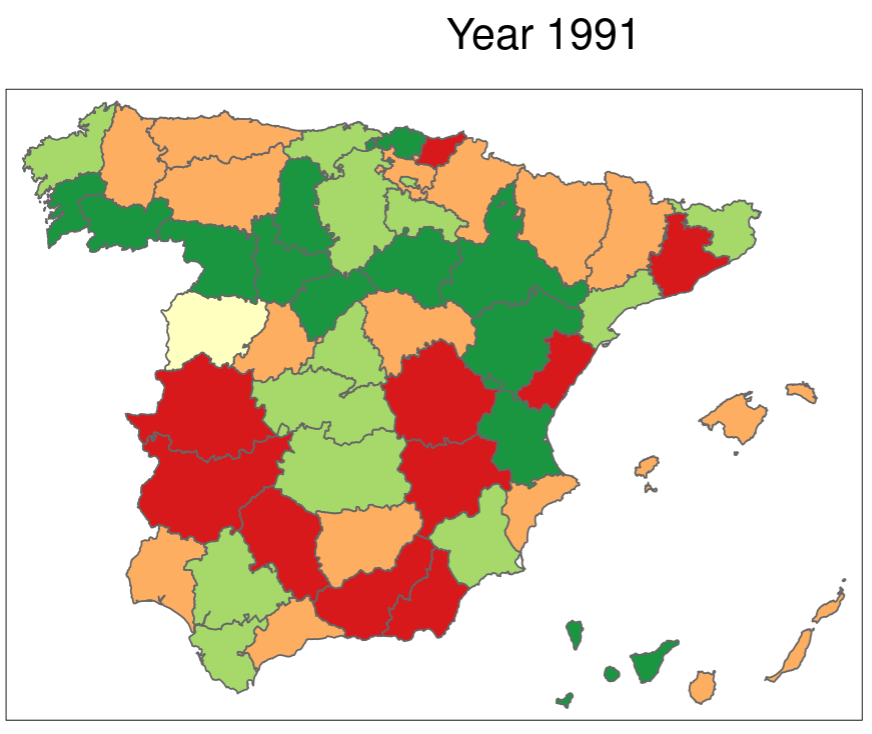}
\end{subfigure}%
\begin{subfigure}{0.20\textwidth}
  \centering
  \includegraphics[width=\linewidth]{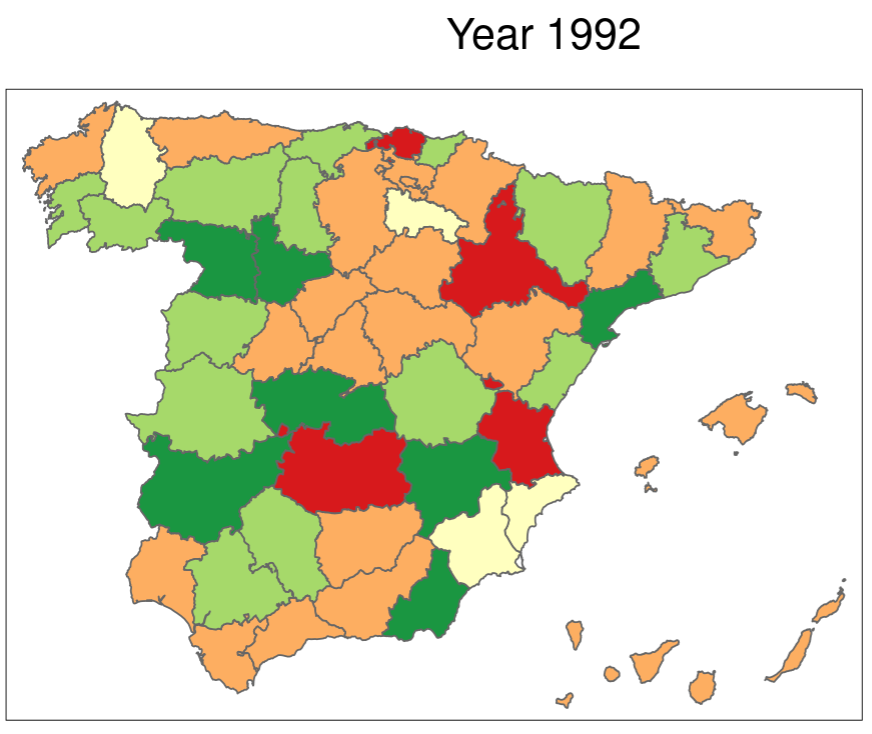}
\end{subfigure}%
\begin{subfigure}{.20\textwidth}
  \centering
  \includegraphics[width=\linewidth]{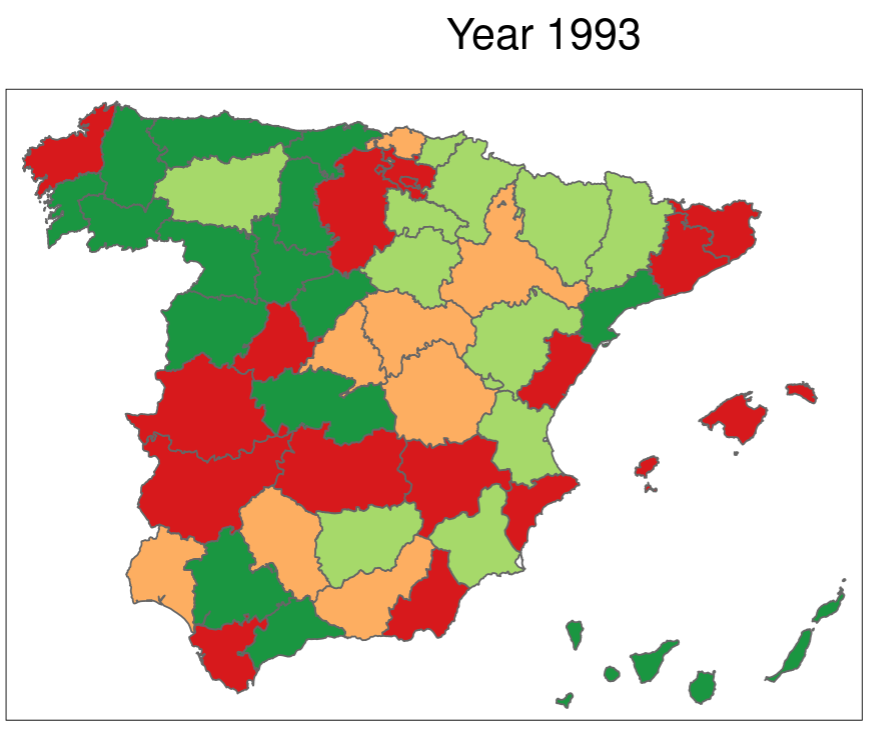}
\end{subfigure}\\
\begin{subfigure}{.20\textwidth}
  \centering
  \includegraphics[width=\linewidth]{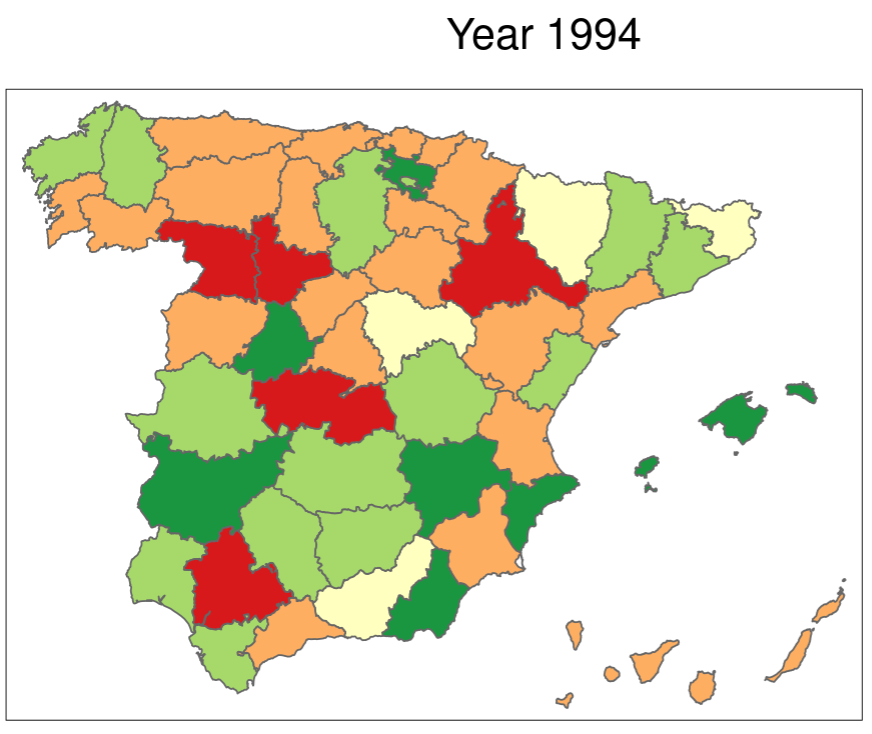}
\end{subfigure}%
\begin{subfigure}{0.20\textwidth}
  \centering
  \includegraphics[width=\linewidth]{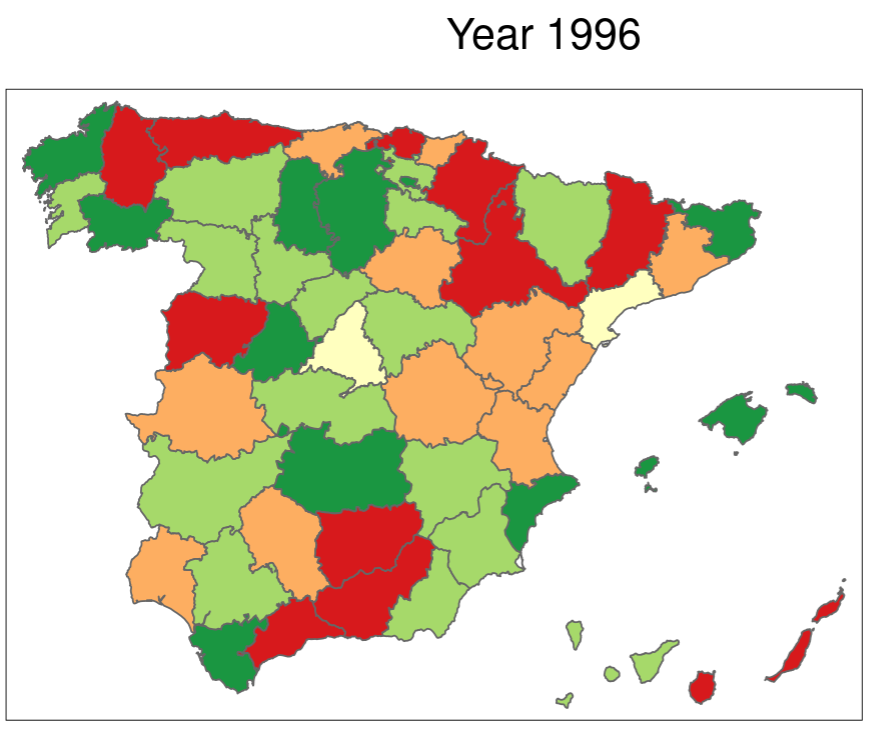}
\end{subfigure}%
\begin{subfigure}{.20\textwidth}
  \centering
  \includegraphics[width=\linewidth]{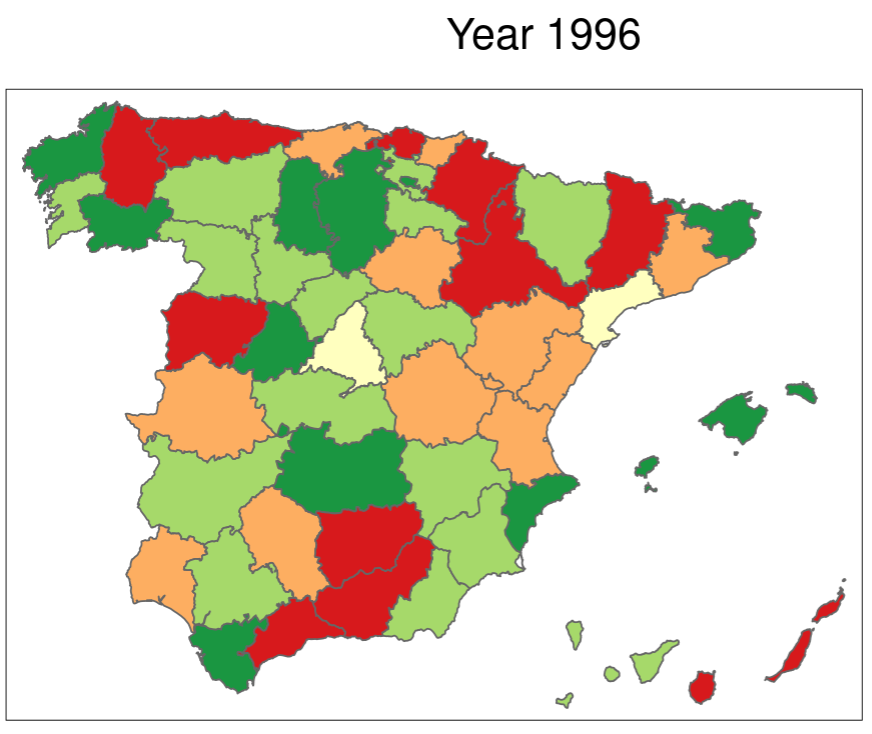}
\end{subfigure}%
\begin{subfigure}{.20\textwidth}
  \centering
  \includegraphics[width=\linewidth]{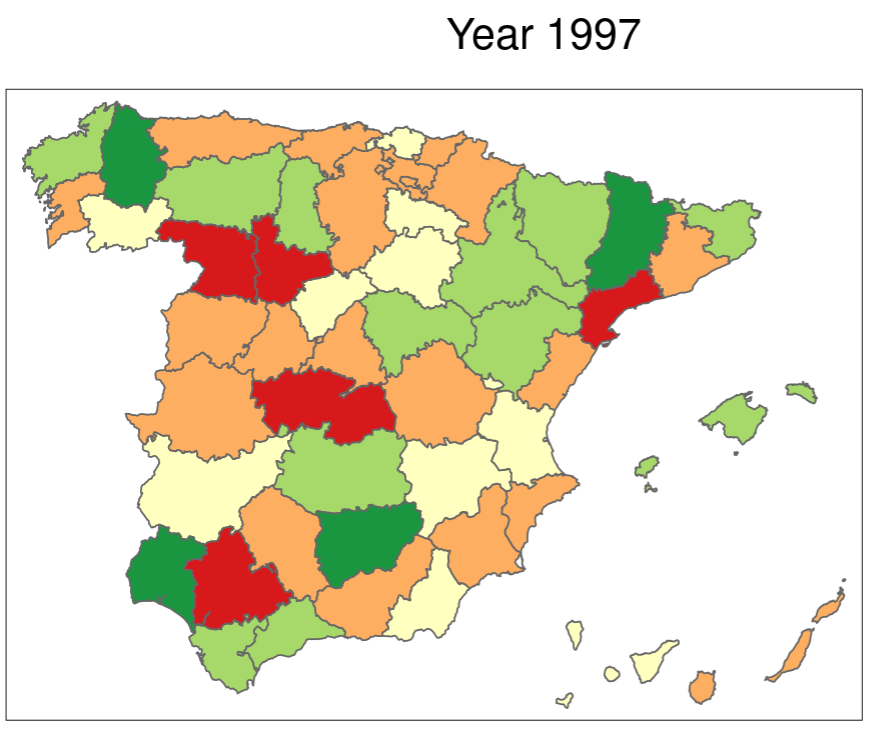}
\end{subfigure}\\
\begin{subfigure}{0.20\textwidth}
  \centering
  \includegraphics[width=\linewidth]{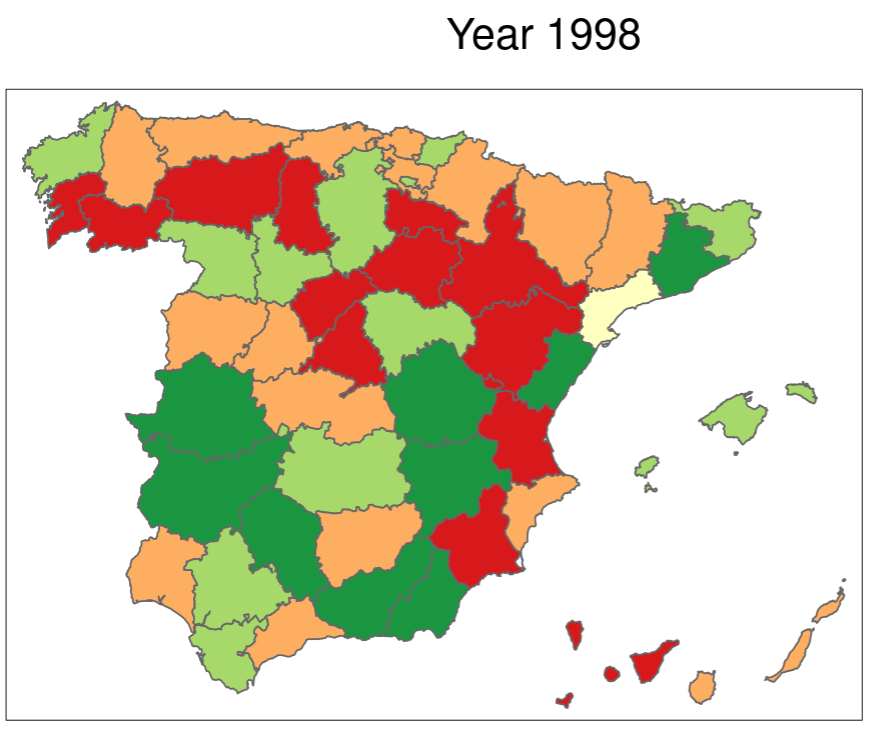}
\end{subfigure}%
\begin{subfigure}{.20\textwidth}
  \centering
  \includegraphics[width=\linewidth]{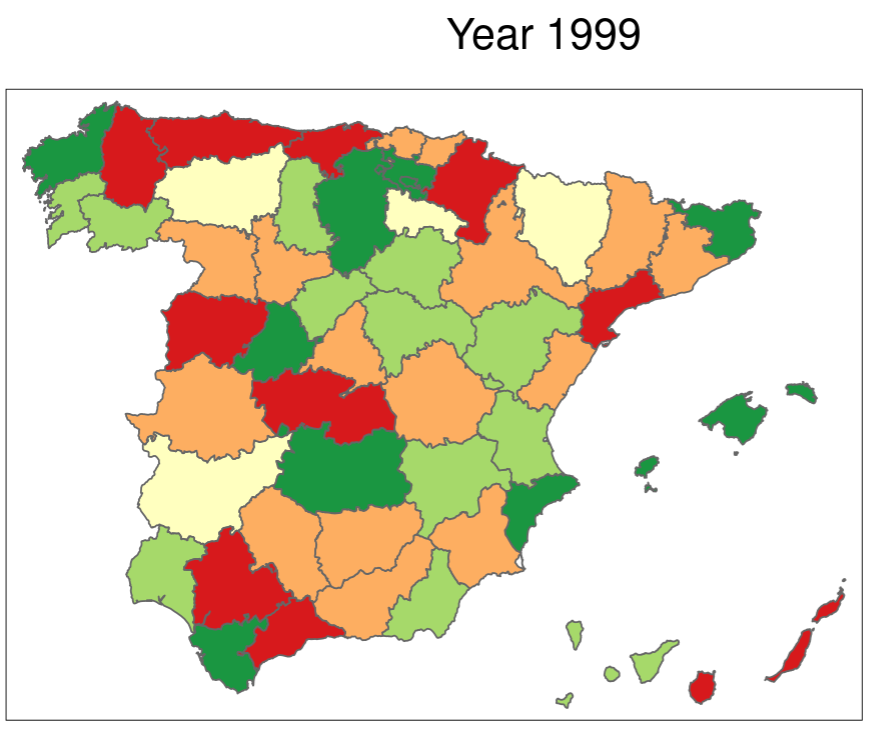}
\end{subfigure}%
\begin{subfigure}{.20\textwidth}
  \centering
  \includegraphics[width=\linewidth]{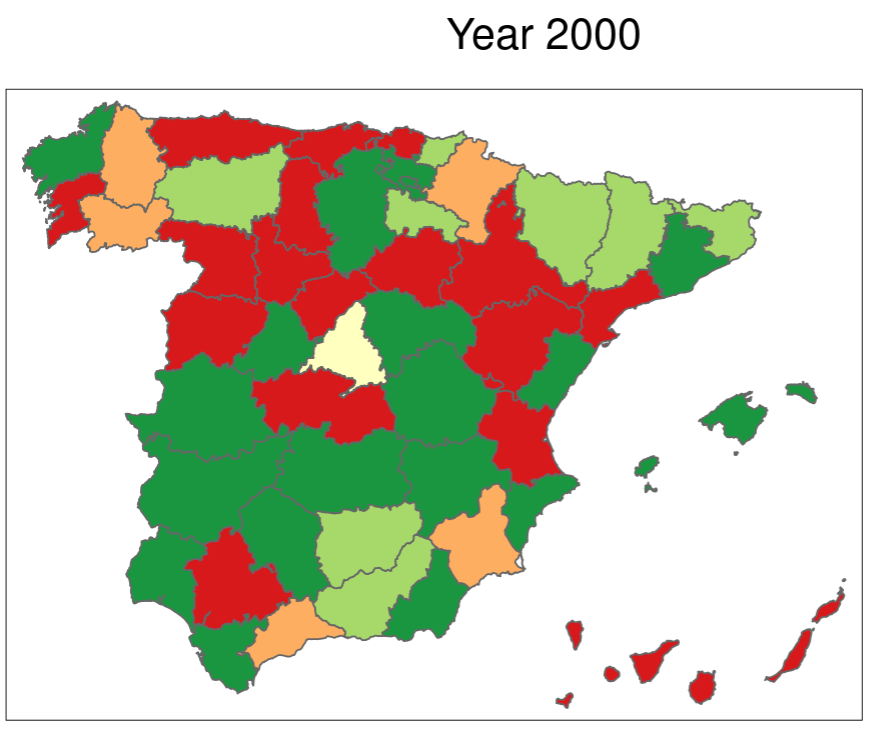}
\end{subfigure}%
\begin{subfigure}{0.20\textwidth}
  \centering
  \includegraphics[width=\linewidth]{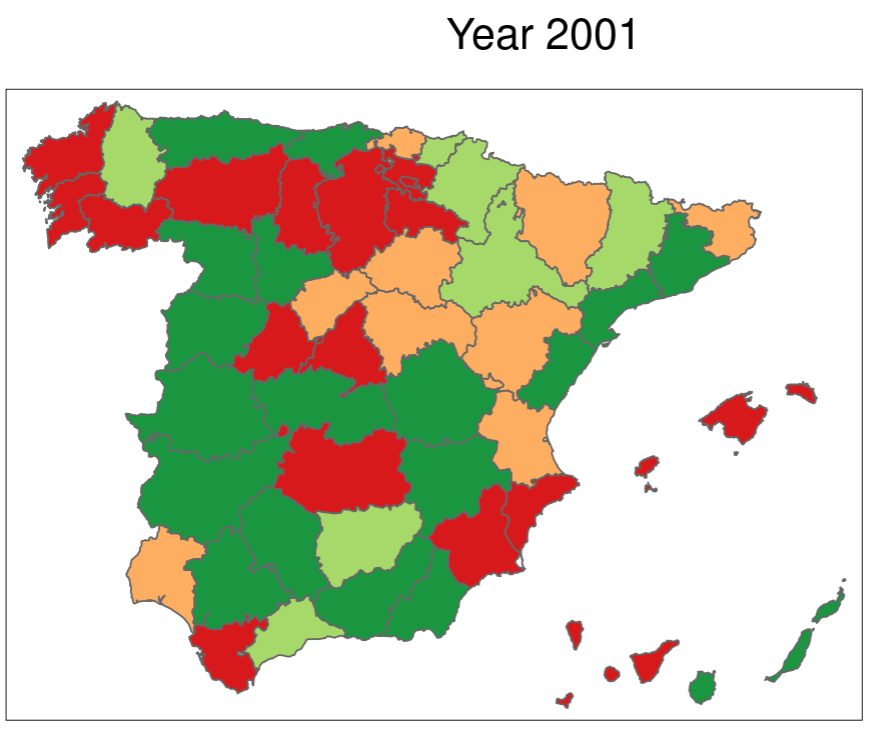}
\end{subfigure}\\
\begin{subfigure}{.20\textwidth}
  \centering
  \includegraphics[width=\linewidth]{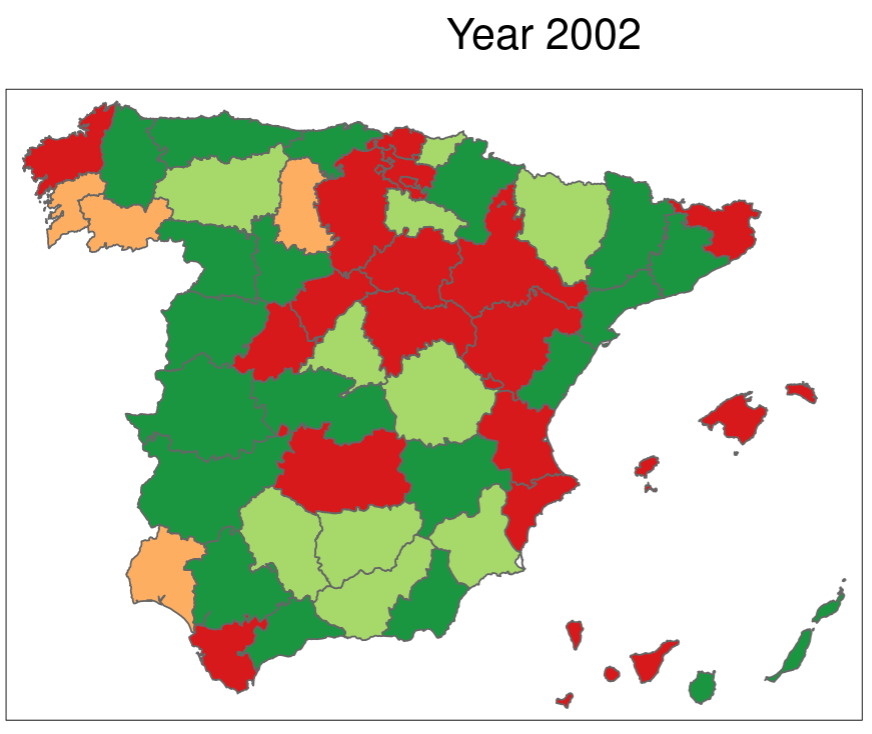}
\end{subfigure}%
\begin{subfigure}{.20\textwidth}
  \centering
  \includegraphics[width=\linewidth]{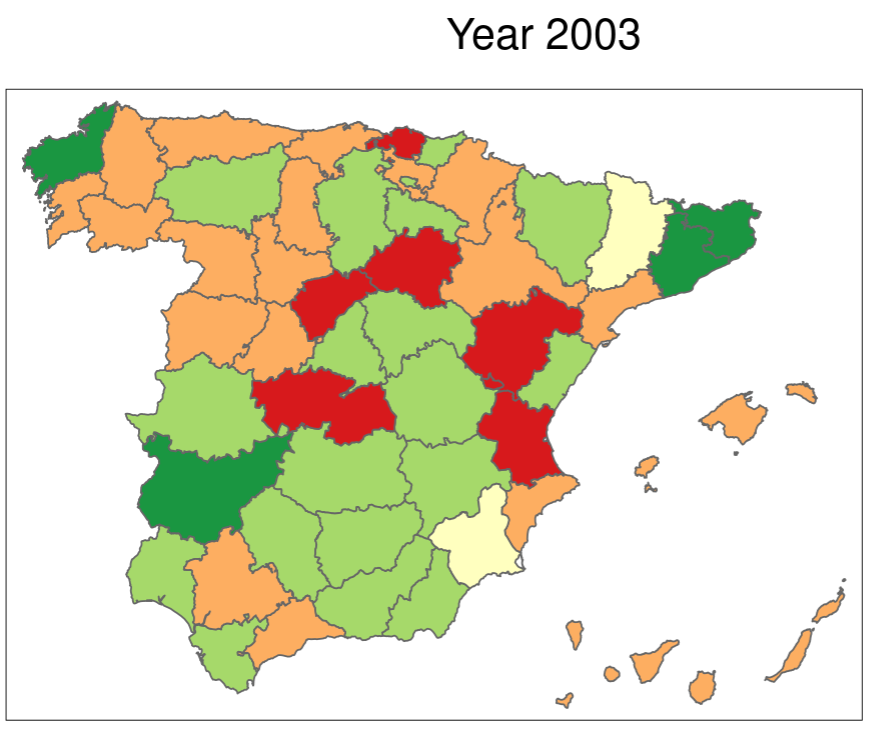}
\end{subfigure}%
\begin{subfigure}{0.20\textwidth}
  \centering
  \includegraphics[width=\linewidth]{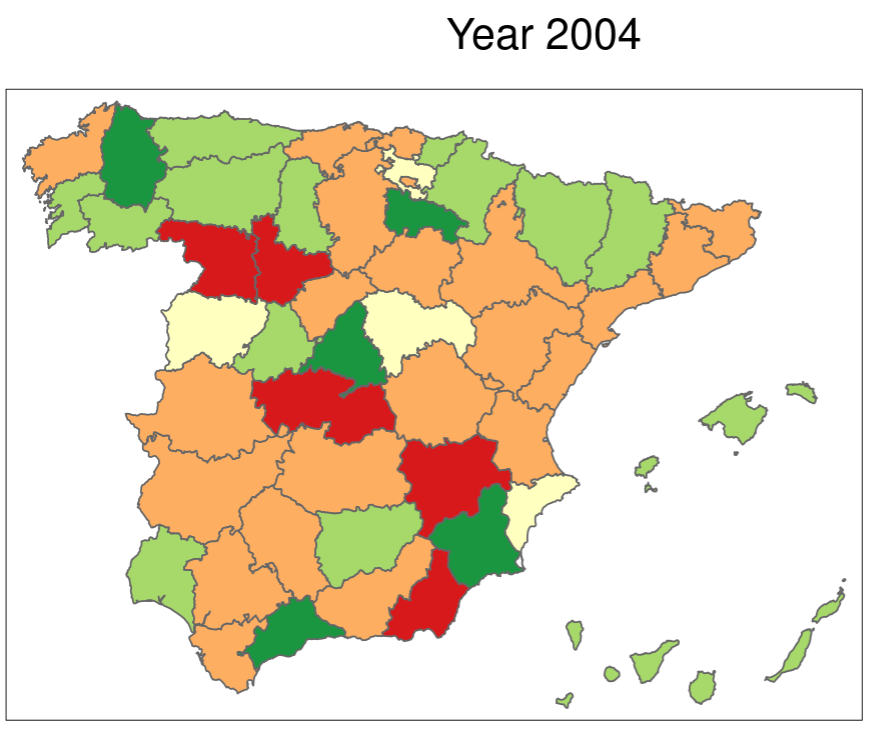}
\end{subfigure}%
\begin{subfigure}{.20\textwidth}
  \centering
  \includegraphics[width=\linewidth]{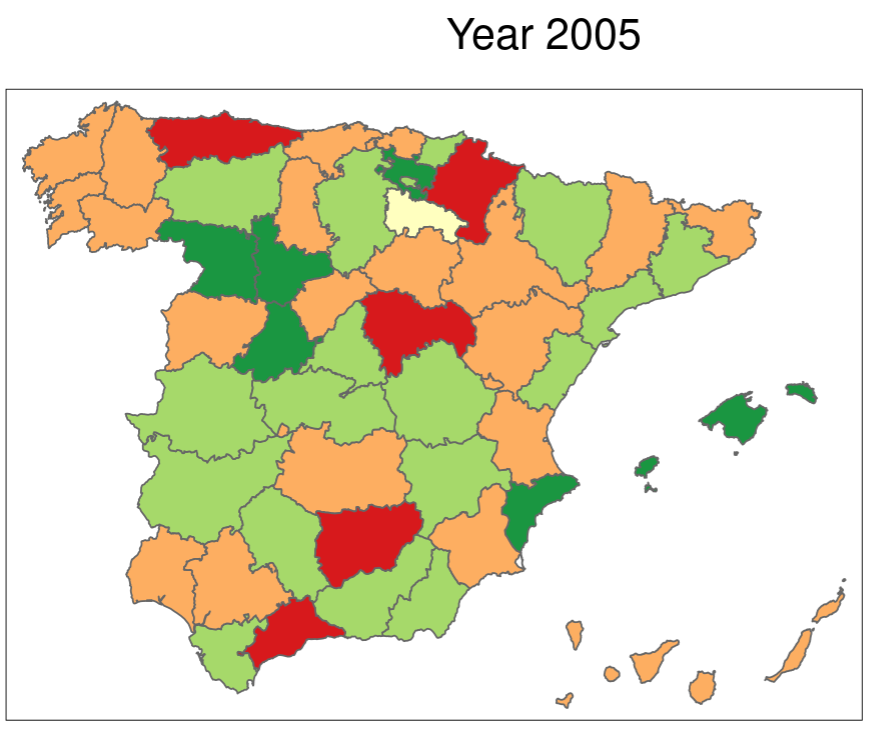}
\end{subfigure}\\
\begin{subfigure}{.20\textwidth}
  \centering
  \includegraphics[width=\linewidth]{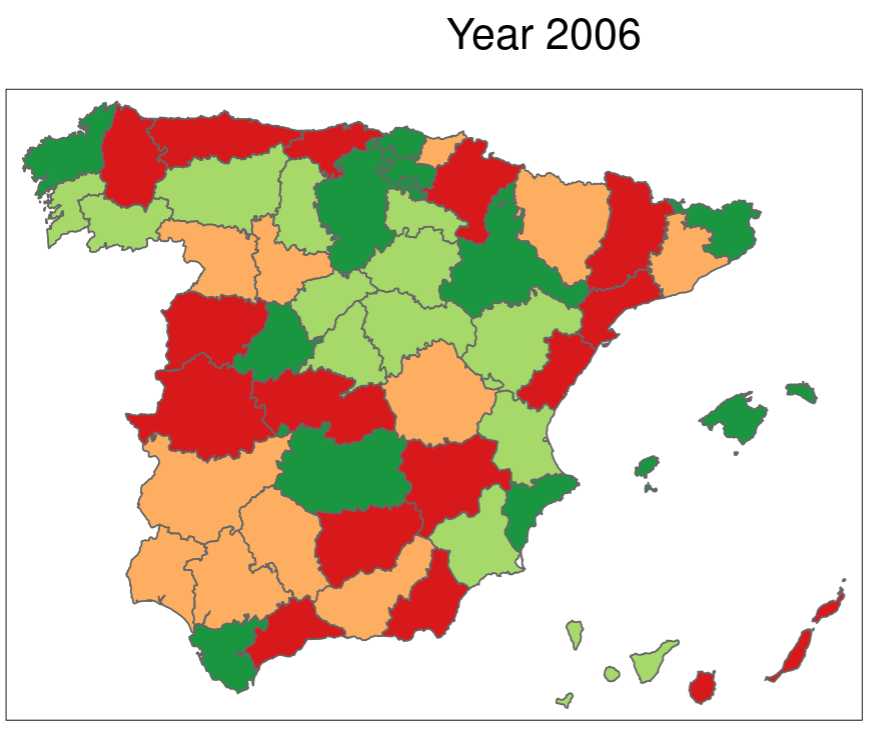}
\end{subfigure}%
\begin{subfigure}{.20\textwidth}
  \centering
  \includegraphics[width=\linewidth]{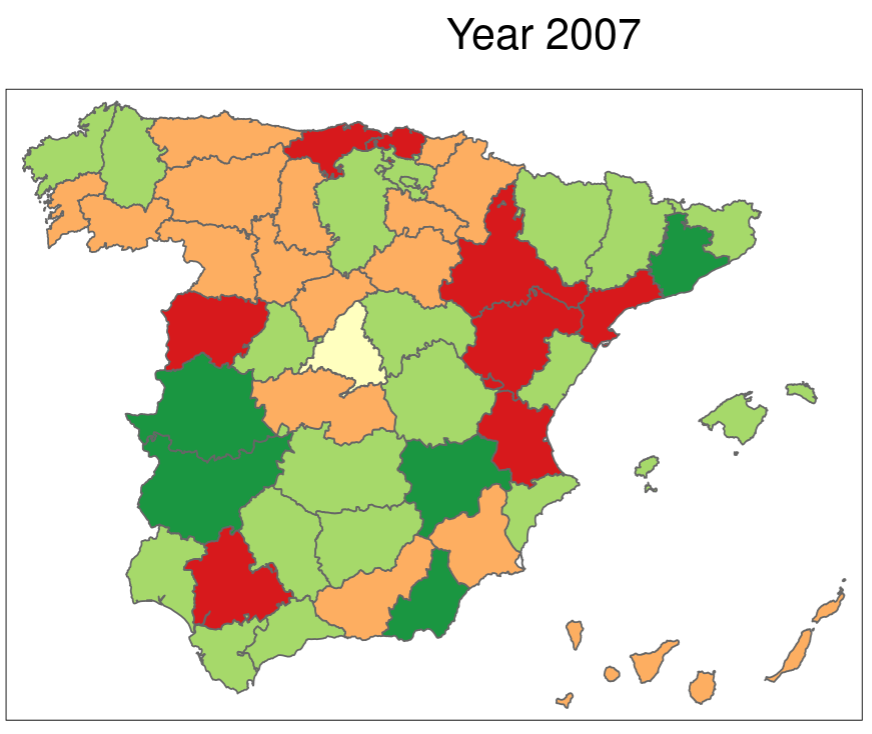}
\end{subfigure}%
\begin{subfigure}{0.20\textwidth}
  \centering
  \includegraphics[width=\linewidth]{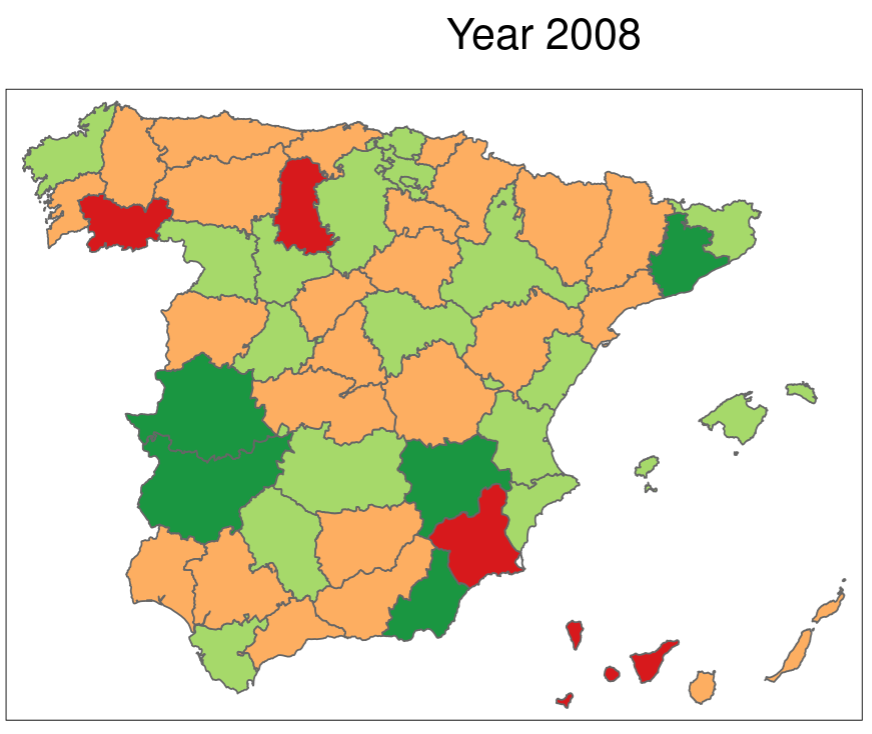}
\end{subfigure}%
\begin{subfigure}{.20\textwidth}
  \centering
  \includegraphics[width=\linewidth]{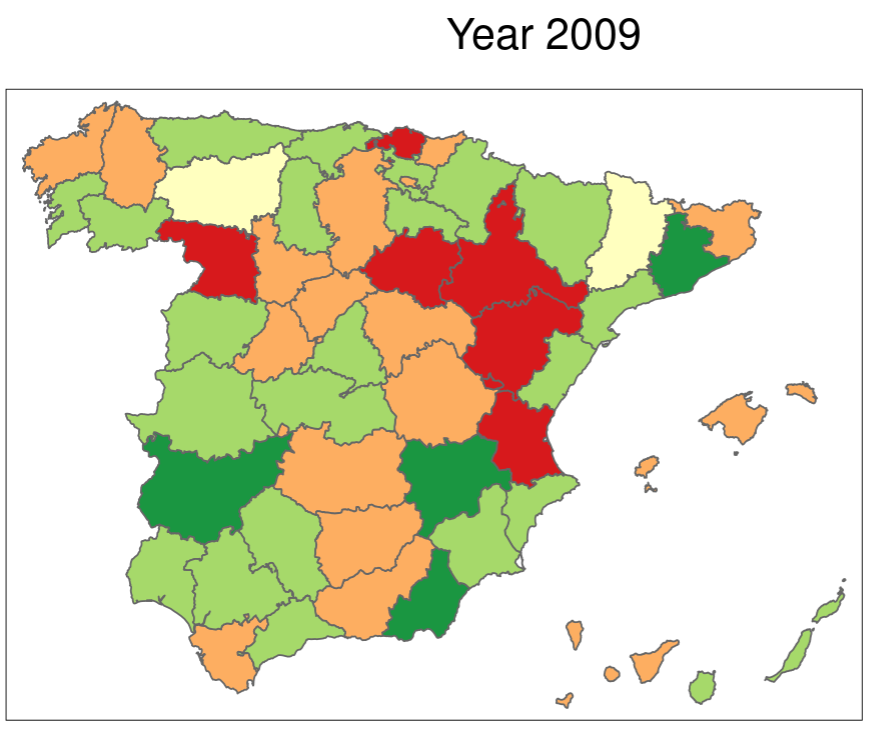}
\end{subfigure}\\
\begin{subfigure}{.20\textwidth}
    \centering
  \includegraphics[width=\linewidth]{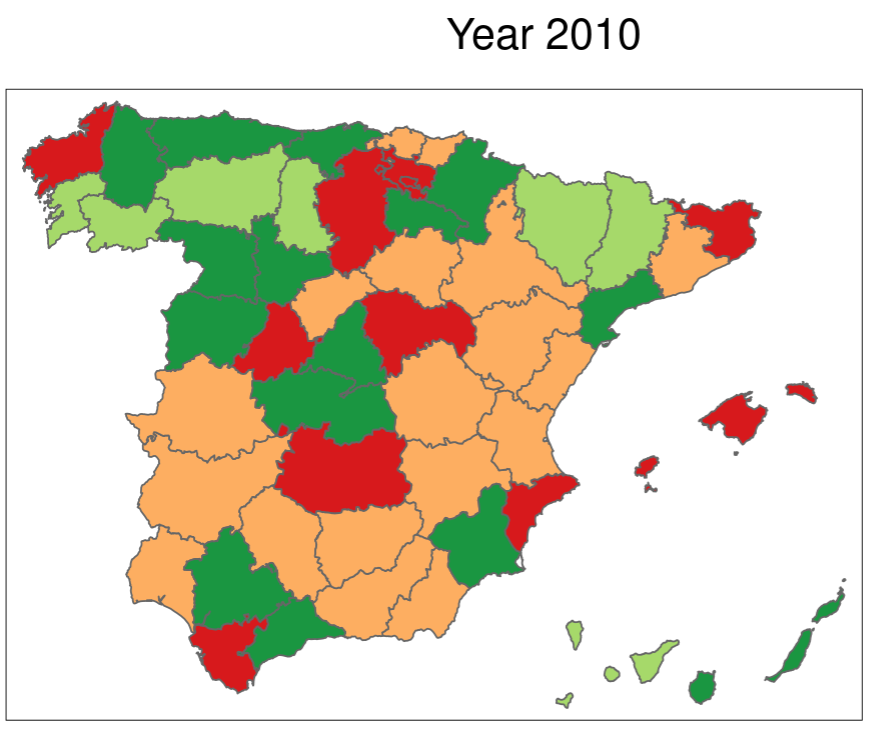}
\end{subfigure}%
\begin{subfigure}{.60\textwidth}
    \raggedright
  \includegraphics[scale=0.3]{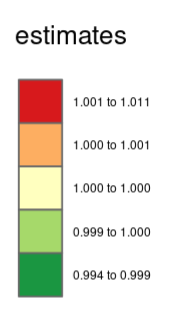}
\end{subfigure}%
\begin{subfigure}{.1\textwidth}
\end{subfigure}%
\begin{subfigure}{.1\textwidth}
\end{subfigure}
\caption{Space x time interaction (exponential scale) in prostate cancer mortality (regional effects for different years).}
\label{results:4}
\end{figure}


\newpage

\begin{thebibliography}{9}

    \bibitem{Rue2009ApproximateBI}Rue, H., Martino, S. \& Chopin, N. Approximate Bayesian inference for latent Gaussian models by using integrated nested Laplace approximations. {\em Journal Of The Royal Statistical Society: Series B (statistical Methodology)}. \textbf{71}, 319-392 (2009)
    \bibitem{Chapman2007UsingO}Chapman, B., Jost, G. \& Van Der Pas, R. Using OpenMP: portable shared memory parallel programming. (MIT press,2007)
    \bibitem{Pacheco1996ParallelPW}Pacheco, P. Parallel programming with MPI. (Morgan Kaufmann,1997)
    \bibitem{papaspiliopoulos2020scalable}Papaspiliopoulos, O., Roberts, G. \& Zanella, G. Scalable inference for crossed random effects models. {\em Biometrika}. \textbf{107}, 25-40 (2020)
    
    \bibitem{Goicoa2018InSD}Goicoa, T., Adin, A., Ugarte, M. \& Hodges, J. In spatio-temporal disease mapping models, identifiability constraints affect PQL and INLA results. {\em Stochastic Environmental Research And Risk Assessment}. \textbf{32} pp. 749-770 (2018)
    
    \bibitem{Stringer2020ApproximateBI}Stringer, A., Brown, P. \& Stafford, J. Approximate Bayesian inference for case-crossover models. {\em Biometrics}. \textbf{77}, 785-795 (2021)
    \bibitem{Shalf2020TheFO}Shalf, J. The future of computing beyond Moore’s law. {\em Philosophical Transactions Of The Royal Society A}. \textbf{378}, 20190061 (2020)
    \bibitem{KnorrHeld2000BayesianMO}Knorr-Held, L. Bayesian modelling of inseparable space-time variation in disease risk.. {\em Statistics In Medicine}. \textbf{19 17-18} pp.  2555-67  (2000)
    
    \bibitem{goicoa2018spatio}Goicoa, T., Adin, A., Ugarte, M. \& Hodges, J. In spatio-temporal disease mapping models, identifiability constraints affect PQL and INLA results. {\em Stochastic Environmental Research And Risk Assessment}. \textbf{32} pp. 749-770 (2018)
    \bibitem{Schrdle2011SpatiotemporalDM}Schrödle, B. \& Held, L. Spatio-temporal disease mapping using INLA. {\em Environmetrics}. \textbf{22}, 725-734 (2011)
    \bibitem{OrozcoAcosta2021ScalableBM}Orozco-Acosta, E., Adin, A. \& Ugarte, M. Scalable Bayesian modelling for smoothing disease risks in large spatial data sets using INLA. {\em Spatial Statistics}. \textbf{41} pp. 100496 (2021)
    
    \bibitem{aanes2023faster}Aanes, F. \& Storvik, G. Faster estimation of the Knorr-Held Type IV space-time model. {\em ArXiv Preprint ArXiv:2304.11851}. (2023)
    \bibitem{Riedel1992ASI}Van Loan, C. \& Golub, G. Matrix computations (Johns Hopkins studies in mathematical sciences). {\em Matrix Computations}. \textbf{5} (1996)
    
    \bibitem{GaedkeMerzhuser2022ParallelizedIN}Gaedke-Merzhäuser, L., Niekerk, J., Schenk, O. \& Rue, H. Parallelized integrated nested Laplace approximations for fast Bayesian inference. {\em Statistics And Computing}. \textbf{33}, 25 (2023)
    \bibitem{Martins2013BayesianCW}Martins, T., Simpson, D., Lindgren, F. \& Rue, H. Bayesian computing with INLA: new features. {\em Computational Statistics \& Data Analysis}. \textbf{67} pp. 68-83 (2013)
    \bibitem{nocedal2006numerical}Nocedal, J. \& Wright, S. Numerical Optimization.  Springer Verlag. {\em New York}. (2006)
    
    \bibitem{fattah2022smart}Abdul Fattah, E., Van Niekerk, J. \& Rue, H. Smart Gradient-An adaptive technique for improving gradient estimation. {\em Foundations Of Data Science}. \textbf{4}, 123-136 (2022)
    \bibitem{Box1951OnTE}Box, G. \& Wilson, K. On the experimental attainment of optimum conditions. {\em Breakthroughs In Statistics: Methodology And Distribution}. pp. 270-310 (1992)
    \bibitem{Niekerk2022ANA}Van Niekerk, J., Krainski, E., Rustand, D. \& Rue, H. A new avenue for Bayesian inference with INLA. {\em Computational Statistics \& Data Analysis}. pp. 107692 (2023)
    \bibitem{blaze2}Iglberger, K. Blaze C++ Linear Algebra Library. (https://bitbucket.org/blaze-lib,2012)
    \bibitem{Rustand2022FastAF}Rustand, D., Niekerk, J., Krainski, E., Rue, H. \& Proust-Lima, C. Fast and flexible inference approach for joint models of multivariate longitudinal and survival data using Integrated Nested Laplace Approximations. {\em ArXiv Preprint ArXiv:2203.06256}. (2022)
    \bibitem{Hadri2015OverviewOT}Hadri, B., Kortas, S., Feki, S., Khurram, R. \& Newby, G. Overview of the KAUST’s Cray X40 system–Shaheen II. {\em Proceedings Of The 2015 Cray User Group}. (2015)
    \bibitem{goicoa2016age}Goicoa, T., Ugarte, M., Etxeberria, J. \& Militino, A. Age–space–time CAR models in Bayesian disease mapping. {\em Statistics In Medicine}. \textbf{35}, 2391-2405 (2016)
    \bibitem{Simpson2014PenalisingMC}Simpson, D., Rue, H., Martins, T., Riebler, A. \& Sørbye, S. Penalising Model Component Complexity: A Principled, Practical Approach to Constructing Priors. {\em Statistical Science}. \textbf{32} pp. 1-28 (2014)
\bibitem{davies1947}Davies, O.L. Statistical methods in research and production. Oliver and Boyd. {\em London} (1947).
    
\end{thebibliography}
\end{document}